\PassOptionsToPackage{pdfpagelabels=false}{hyperref}
\documentclass[useAMS,usenatbib]{mnras}
\pdfoutput=1
\usepackage[pdftex]{graphicx}
\usepackage{url}
\usepackage{amsmath}
\usepackage{natbib}
\usepackage{geometry}
\usepackage{txfonts}
\usepackage{times}

\def \MSUN{{\rm M}_{\odot}}
\def \RVIR{\rm{R}_{\rm{200c}}}

\title[The CGM of TNG50 Milky Way-like galaxies]{The Circumgalactic Medium of Milky Way-like Galaxies in the TNG50 Simulation -- I: Halo Gas Properties and the Role of SMBH Feedback}
\author[R. Ramesh, D. Nelson \& A. Pillepich]{Rahul Ramesh$^{1}$\thanks{E-mail: rahul.ramesh@stud.uni-heidelberg.de}, Dylan Nelson$^{1}$ and Annalisa Pillepich$^{2}$\\
$^{1}$ Universität Heidelberg, Zentrum für Astronomie, Institut für theoretische Astrophysik, Albert-Ueberle-Str. 2, 69120 Heidelberg, Germany\\
$^{2}$ Max-Planck-Institut f\"{u}r Astronomie, K\"{o}nigstuhl 17, 69117 Heidelberg, Germany\\
}

\date{}
\pubyear{}

\begin{document}

\maketitle

\begin{abstract}
We analyze the physical properties of gas in the circumgalactic medium (CGM) of 132 Milky Way (MW)-like central galaxies at $z=0$ from the cosmological magneto-hydrodynamical simulation TNG50, part of the IllustrisTNG project. The properties and abundance of CGM gas across the sample are diverse, and the fractional budgets of different phases (cold, warm, and hot), as well as neutral HI mass and metal mass, vary considerably. Over our stellar mass range of $10^{10.5} < M_\star / \rm{M}_\odot < 10^{10.9}$, radial profiles of gas physical properties from $0.15 < R/\rm{R_{\rm 200c}} < 1.0$ reveal great CGM structural complexity, with significant variations both at fixed distance around individual galaxies, and across different galaxies. CGM gas is multi-phase: the distributions of density, temperature and entropy are all multimodal, while metallicity and thermal pressure distributions are unimodal; all are broad. We present predictions for magnetic fields in MW-like halos: a median field strength of $|B|\sim\,1\mu$G in the inner halo decreases rapidly at larger distance, while magnetic pressure dominates over thermal pressure only within $\sim0.2 \times \RVIR$. Virial temperature gas at $\sim 10^6\,$K coexists with a sub-dominant cool, $< 10^5\,$K component in approximate pressure equilibrium. Finally, the physical properties of the CGM are tightly connected to the galactic star formation rate, in turn dependent on feedback from supermassive black holes (SMBHs). In TNG50, we find that energy from SMBH-driven kinetic winds generates high-velocity outflows ($\gtrsim 500-2000$ km s$^{-1}$), heats gas to super-virial temperatures ($> 10^{6.5-7}$ K), and regulates the net balance of inflows versus outflows in otherwise quasi-static gaseous halos. 
\end{abstract}


\section{Introduction}

The circumgalactic medium (CGM) is believed to play a vital role in the evolution of galaxies. It is the region through which gas accreting from larger scales and from the intergalactic medium must flow. It harbours gas previously ejected from the galaxy due to feedback processes, as well as ‘recycling’ or fountain flows that circulate in the halo. Each of these gaseous flows will have distinct kinematics, producing complex and evolving dynamics in the CGM. In addition, gaseous halos are thought to have complex spatial and phase structure, being both multi-scale and multi-phase. In particular, gas co-exists in both cool $\sim 10^4\,$K and volume-filling hot components, the latter at or near the halo virial temperature (see \citealt{tumlinson2017} for a recent review, and \citealt{fielding2020} for a recent cross-simulation comparison of CGM properties).

Our own Milky Way galaxy is embedded in a multi-phase and multi-scale CGM. Observations of the Milky Way halo have detected cool gas via H$\alpha$ emission at T $\lesssim 10^{4.5}\,$K \citep{putman2003}, as well as through the HI 21-cm line at T $\lesssim 10^{4}\,$K \citep{peek2011}. At the same time, observations of X-ray lines such as OVI, OVII and OVIII suggest the presence of a warm-hot phase at T $> 10^{5}\,$K, with XMM-Newton \citep{henley2010, miller2015, miller2016}; Chandra \citep{gupta2012, fang2013}; and HaloSat \citep{kaaret2020}. In addition, absorption features from HST/COS spectroscopy towards nearby stars at varying distances ($\sim 4 - 15$ kpc) reveals multiple cool to warm metal ionisation species, including CaII, FeII, SiIV, CIV \citep{werk2019}.

The CGM of the Milky Way exhibits a prominent, large-scale feature. Namely, the Fermi \citep{su2010} and and eROSITA \citep{predehl2020} bubbles: these are two bipolar cocoon-like structures emerging from the Galactic center, extending below and above the Galaxy's disk up to $10-14$ kpc, and emitting in gamma-ray and X-ray, respectively. Recently, \cite{ashley2022} observed the existence of multi-phase gas clouds embedded within the two Fermi bubbles on either side of the galactic plane, adding support to the picture of a multi-phase structure of the Milky Way CGM.

The Milky Way CGM is highly multi-scale. It is believed to host small structures of the order of $\lesssim 1-10$ kpc in the form of neutral hydrogen rich, high-velocity clouds \citep{wakker1997}. The resulting kinematics have also been observed to be complex. While a non-negligible amount of cold gas is both inflowing and outflowing \citep{clark2022}, at least a fraction of hot gas is believed to rotate around the center of the halo, with the amount of angular momentum comparable to that of the stellar disk \citep{kluck2016}. Furthermore, the gas within the Fermi/eROSITA bubbles has been measured to have radial velocities ranging from 330 km s$^{-1}$ \citep[for HI clouds close to the disk:][]{diteodoro2018} to $900-1300$ km s$^{-1}$ for UV clouds at higher altitudes \citep{fox2015, bordoloi2017, ashley2020}.

This picture of a multi-phase, multi-scale CGM has also gained support from cosmological hydrodynamical simulations. For instance, using cosmological simulations, \cite{keres2009}, \cite{fernandez2012} and \cite{juong2012} showed the existence of cold gas clouds in the CGM of Milky Way-like (MW-like) galaxies, and \cite{sokolowska2016} pointed out the existence of a hot diffuse halo around similar galaxies. With the HESTIA simulations, \cite{damle2022} noted the presence of both the cold (H I and Si III) and warm-hot phases (O VI, O VII, and O VIII) in the CGM of halos resembling the MW-M31 system. With the FOGGIE simulations, a set of simulations that preferentially increase resolution in the CGM, \cite{peeples2019} and \cite{corlies2020} noted the existence of a multi-phase structure in the CGM of Milky-Way like halos, as did \cite{hani2019} with the Auriga suite of simulations, a set of high-resolution zoom-in simulations of Milky-Way like galaxies. \cite{hafen2019} used a sample of FIRE-2 simulations to show that gas in the CGM has multiple origins, leading to a diversity of gas properties. Collectively, these studies reveal how the Milky Way gaseous halo assembles to its redshift zero state.

While cosmological simulations are important tools to study properties of realistic CGM realizations, they are computationally expensive. As an alternative, various semi-analytic models have been developed. For instance, \cite{stern2016} used measurements of ionic column densities to develop a model describing cool gas in the CGM, deriving a mean cool gas density radial density profile. This model was extended to include steady-state cooling flows as a single-parameter family of solutions \citep{stern2019}. On the other end of the temperature spectrum, \cite{faerman2017} used observations of OVI, OVII and OVIII lines to build a model for the warm-hot phases of the CGM, thereby suggesting that hot galactic coronae can contain significant amounts of gas, possibly accounting for the previously `missing baryons' \citep[see also][]{faerman2020}. \cite{voit2017} proposed a global model to describe the condensation and precipitation of gas in the CGM, arguing that precipitation can proceed in two modes: either by gas `uplift' in galactic outflows for $\rm{t_{cool}/t_{ff}} \lesssim 10$, or due to the slope of the entropy gradient \citep[see also][]{sharma2012}. Using observations of SiIV, OVI and CIV, \cite{qu2019} built two dimensional models for the distribution of warm gas (T $\sim 10^5$K), suggesting that the disk components of SiIV and OVI have similar density profiles, while the distribution and kinematics of CIV is similar to SiIV \citep{qu2020, qu2022}.

While early observations focused on the nearest CGM -- the gaseous halo of our own Milky Way -- we now have detailed observations of other halos, out to $z \sim 6$ \citep{leclercq2017}. Extragalactic halo samples increase the scope of observational probes, studying features and physics not dominant in MW-like systems. Observations of neutral hydrogen \citep{cai2017, prochaska2017, chen2018}, MgII \citep{zhu2014} and Ly$\alpha$ (MUSE: \citealt{leclercq2017, leclercq2020}; see also \citealt{byrohl2021} for comparisons with the TNG50 simulation) have shown the presence of non-negligible amounts of cold gas in the CGM around galaxies across cosmic epochs. For example, \cite{hennawi2015} reported that the most massive structures in the distant universe ($z \sim 2$; ${M_{\rm{halo}} \sim 10^{13}\,\rm{M}_\odot}$) can have as much as $10^{11}\,\rm{M_\odot}$ of cold gas. Focusing on ionized gas phases, observations of OVI with HST/COS \citep{stocke2013, werk2016} have detected warm gas, while hot X-ray emitting gas in extended halos around other galaxies has also been detected (with e.g. Chandra \citealt{bogdan.2013a, bogdan2013b, goulding2016}; XMM-Newton: \citealt{Li.2016}; CASBaH: \citealt{burchett2019}; and eROSITA via stacking: \citealt{comparat2022, chadayammuri2022}). The CGM of the Andromeda galaxy has been studied in detail with multiple independent quasar sightlines, due to its proximity \citep{lehner2020}. Finally, the warm-hot phase of extragalactic gaseous halos is also accessible through the Sunyaev-Zel'dovich effect \citep[tSZ;][]{degraff2019,lim2021}. 

In addition to the observable tracers of different gas phases mentioned above, non-thermal components including magnetic fields and cosmic rays may play an important role with respect to CGM gas. Using the Auriga suite of simulations, \cite{pakmor2020} showed that the CGM of MW-like halos are highly magnetized ($|B| \gtrsim 0.1~\mu$G) before $z \sim 1$, mainly due to outflows that transport magnetized gas outwards away from the galaxy. \cite{voort2021} showed, with one Auriga halo, that these magnetic fields can alter the physical properties of CGM gas, including inflow velocities, temperatures, and pressures. Magnetic pressure may be important in gaseous halos, including its ability to stabilise dense, cold clouds surrounded by a hot ambient medium \citep{nelson2020}. Pressure support due to cosmic rays may also stabilize cold gas clouds \citep{butsky2018, ji2020, huang2022}, which can change characteristics of cold-phase gas \citep{butsky2020}, and can make the CGM less thermally supported and thus cooler \citep{hopkins2020}. The formation, and survival, of such clouds remains, however, an open theoretical question \citep[e.g.][]{mccourt2015, liZ2020, gronke2020, dutta2022}. 

Finally, the CGM does not exist in isolation. It is influenced by external perturbations, such as galaxy mergers and gas accretion, as well as by feedback-driven outflows from the central galaxy. Recent studies based on current large-scale cosmological hydrodynamical galaxy simulations -- EAGLE and IllustrisTNG -- have quantified the importance of supermassive black hole (SMBH) feedback on properties of the CGM \citep{davies2020, oppenheimer2020, zinger2020, truong2020, truong2021a}. The injection of energy by SMBHs drives gas out of galaxies at high velocities, increases the average entropy of the halo gas, and lengthens typical cooling times, not only preventing future star formation but also directly impacting the thermodynamics of the gaseous halo. Feedback episodes can cause gas to deviate from hydrostatic equilibrium \citep{oppenheimer2018}, ultimately altering the morphology \citep{kauffmann2019} and kinematics \citep{nelson2015} of gas in the CGM.

Furthermore, SMBH feedback contributes to the enrichment of the metal content of the CGM \citep{sanchez2019}, and to the abundance of various ionization species of these metals \citep{segers2017}. For example, using the IllustrisTNG simulations, \cite{nelson2018b} showed that galaxies with more massive SMBHs, which are thus preferentially quenched and red, have a lower abundance (i.e. column density) of OVI in their CGM. The IllustrisTNG simulations have also shown that low-accretion state energy input by the central SMBH can create X-ray eROSITA-like bubbles in disk galaxies \citep{pillepich2021} and anisotropic X-ray signatures \citep{truong2021} in the CGM gas. For less massive galaxies, winds driven by stellar feedback including supernovae are responsible for driving gas out of the galaxy and into the CGM \citep{fielding2017,li2020}.

Despite mounting observational data and ever-improving models, a complete picture of the physical properties of the CGM of galaxies remains remote. For example, the multi-phase nature of the CGM remains a nebulous concept. We do not know the CGM mass and volume fractions in different gas phases, as a function of galaxy mass and redshift, on average, and as a function of morphological and star-formation states, and from galaxy to galaxy. Crucially, it is not currently empirically known whether all disk-like galaxies similar to our Milky Way are surrounded by a hot, X-ray emitting gaseous halo. We have no comprehensive view of the distribution, structure, or kinematics of the bulk of the CGM surrounding galaxies less massive than brightest cluster/group galaxies.

In this work, we provide insights on the properties of the CGM of MW-like galaxies by using the outcome of the TNG50 simulation of the IllustrisTNG project. We quantify the amount and distribution of the CGM gas, and its key physical properties including temperature, density, entropy, pressure, metallicity, and magnetic field strength, as well as observable signatures including HI, various metal ions, and X-ray emission. We investigate the connection between CGM properties and galaxy global properties, including star formation and recent feedback from central SMBHs. Importantly, we characterize and contrast the gaseous halos of a large sample of TNG50 MW-like galaxies, numbering 132 in total. This allows us to quantify the galaxy population-wide median behaviour, as well as study the amount, and origin, of the scatter in trends as galaxies evolve in unique ways.

This paper is organised as follows: Section~\ref{methods} describes the simulations, the galaxy sample, and our definitions and analysis/methodological choices. We present our main results in Section~\ref{results}, and summarize our conclusions in Section~\ref{summary}.


\section{Methods}\label{methods}

\subsection{The TNG50 simulation}\label{TNG}

In this paper, we present results from the TNG50 simulation \citep{pillepich2019, nelson2019} of the IllustrisTNG suite \citep[hereafter TNG: ][]{marinacci2018, naiman2018, nelson2018, pillepich2018b,springel2018}, a set of cosmological magneto-hydrodynamical simulations. IllustrisTNG is \textit{The Next Generation} of the original Illustris simulation \citep{vogelsberger2014b, vogelsberger2014a, genel2014, sijacki2015}, which was one of the first cosmological simulations to reproduce a reasonably realistic, diverse population of galaxies as in the observed universe. IllustrisTNG was run using the \textsc{AREPO} moving-mesh code \citep{springel2010}, with a new underlying physics model including magneto-hydrodynamics \citep{pakmor2014} and modified feedback processes \citep{weinberger2017, pillepich2018a}. 

TNG50 employs a (periodic) box of side length $\sim50$ comoving Mpc (cMpc). Despite the relatively large box, the resolution is comparable to many zoom-in simulations, with an average baryonic (dark matter) mass resolution of $\sim 8 \times 10^4~\rm{M_\odot}~ (4 \times 10^5~\rm{M_\odot})$. This unique combination of resolution and volume allows one to study highly resolved structures over a large sample size of galaxies.

The TNG simulations evolve from $z=127$ to present day ($z=0$), and their output is stored at 100 points in time between $z=20$ and $z=0$. As detailed in \cite{pillepich2018a}, TNG employs an extensive and well-validated galaxy formation model. We briefly mention here the key features of the model related to star formation.  Stars form stochastically from gas cells above a threshold of neutral hydrogen density $n_{\rm{H}} \gtrsim 0.1~\rm{cm^{-3}}$, according to the two-phase ISM model of \cite{springel2003}. Gas colder than 10$^4$\,K is not explicitly modeled in the TNG simulations. Metal enrichment due to stellar evolution and supernovae is also included, as well as the transfer of supernovae feedback energy in the form of a kinetic, galactic-scale decoupled wind \citep{springel2003}.

In the TNG model, SMBHs are `seeded' when the mass of a friends-of-friends halo exceeds $\sim7 \times 10^{10}~\rm{M_\odot}$. Starting with a mass of $\sim1 \times 10^{6}~\rm{M_\odot}$,  SMBHs grow over time by merging with others and by accreting nearby gas, with the accretion rate set as the minimum of the Bondi and Eddington accretion rates. As described in detail in \cite{weinberger2017}, the accretion rate then determines the mode in which active galactic nuclei (AGN) feedback energy is injected into the surroundings. At high-accretion rates, SMBHs continuously deposit thermal energy into neighbouring gas cells. At low-accretion rates, they instead impart kinetic energy in discrete events after a certain amount of energy is accumulated. In both modes, energy injection is isotropic at the injection scale. The former is the dominant state for less massive SMBHs, while the latter dominates for high-mass SMBHs, with the transition occurring roughly at $M\rm{_{BH}} \sim 10^8 \rm{M_\odot}$ \citep{weinberger2017}. Additional details on the effective functioning of the SMBH feedback in TNG and the ensuing flows can be found in \citet{pillepich2021}.  

TNG adopts a cosmology consistent with the Planck 2015 analysis \citep{planck2016}, with: $\Omega_\Lambda = 0.6911$, $\Omega_{\rm m} = 0.3089$, $\Omega_{\rm b} = 0.0486$ and $h = 0.6764$ [100 km s$^{-1}$ Mpc$^{-1}$].

\subsection{The TNG50 Milky Way-like galaxy sample}

Among the thousands of galaxies realized within TNG50 at $z=0$, in this paper we focus on a sample of Milky Way-like systems.

To this aim, we adopt the MW/M31 selection and sample as presented in \textcolor{blue}{Pillepich et al. in prep}, and first used in \cite{engler2021} and \cite{pillepich2021}. At $z=0$, this sample is based on galaxies (1) having a stellar mass, measured within a 3D aperture of 30kpc, in the range $10^{10.5}\,\rm{M}_\odot$ to $10^{11.2}\,\rm{M}_\odot$, (2) being disky, either based on a constraint on the minor-to-major axis ratio of the stellar mass distribution ($s < 0.45$) or through a visual inspection by-eye, of stellar light maps (see \textcolor{blue}{Pillepich et al. in prep} for details), (3) not being in overly massive halos ($M_{\rm{200,c}} < 10^{13}~\rm{M}_\odot$), and (4) being reasonably isolated, with no other galaxy having $M_\star > 10^{10.5}~\rm{M}_\odot$ within a distance of $500$\,kpc.

In this work, to select MW-like galaxies only, we further restrict the mass range to $10^{10.5} < M_\star / \rm{M}_\odot < 10^{10.9}$. These criteria return a diverse sample of 138 galaxies, of which we consider only those that are the central galaxies of their underlying dark-matter and gaseous halo. We hence have a sample of 132 MW-like central galaxies, with total halo masses M$_{\rm{200,c}}$ between $10^{11.7} \rm{M}_\odot - 10^{12.5} \,\rm{M}_\odot$, and virial radii R$_{\rm{200,c}}$ spanning $\sim 150-300$ kpc. These are all the central galaxies of their underlying dark-matter and gaseous halo. Most of these galaxies have a stellar bar (\textcolor{blue}{Pillepich et al. in prep}), exhibit eROSITA-like X-ray bubbles in their CGM \citep{pillepich2021}, and are surrounded by a number of classical satellites consistent with the observations of the Milky Way and similar nearby galaxies \citep{engler2021}. Moreover, they are the result of a variety of merger and assembly histories, including several with gas-rich major mergers in the last 5 billion years \citep{sotillo2022}. As a reuslt, they exhibit a diversity of stellar bulge, disk, and halo properties, as well as star formation histories. Throughout the rest of the paper, we refer to this sample as `TNG50 MW-like galaxies'.

\subsection{Physical definitions}\label{definitions}

While the term circumgalactic medium (CGM) is widely understood as the region between the galaxy and the inter-galactic medium (IGM), there is no consensus for the definition of the CGM. In this paper, unless otherwise stated, we define the CGM to be the region bounded by $[0.15, 1.0] \times \RVIR$ of the corresponding halo. Our analysis always considers all gas in the simulation volume, and is not restricted to gas cells which belong to the corresponding friends-of-friends halo. To focus on the CGM of the central galaxy itself, we always exclude all gas that is gravitationally bound to satellite galaxies according to the \textsc{SUBFIND} algorithm \citep{springel2001}.

Unless otherwise stated, we adopt the following definitions:

\begin{enumerate}
    \item Temperature of star forming gas: the TNG ISM model employs a sub-grid pressurization model for star forming gas \citep{springel2003}. Due to this, the temperature of such gas is `effective' and not directly physical. We therefore always set the temperature of star forming gas to $10^3$ K, its cold-phase value.
    \item Stellar mass of the galaxy: total mass of all stellar particles within an aperture of 30 physical kpc.
    \item Star formation rate (SFR): rather than summing up the instantaneous SFR of gas, we instead quote the SFR as the total stellar mass formed within an aperture of 30 physical kpc, averaged over the last 1 Gyr \citep[see][]{donnari2019,pillepich2019}.
    \item Specific star formation rate (sSFR): ratio of the star formation rate to the stellar mass of the galaxy, both measured as above.
    \item Neutral atomic hydrogen (HI): while the simulation directly tracks and outputs the neutral hydrogen content of gas, we use the H$_2$ model of \cite{gnedin2011} to compute the fraction of neutral hydrogen that is molecular, to remove this component, in order to derive the atomic HI mass, following \cite{popping2019}.
    \item OVI density: similarly, while the total oxygen mass is directly tracked and output by the simulation, the relative abundances of different ionization species are not calculated. Following \cite{nelson2018b}, we model these metal ionization states using \textsc{Cloudy} (\cite{ferland2013}, v13.03): accounting for (a) photo-ionization in the presence of a high energy (X-ray + UV) background \citep[the 2011 update of][]{fg2009}, (b) collisions, and (c) self-shielding by highly dense gas \citep{rahmati2013}, we iterate to equilibrium using \textsc{Cloudy}'s single zone mode.
    \item X-ray luminosity: we compute X-ray emission in the energy range $0.5-2.0$ keV (soft X-rays) using the APEC emission model \citep{smith2001}. This accounts for both emission lines by highly ionized species of metals, and the continuum emission produced through bremsstrahlung.
    \item Mass flow rate: following \cite{nelson2019}, the mass flow rate at a radius $r_0$ across a shell of thickness $\Delta r$ is:
    \begin{equation}
        \dot{M} = \frac{1}{\Delta r} \sum\limits_{\substack{i=0 \\ |r_i - r_0| \leq \Delta r / 2}}^n m_i \times v_{\rm{rad},i}
    \end{equation}
    where $m_i$ and $v_{\rm{rad},i}$ are the mass and radial velocity of the $i^{\rm{th}}$ gas cell. Unless otherwise stated, we consider gas to be outflowing if it has positive radial velocity (i.e. $v_{\rm{rad},i} > 0$), and inflowing otherwise.
    Throughout the paper, we adopt a fixed value of $\Delta r$ = 5kpc.
\end{enumerate}


\begin{figure*}
\centering 
\includegraphics[width=16cm]{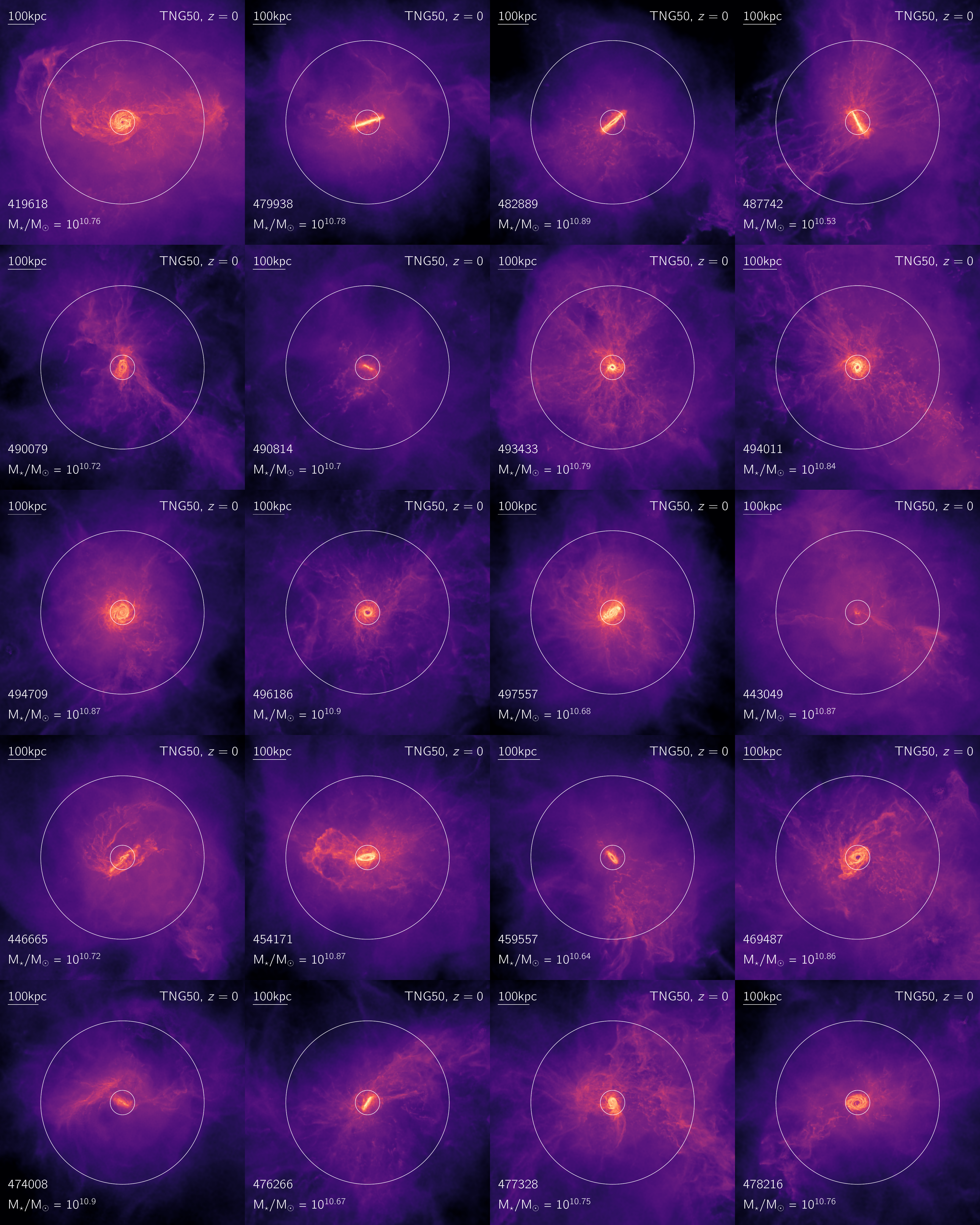}
\includegraphics[width=7cm]{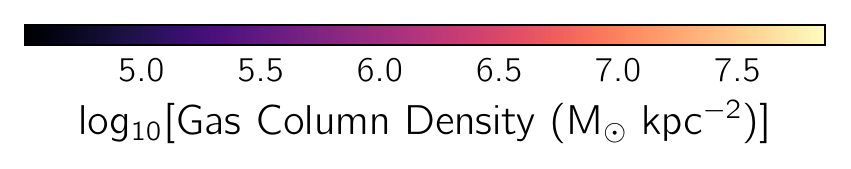}
\includegraphics[width=7cm]{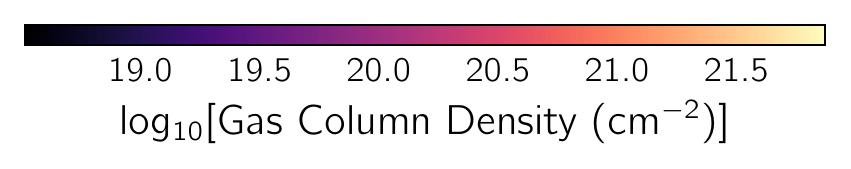}
\caption{Gas density projections at $z=0$ (along random orientations; in a box that extends $\pm1.5\RVIR$ along the perpendicular axis) of 20 random TNG50 Milky Way-like (MW-like) galaxies and their surrounding circumgalactic medium (CGM), excising satellites. In each panel, the two circles mark $(0.15, 1.0) \times \RVIR$ of the parent halo -- we refer to the region between the two circles as the CGM. On the bottom-left of each panel, the numbers indicate the subhalo ID and stellar mass of the corresponding (central) galaxy. These halo-scale gas density projections reflect the wide variety we find across the 132 MW-like TNG50 galaxies in our sample.}
\label{fig:collageMW}
\end{figure*}

\section{Results}\label{results}

\subsection{Global properties of the CGM of TNG50 MW-like galaxies}\label{sec:cgm_global}

The TNG50 simulation produces a diverse set of more than 100 MW-like galaxies (at $z=0$), which in turn exhibit diverse CGM properties both across and within galaxies. 

\subsubsection{CGM maps}

We begin by visualizing this diversity in Figure~\ref{fig:collageMW}, where we show gas column density projections for a random subset of 20 galaxies/halos\footnote{See \url{www.tng-project.org/ramesh22} for the CGM maps of all our TNG50 Milky Way-like galaxies.}. Note that we show random projections of galaxies, i.e. the central gaseous disks are occasionally seen edge-on, face-on, or at intermediate inclinations. The projections extend $\pm 1.5 \rm{R_{200,c}}$ from edge to edge, as well as along the line of sight direction. The two white circles mark $(0.15, 1.0) \times \RVIR$ of the corresponding halo, which we adopt as our fiducial definition for the spatial region of the CGM (see Section~\ref{definitions}). 

We see that the sample includes a variety of (a) galaxy/halo sizes, as can be inferred from the scale bar in the top-left of each panel, (b) stellar masses, bottom-left of each panel, and (c) most importantly for the sake of this paper, CGM gas distributions. To set a scale, the virial radii of TNG50 MW-like galaxies vary in the range $\RVIR  \sim 150-300$ kpc, with a median value of $\sim 210$ kpc. The distribution of CGM gas is, in general, not azimuthally symmetric \citep{peroux2020, truong2021, pillepich2021}, occasionally due to CGM gas that is directly connected to the central galactic disk. At the same time, there are often tidal features including cold gas tails and debris in the CGM, due to ram-pressure stripping of satellite gas \citep{yun2019,ayromlou2019}.\footnote{Recall that satellites are intentionally excluded in our analysis (Section \ref{definitions}), and hence are missing in these gas projections, but several would in general be expected in $z=0$ MW-like halos \citep{engler2021}.}

\begin{figure*}
\centering 
\includegraphics[width=16cm]{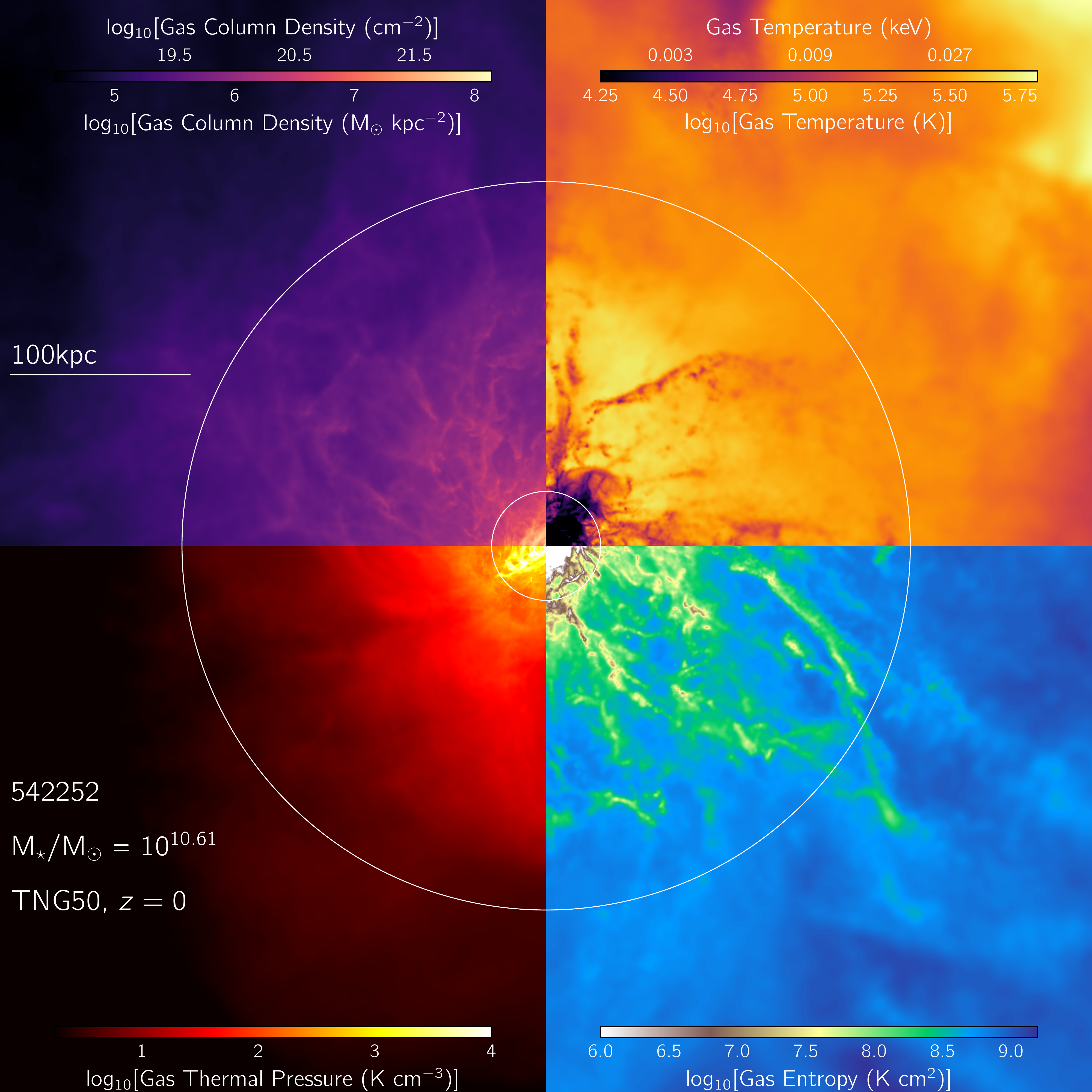}
\caption{Visualization of four different physical quantities, each shown in one quadrant, for a representative gaseous halo around a MW-like galaxy (subhalo ID 542242) from the TNG50 simulation at $z=0$. The orientation is random, the projections extend $\pm 1.5 \rm{R_{200,c}}$ perpendicular to the image plane, each quadrant extends $\pm 0.75 \rm{R_{200,c}}$ in extent from side to side, and satellite gas is excised. As above, the two circles mark $(0.1, 1.0) \times \RVIR$ of the corresponding halo. The panels show: gas column density (top left), gas temperature (top right), gas pressure (bottom left), and gas entropy (bottom right). While the galaxy at the center of the halo is dominated by dense cold gas with large pressure, the CGM has different physical properties -- halo gas is, on average, hotter and less dense. There are strong inhomogeneities on top of the background state, which are overdense, cool, and low entropy, i.e. 'clouds' of cool, dense gas.}
\label{fig:theoryMwImages}
\end{figure*}

In Figure~\ref{fig:theoryMwImages} we take a closer look at a single gaseous halo, selected as a relaxed example without significant asymmetric features. We show four different physical properties in the different quadrants. 
In the top-left quadrant, we show the column gas density projection: gas is most dense close to the disk, reaching $\gtrsim 10^8~\rm{M_\odot~kpc^{-2}} (\gtrsim 10^{21}~\rm{cm^{-2}})$, and becomes more rarified at larger distances, into the halo outskirts and the IGM. While gas outside of the virial radius (i.e. outside the upper limit of the CGM) shows minimal (visible) structure, i.e. is mostly smooth, we observe noticeable `clouds' of gas in the CGM. These gas density fluctuations are local overdensities that are typically cooler than their surroundings. 
The top-right panel shows the gas temperature: the disk is dominated by cold gas, while more extended gas is much warmer. Halo gas has a complex temperature structure, with hot gas at or near the halo virial temperature ($10^{5.8-6}$ K) co-existing with colder gas structures.
In the bottom-left quadrant, we show the thermal pressure $P_{\rm th} = (\gamma-1)\,\rho\,U_{\rm th} $ of the gas, where $\rho$ and $U_{\rm th}$ are the density and internal energy of gas, respectively. For star forming gas, note that this is the pressure based on the effective equation of state. A clear radial pressure gradient exists, moving from the galaxy (dominated by high thermal pressure) to the IGM (low pressure). Finally, in the bottom-right panel, we show the gas entropy ($\sim P_{\rm th} / \rho^\gamma$). Gas in the disk is at relatively low entropy, while the IGM hosts gas at higher entropies. Similar to the other quantities discussed above, gas in the CGM shows a mix of the two: while most of the volume is filled by gas at high entropy, small overdensities at lower values of entropy are realized by the TNG50 model \citep[see also][]{nelson2020}.

\begin{figure*}
\centering 
\includegraphics[width=16cm]{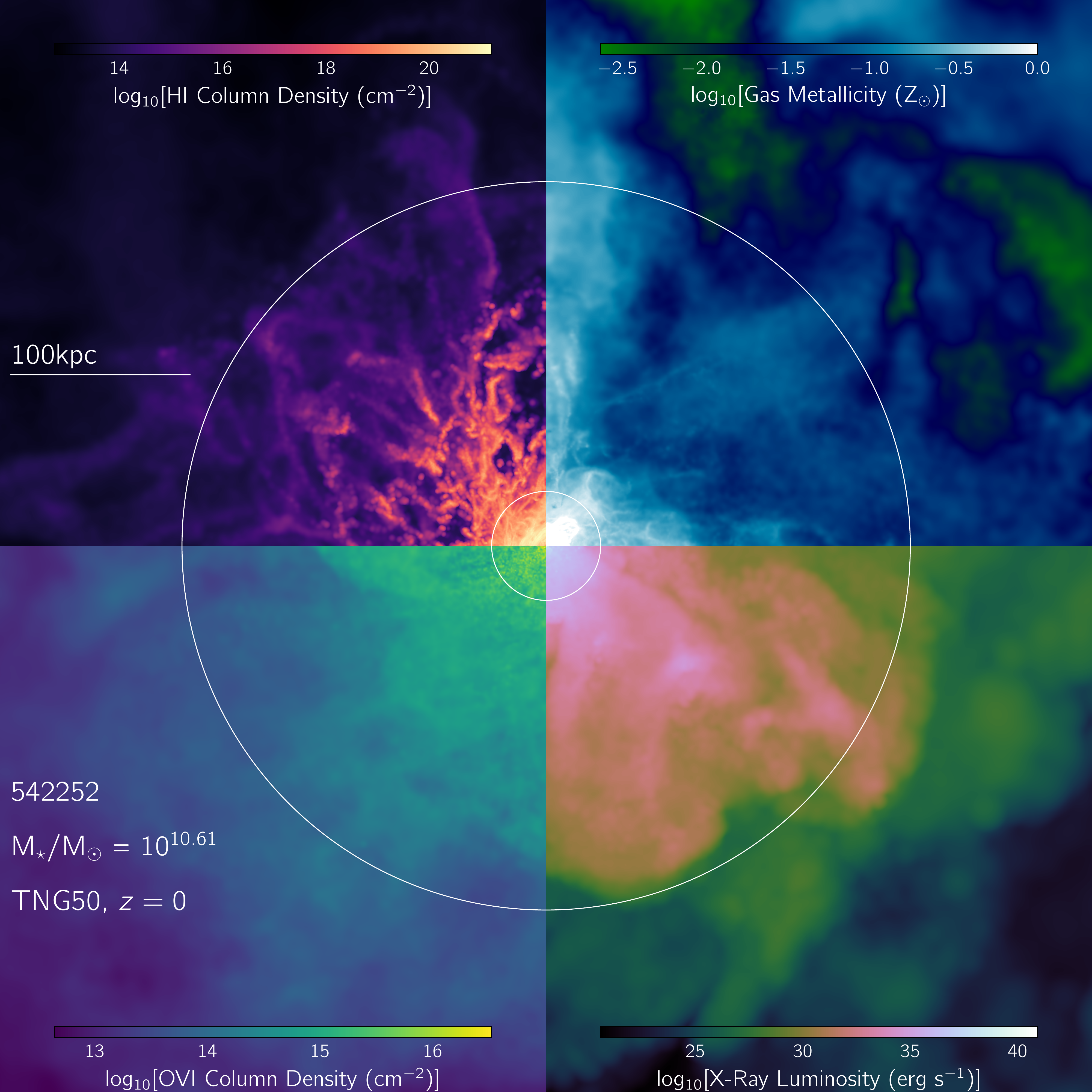}
\caption{Visualizations of four additional physical properties, for the same gaseous halo shown in Figure~\ref{fig:theoryMwImages}. Here we focus on more directly observable quantities. We show: HI column density (top left), gas metallicity (top right), OVI column density (bottom left), and X-ray luminosity in the 0.5-2.0 keV band (bottom right). In general, the CGM of MW-like galaxies can host regions of much lower metallicity than the galaxy (also at less than one percent $\rm{Z_\odot}$: green). Cold gas, as traced by neutral hydrogen, is particularly clumpy, while the warm-hot phase traced through OVI and X-rays is much more spatially smooth.
}
\label{fig:obsMwImages}
\end{figure*}

Figure~\ref{fig:obsMwImages} shows four additional views of the same halo, focusing on quantities that are either directly observable, or commonly inferred from observables of the gas. In the top-left quadrant, we show the HI column density: while $N_{\rm HI}$ is qualitatively similar to total gas column density, it decreases more rapidly as one transitions from the disk towards the IGM because self-shielding from the metagalactic background radiation field becomes negligible at densities characteristic of the outer CGM \citep{rahmati2013}. The top-right quadrant shows the gas metallicity: while the disk is dominated by metal-enriched gas $\gtrsim \rm{Z_{\odot}}$, the CGM host regions that are significantly less metal-enriched, at $0.1\,\rm{Z_{\odot}}$ and lower; in fact, columns of highly metal-enriched gas typically emerge above and below the galactic disk \citep{pillepich2021}, as it can be seen at the edge of the quadrant. The bottom-left (bottom-right) quadrant shows OVI column density (X-ray luminosity), which is an indirect tracer for warm (hot) gas. $\rm{N_{OVI}}$ declines from $\sim10^{16}$ to $\sim10^{13}$ cm$^{-2}$ from 0.15 to 1 $\rm{R_{200,c}}$, while $\rm{L_X}$ declines rapidly with distance primarily due to its $n^2$ dependence, dropping ten orders of magnitude, from $\sim 10^{40}$ erg s$^{-1}$ to $\sim 10^{30}$ erg s$^{-1}$, over the same distance. Both of these tracers of the warm and hot phases are more diffuse/volume-filling than the cold gas traced by neutral hydrogen. Although this example is typical, significant variation across the sample is evident.

\begin{figure*}
\centering 
\includegraphics[width=17cm]{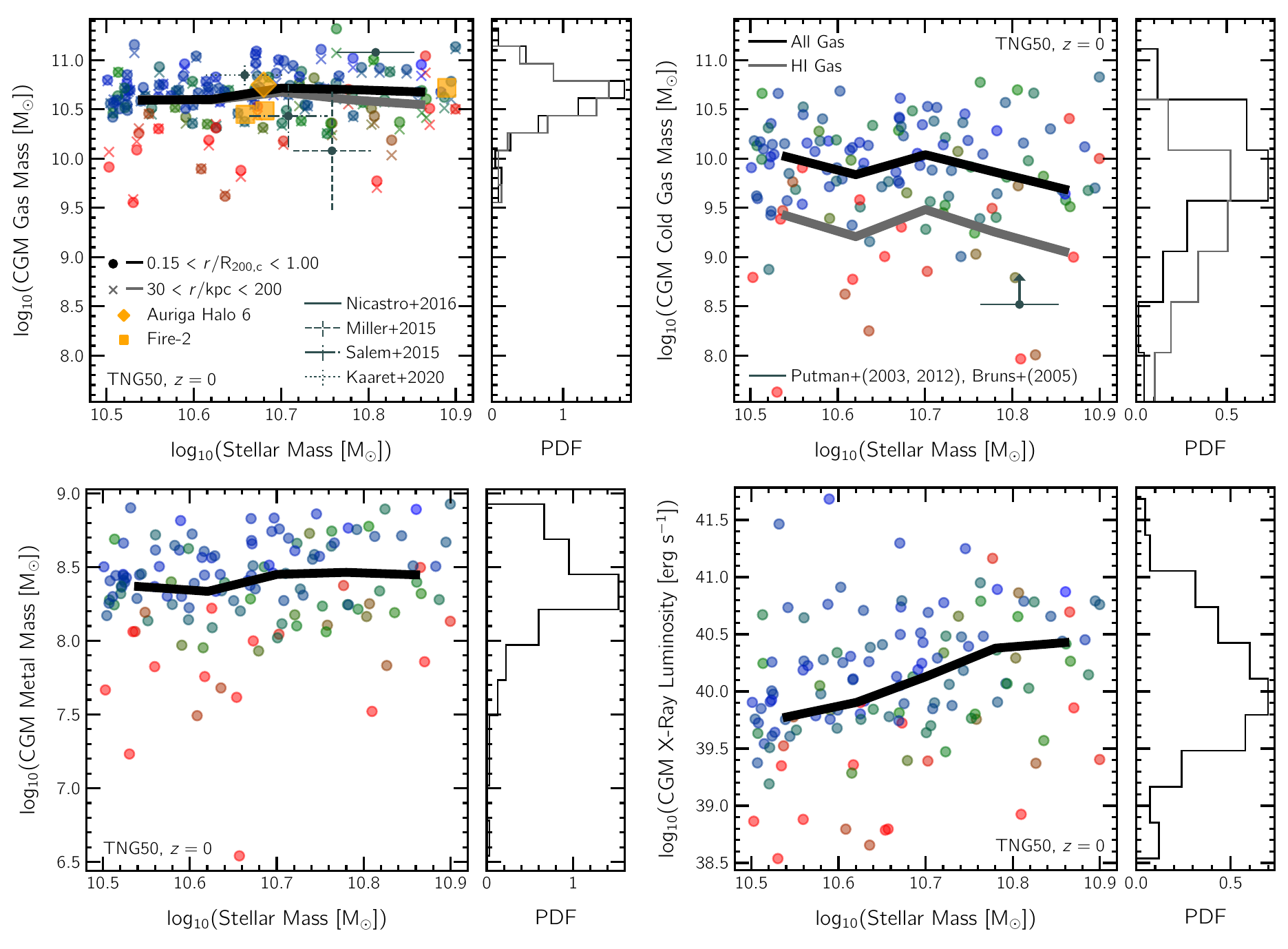}
\includegraphics[width=6cm]{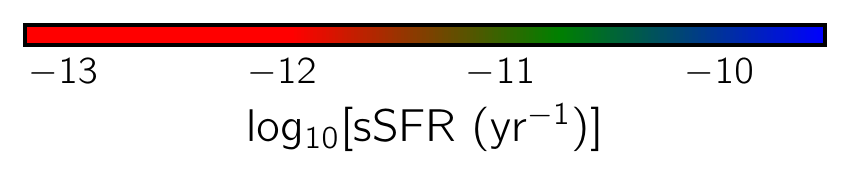}
\caption{Selected integral properties of the CGM plotted as a function of galaxy stellar mass for all 132 MW-like galaxies from the TNG50 simulation, at $z=0$: total CGM gas mass (top-left), CGM cold gas (T $<10^{4.5}$K) mass (along with the median line of CGM HI mass; top-right panel), CGM metal mass (bottom-left), and CGM X-ray luminosity (0.5-2.0 keV; bottom right panel). The first three quantities depend only weakly on galaxy stellar mass, while X-ray luminosity rises sharply with increasing stellar mass of the galaxy. In each panel, the circular markers are color coded by the sSFR of the galaxy -- in all cases, we note that galaxies with a lower sSFR have smaller values of the corresponding quantity. In the top-left panel, we consider two different definitions for the CGM: variable volume (extends between 0.15 and 1.0 R$_{\rm{200,c}}$) and constant volume (extends between 30 and 200 kpc) -- the two definitions yield statistically similar results at the low mass end, while differences are apparent towards the high mass end. In all other panels, we adopt the variable volume definition, which is our fiducial choice.}
\label{fig:cgm_props_vs_mass}
\end{figure*}

\subsubsection{CGM integrated properties}

To explore this diversity in CGM gas properties among MW-like galaxies, Figure~\ref{fig:cgm_props_vs_mass} presents trends of integrated properties of CGM gas as a function of stellar mass. The markers show the values of individual galaxies, colored by specific star formation rate (sSFR), while lines plot the median as a function of mass. 

The top-left panel of Figure~\ref{fig:cgm_props_vs_mass} shows the total CGM gas mass, where the circles/black median curve correspond to our fiducial definition of the CGM, i.e. the region bounded by $[0.15, 1.0] \times \RVIR$. For comparison, in crosses/gray median line, we also include an alternate definition of the CGM: the region bounded by $[30, 200]$ kpc. While differences between the two definitions are negligible at low stellar masses, more massive galaxies, on average, are surrounded by more extended halos, and defining the outer boundary of the halo at 200 kpc reduces the amount of gas that we can measure in the CGM. To best capture the CGM across our sample, which covers a range of dark matter halo masses and thus sizes, we therefore adopt the $\RVIR$-relative definition as our fiducial choice.

The median total CGM gas mass is almost constant as a function of stellar mass, with a difference of $\sim 0.1$ dex across the stellar mass range. Two more important trends emerge: (a) at fixed stellar mass, there is a significant diversity in amount of gas present in the CGM, as visible in the scatter, and (b) galaxies with a higher sSFR preferentially have more gas in their CGM, i.e. blue points are systematically above their green and red counterparts. Conversely, TNG50 MW-like galaxies below the star-forming main sequence are surrounded by less massive CGMs than their more star-forming counterparts. We have checked that this trend is not driven by halo mass, i.e. it is in place even though more massive halos tend to host more massive CGMs and even though, in the TNG model, at fixed stellar mass, quiescent galaxies tend to reside in somewhat more massive halos.

To make a comparison with other models we show a similar measurement of the (total) CGM gas mass (at $z=0$) from the Auriga simulations, a set of zoom-in simulations run with \textsc{AREPO}, albeit with a different underlying galaxy physics model. Shown in the orange diamond, the CGM of Auriga Halo 6 \citep[][hereafter, Au6]{voort2021} -- a relatively isolated MW-like galaxy -- is comparable to the median behaviour of the TNG50 sample at that stellar mass.\footnote{Note that \cite{voort2021} defines CGM gas mass as the total mass of all non-star forming gas within the virial radius, a minor difference.} We also show properties of three halos from the FIRE-2 simulations, run with the code Gizmo \citep{hopkins2015} and with rather different physical models compared to TNG -- the (total) CGM gas mass of m12b, m12c, and m12w, all at $z=0.25$, are shown in orange squares \citep{esmerian2021}.\footnote{\cite{esmerian2021} considers the outer boundary of the CGM to be 1.0 R$_{\rm{vir}}$, while the inner boundary is set to max(1.2 R$_{\rm{gal}}$, 0.1 R$_{\rm{vir}}$).} The measurements from FIRE-2 are roughly comparable with TNG50, although the two galaxies at $\sim 10^{10.65}~\rm{M_\odot}$ are lower by $\sim 0.2$ dex in their CGM gas mass content with respect to the TNG50 median, albeit well within the TNG50 scatter, which could be due to more powerful/ejective stellar feedback.

On the observational side, the total gas mass of the CGM of the Milky Way is highly uncertain: while \cite{nicastro2016} report that this value may be as high as $1.2 \times 10^{11} \rm{M_\odot}$, \cite{miller2015} estimate the total mass of CGM gas out to 250 kpc to be $1.2^{+0.9}_{-0.8} \times 10^{10} \rm{M_\odot}$. Through a study of the orbital motion of the Large Magellanic Cloud (LMC), \cite{salem2015} infer the mass of CGM gas out to 300 kpc to be $2.7^{+1.4}_{-1.4} \times 10^{10} \rm{M_\odot}$. Using X-ray observations, \cite{kaaret2020} estimate the mass out to 260 kpc to lie in the range $(5.5 - 8.6) \times 10^{10} \rm{M_\odot}$. We show these measurements with dark gray solid, dashed, dot-dashed and dotted errorbars, respectively. In each case, for the stellar mass of the Milky Way, we use the measurement reported by \cite{mcmillan2011}: $M\rm{_{\rm{\star, MW}}} \sim (6.43 \pm 0.63) \times 10^{10}~\rm{M_\odot}$, and spread these four data points somewhat along the x-axis for visibility. All four of these observational inferences are well bracketed within the TNG50 scatter. 

In the top-right panel of Figure~\ref{fig:cgm_props_vs_mass} we show the cold gas mass (T $<10^{4.5}$K) in the CGM. In addition, the gray curve shows the median HI gas mass, which resembles a down-scaled version of the black curve. As with the total CGM gas mass, the trends of cold gas mass are more or less constant as a function of stellar mass. There is again a noticeable gradient in the sSFR colors across the vertical scatter: galaxies with a higher sSFR have substantially more cold gas in their CGM. We have checked and the trend with sSFR is present also when adopting a larger inner boundary of the CGM (0.3 instead of 0.15 $\times$ R$_{\rm{200,c}}$), i.e. by ensuring that extended cold gaseous disks -- which may be expected in highly star-forming galaxies -- do not ``contaminate'' our definition of CGM.\footnote{This is the case for all four CGM integrated properties of Figure~\ref{fig:cgm_props_vs_mass}. While setting the inner boundary of the CGM to $0.15 \times \rm{R_{vir}}$ sometimes leads to a part of the disk being included in the CGM (as can be seen in some panels of Figure~\ref{fig:collageMW}), we confirm here that this does not strongly impact the conclusions of Figure~\ref{fig:cgm_props_vs_mass}. For instance, setting the inner boundary of the CGM instead to $0.3 \times \rm{R_{vir}}$ leads to a systematic drop of $\sim 0.5$ dex in values of cold gas mass, metal mass and X-ray luminosity. However, the median trends and the dependencies on star formation rates remain almost unchanged, albeit somewhat weakened in the case of X-ray luminosity.} Moreover, although not shown here, such a gradient is not present when points representing MW-like galaxies are colored by either the halo mass or SMBH mass but it does appear to depend on the cumulative energy injected via the kinetic mode of SMBH feedback. 

Observationally, \cite{putman2012} estimate the cold (HI) gas mass contained in high velocity clouds (HVCs) using 21-cm emission to be $\sim 2.6 \times 10^7~\rm{M_\odot}$. \cite{putman2003} and \cite{bruns2005} report that the Magellanic System (i.e the Magellanic Stream, Leading Arm and Magellanic Bridge) contributes at least $\sim 3 \times 10^8~\rm{M_\odot}$ to the HI mass of the Milky Way, resulting in a \textit{lower limit} to the total HI mass of $\gtrsim 3.3 \times 10^8~\rm{M_\odot}$ for the Milky Way CGM. This measurement is shown with the gray solid errorbar. Once again, we use the measurement reported by \cite{mcmillan2011} for the stellar mass. Given this lower limit, the TNG50 median is $\sim 0.5$ dex above, although some outliers from our sample have consistent HI masses, as apparent with the marginalized distributions on the right.

The total amount of metals in CGM gas is shown in the bottom-left panel of Figure~\ref{fig:cgm_props_vs_mass}: while galaxies with a higher (lower) sSFR have more (fewer) metals in the CGM gas, the metallicity content in CGM gas shows no strong trend with respect to stellar mass. Since the stellar mass range investigated is rather small ($0.4$ dex), such a flat trend may not be altogether surprising even though there is a slight trend of higher CGM metallicity in higher mass halos within this galaxy sample (not shown). 

Lastly, in the bottom-right panel of Figure~\ref{fig:cgm_props_vs_mass}, we show the X-ray luminosity of CGM gas. We see a clear increase in luminosity as a function of stellar mass, by a factor of $\sim 10$ across the stellar mass range explored. Additionally, at fixed stellar mass, higher sSFR galaxies have more luminous gaseous halos in the 0.5 - 2.0 keV (soft) X-ray band. Similar to other quantities, there is a large scatter around the median value, i.e larger than the mass trend itself. While the computed X-ray luminosity includes continuum emission, plasmas at temperatures typical of MW-like systems are dominated by metal line emission \citep{anderson2015}, making the metal content of the CGM and the enrichment history of the galaxy critically important for extended X-ray halos. The bottom-right panel of Figure~\ref{fig:cgm_props_vs_mass} confirms, also for the case of disky TNG50 MW-like galaxies, the predictions previously put forward by \cite{truong2020} and \cite{oppenheimer2020} who showed that, according to both TNG and EAGLE, at the transitional mass scale of $\sim 10^{10.5-11}\, \MSUN$ in stars, star-forming galaxies are surrounded by X-ray brighter halos than similarly-massive more quiescent galaxies -- a phenomenon dubbed the X-ray luminosity halo dichotomy.

Overall, Figure~\ref{fig:cgm_props_vs_mass} demonstrates that, according to TNG50 and even within the relatively narrow selection of MW-like galaxies, the abundance and physical properties of gas in the CGM of MW-like halos exhibit a large diversity. Although three of the four integral properties studied here have a weak dependence on stellar mass, across the stellar mass range of $10^{10.5}\,\rm{M}_\odot$ to $10^{10.9}\,\rm{M}_\odot$, the star formation activity or status of the central galaxy appear closely linked to the state of its CGM.

Following previous works \citep[see e.g.][]{davies2020, zinger2020, terrazas2020, truong2020, truong2021}, we have checked that, particularly in the intermediate stellar mass range of MW-like galaxies, a connection of CGM gas properties also exists with the (cumulative, in time) SMBH energy injection in the kinetic mode. We find that more energetic SMBH kinetic feedback implies lower CGM mass, CGM cold mass, CGM metal mass and X-ray luminosity (not explicitly shown). In fact, within the TNG model, SMBH kinetic feedback -- rather than SMBH thermal energy injection at high-accretion rates or stellar feedback -- is the cause for the suppression of star-formation in massive galaxies \citep{weinberger2017, nelson2018, terrazas2020}. In fact, there would be no low SFR galaxies in our TNG50 MW-like sample if SMBH kinetic feedback was switched off.

We therefore conclude that the modulations in total CGM mass, CGM metal mass, and X-ray luminosity of MW-like galaxies \cite[see][]{truong2020}, are primarily due to the ejection and heating of CGM gas driven by the SMBH kinetic feedback in TNG \citep{nelson2019b, zinger2020}. Star formation status is, to zeroth order, a proxy for this SMBH activity. On the other hand, we speculate that the gradient with sSFR in the {\it cold} gas mass in the CGM may arise due to a physical connection between the CGM and galactic star formation. This could be caused by a link with the flow of cold gas around galaxies, although the direction of causation is unclear. A greater amount of cold halo gas may, for example, imply larger gas accretion rates on to the galaxy, resulting in higher sSFRs.

\begin{figure*}
\centering 
\includegraphics[width=17cm]{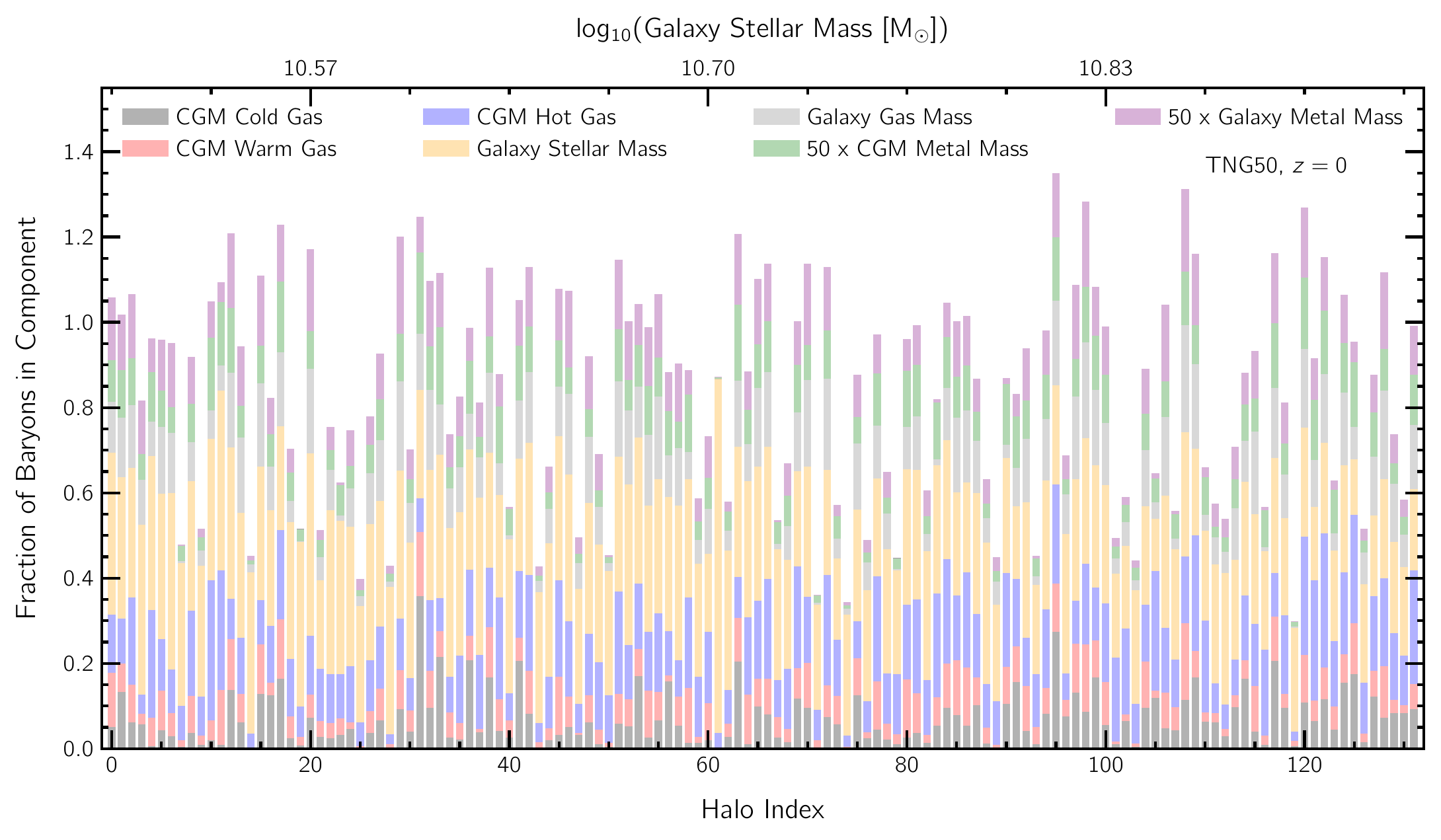}
\caption{Fraction of baryonic matter in different components and phases, relative to the expected baryon budget of the halo, i.e. $M_{\rm{comp}}/(f_{\rm b}\,{\rm{M_{200c}}})$, at $z=0$. Each of the 132 MW-like galaxies from TNG50 is represented as an individual vertical bar. The galaxies/halos are arranged in \textit{ascending order} of their galaxy stellar mass, from left to right (see top x-axis). The different components shown are: CGM cold gas (gray), CGM warm gas (red), CGM hot gas (blue), galaxy stellar mass (orange), galaxy gas mass (grey), CGM metal mass (green) and galaxy metal mass (purple). Note that the latter two quantities have been multiplied by a factor of fifty for better visibility. We find a large diversity across the sample, with no clear trend in fractional abundances as a function of galaxy stellar mass. For this Figure, `galaxy' refers to material within the inner CGM boundary, i.e. $<0.15 \rm{R_{200c}}$.}
\label{fig:bary_comp}
\end{figure*}

\subsubsection{The baryon budget of the CGM}

Inspired by efforts to make an empirical consensus of the baryonic content of halos \citep{peeples2014, werk2016}, we explore the budget of baryons across different components and gas phases. Figure~\ref{fig:bary_comp} shows the baryonic components of galaxies/halos, split into seven different categories. We calculate each as a fraction with respect to the halo virial mass multiplied by the cosmic baryon fraction, $M_{\rm{comp}}/(f_{\rm b}\,{\rm{M_{200c}}})$. We consider seven different components: CGM gas split into cold (T $<10^{4.5}$K), warm ($10^{4.5}$K $<$ T $<10^{5.5}$K) and hot (T $>10^{5.5}$K) phases, galaxy stellar mass, galaxy gas mass, CGM metal mass, and galaxy metal mass. Each MW-like galaxy in our sample is represented by a single bar, with its components stacked vertically, and arranged in ascending stellar mass, from left to right (see top x-axis). For better visibility, the latter two components have been multiplied by a factor of fifty. To avoid double-counting gas cells between the galaxy and the CGM, in this figure (only) we use the term galaxy to refer to the part of the halo within the inner boundary of the CGM, i.e $<0.15 \times \rm{R_{200c}}$.

Galaxies with similar stellar masses can vary significantly in terms of how their baryons are distributed across different components: while one galaxy may be surrounded by a CGM that is dominated by hot gas, a different galaxy with a similar stellar mass can instead be surrounded by a largely cold gas dominated CGM. The median fractional mass of gas in the CGM with respect to $f_{\rm b} \rm{M_{200c}}$ is $\sim 29\%$, although this value varies considerably across the sample, from $\sim 15\%$ (16$^{\rm{th}}$ percentile) to $\sim 41\%$ (84$^{\rm{th}}$ percentile). This variability is also visible in the fractions of the different components: the median values and percentiles for the fractional mass of cold gas, warm gas, hot gas and metals, all in the CGM, are $5^{+7}_{-4}$, $6^{+5}_{-4}$, $14^{+8}_{-6}$ and $0.17^{+0.08}_{-0.10}$ \%, respectively. Because of this significant galaxy to galaxy variation, no strong trends as a function of stellar mass are evident. However, generally, the CGM mass of MW-like galaxies is largely dominated by hot, and thus X-ray emitting, gas.

\subsection{Spatial distribution of gas around TNG50 MW-like galaxies}

Having explored various global properties of the CGM, we now consider how gas in the CGM is distributed, and how its properties vary across the CGM.

\begin{figure*}
\centering 
\includegraphics[height=20.5cm]{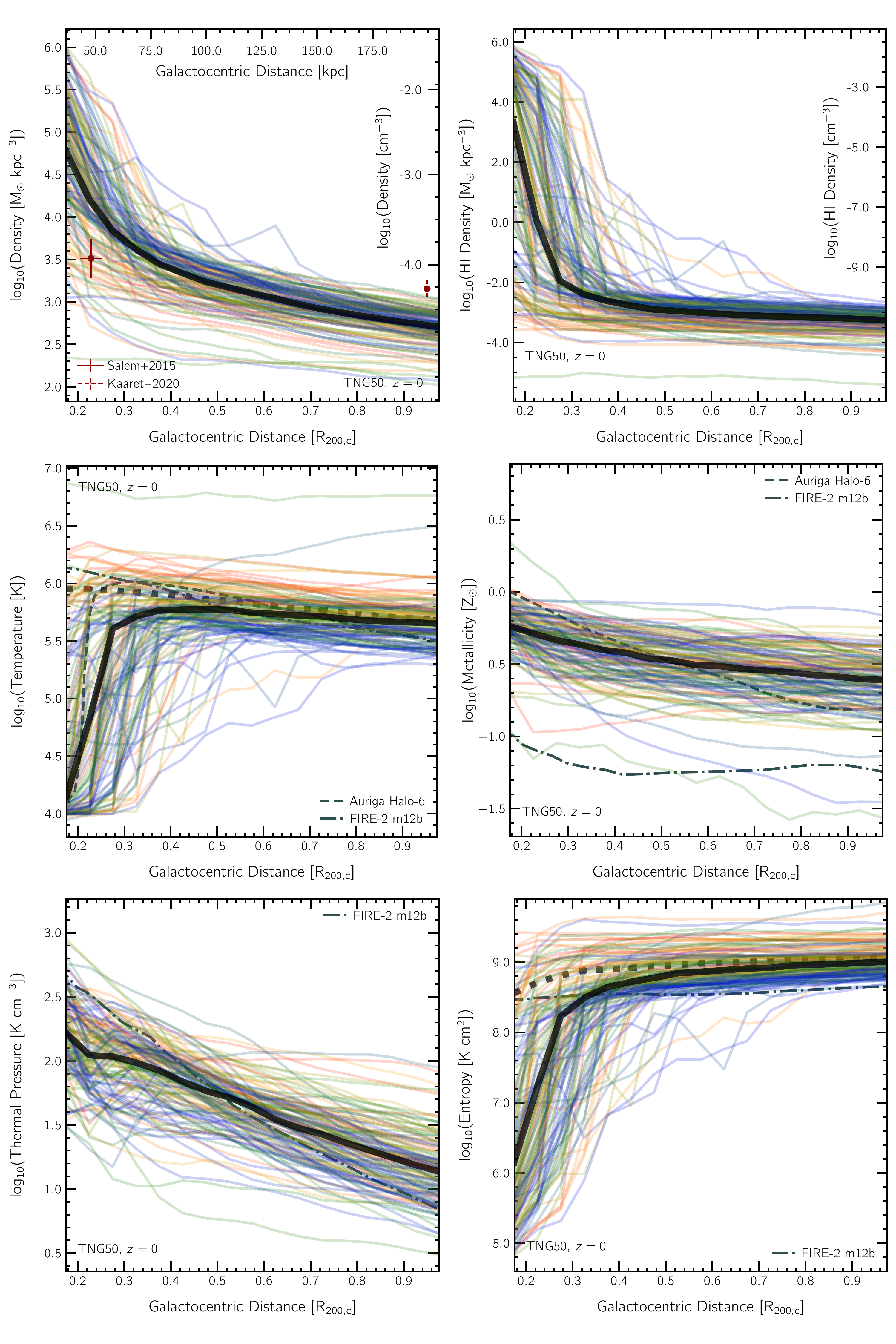}
\includegraphics[width=6cm]{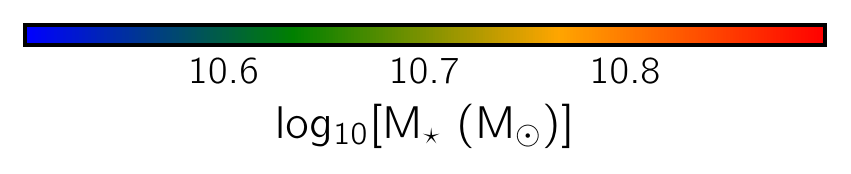}
\caption{Spherically-averaged radial profiles of six physical gas quantities as a function of (3D) galactocentric distance around TNG50 MW-like galaxies. We consider density (upper left), HI density (upper right), temperature (center left), metallicity (center right), thermal pressure (lower left) and entropy (lower right). Each panel shows the median profile across the galaxy sample (black line), while the thinner curves correspond to profiles of individual halos, colored by galaxy stellar mass. Such profiles represent the spherically-averages of the depicted integrated or mass-weighted gas property at the given radius. Whereas the temperature and entropy of gas is generally lower in the inner halo as compared to the outskirts, due to the presence of a second cooler component becoming dominant, no such trend is clearly seen in the other four panels. The galaxy-to-galaxy scatter of individual curves about the median profile is large, and often correlates well with stellar mass.}
\label{fig:radial_profiles}
\end{figure*}

To begin, Figure~\ref{fig:radial_profiles} shows spherically-averaged, radial profiles as a function of galactocentric distance. We include density (upper left), HI density (upper right), temperature (center left), metallicity (center right), thermal pressure (lower left) and entropy (lower right). Each (thin) curve shows a single profile of an individual halo (mass-weighted median for temperature, metallicity, pressure and entropy), whereas the black curve shows the median behaviour across all halos in the sample. The individual curves are colored according to the stellar mass of the corresponding galaxy, as shown by the colorbar at the bottom. While the (lower) x-axes of all panels show the galactocentric distance normalised by the virial radius of the corresponding halo, the upper x-axis on the top-left panel shows the (median) galactocentric distance in physical kpc units, for reference.

The total gas density profile (top left panel of Figure~\ref{fig:radial_profiles}) peaks close to the disk, i.e. at the inner radius of the CGM, and decreases rapidly outwards into the halo, where the average CGM density can become as low as $10^{2-3}\,\MSUN$ kpc$^{-3}$. The median density profile is well fit by a function $r^{-\alpha}$, where $r$ is the galactocentric distance, and $\alpha \sim 0.36$. Perhaps unexpectedly (see previous Section), galaxies with larger stellar masses tend to have smaller gas densities in their CGM: a difference of $\sim0.1$ dex (outer halo) to $\sim0.3$ dex (inner halo) when the sample is split into two bins based on stellar mass (not explicitly shown). Observationally, based on ram pressure stripping of the LMC, \cite{salem2015} constrain the CGM density of the Milky Way at $48 \pm 5$ kpc to be $1.1^{+0.44}_{-.45} \times 10^{-4}~\rm{cm^{-3}}$, while \cite{kaaret2020} estimate the density of the Milky Way CGM at the virial radius to be $(4.8 \pm 1.0) \times 10^{-5}~\rm{cm^{-3}}$. These are shown in red solid and dashed errorbars, respectively. The inner halo measurement is below the median TNG50 profile, though well within the scatter, while the outer halo measurement is $\sim 0.3$ dex higher than that expected average across the TNG50 MW-like sample.

The HI density (top right panel, Figure~\ref{fig:radial_profiles}) behaves qualitatively similarly to the total gas density, although the drop in density is much sharper as self-shielding becomes ineffective in the low-density environment of the outer halo, as discussed in \S\ref{sec:cgm_global}. Galaxies with larger stellar masses generally have lower HI density profiles, and the trend is stronger in comparison to the total gas density: a difference from $\sim0.2$ dex (outer halo) to $\sim1$ dex (inner halo) when splitting the sample into two bins at the median.

\begin{figure*}
\centering
\includegraphics[height=20.5cm]{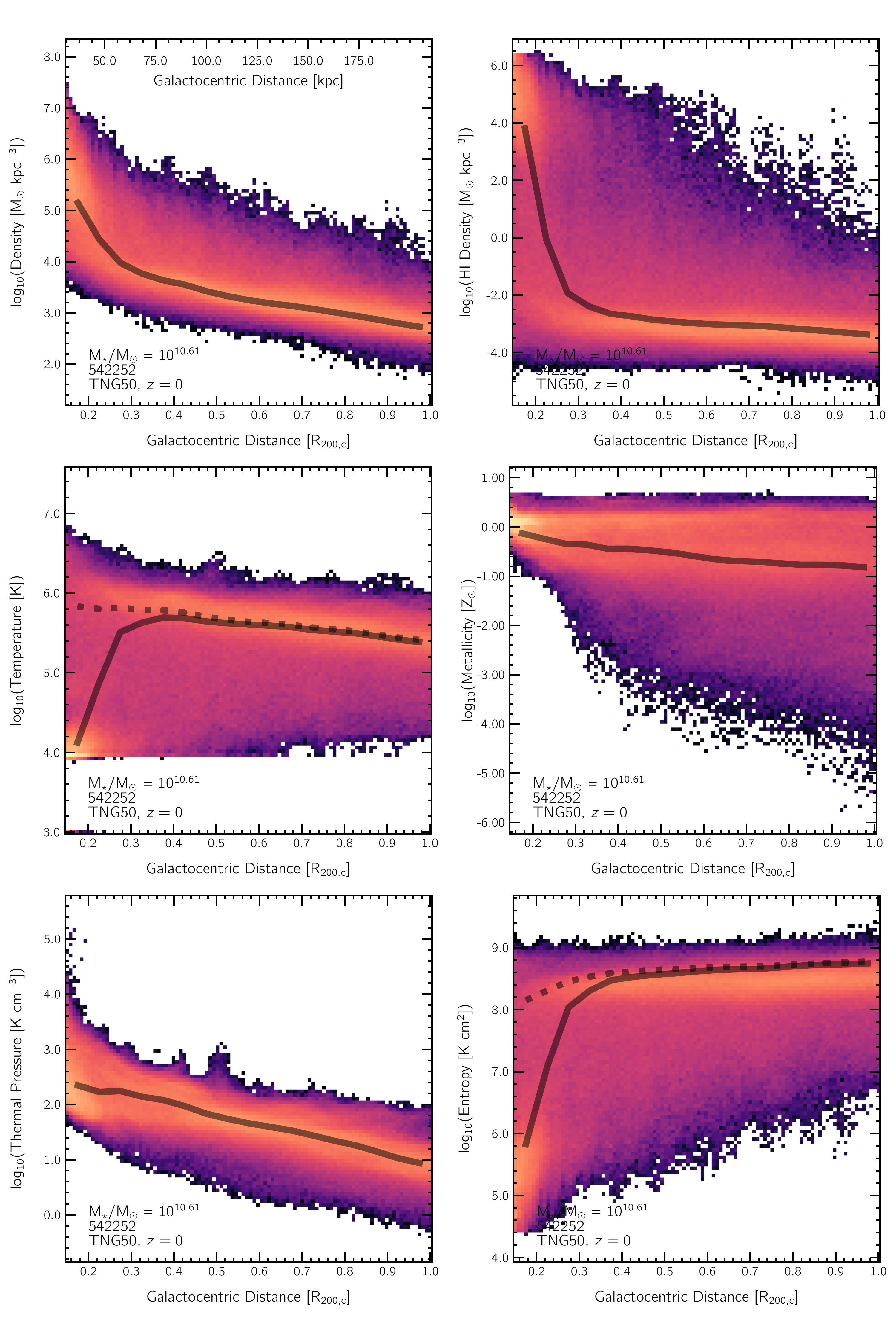}
\includegraphics[width=6cm]{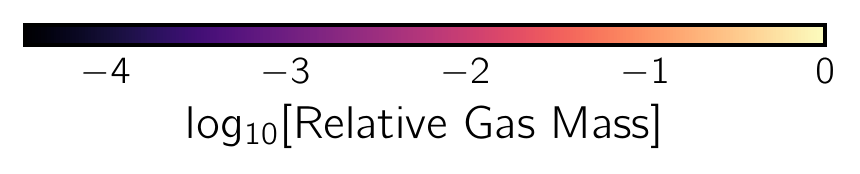}
\caption{Two-dimensional radial profiles of six physical gas properties for a single TNG50 MW-like galaxy (subhalo 542252), showing the same quantities as Figure~\ref{fig:radial_profiles}. The colors correspond to the relative distribution of gas mass, normalized to the maximum value (= $M_{\rm{pixel}}$/$M_{\rm{pixel,max}}$). In each panel, the black curves show the median profiles of the galaxy's CGM, i.e. the spherically-averaged gas quantity as in Figure~\ref{fig:radial_profiles}, thin curves (there, one curve per galaxy). Satellites are excised. Beyond spherically-averaged radial profiles, we see that gas properties in the CGM can have significant diversity at fixed distance, depending on where they are within the halo, i.e. their angular coordinates. This is particularly true in the inner halo, where distinct (multi-modal) gas populations are evident.}
\label{fig:2d_radial_profiles}
\end{figure*}

The center left panel of Figure~\ref{fig:radial_profiles} shows gas temperature: while gas close to the disk is predominantly cold ($< 10^{4.5}$ K), typically (see the solid black curve representing the population-wide average), the temperature rises sharply to the virial temperature (around $0.25 - 0.3 \rm{R_{200c}}$), before slowly declining outwards to larger galactocentric distances. This behavior is a result of cold gas dominating the disk-halo interface. To demonstrate this, we also show the median trend including only gas hotter than $10^5$\,K (thick dotted black curve), a lower cutoff of roughly $0.1 {\rm T_{vir}}$ which excludes the cold phase. In the inner halo, this curve is higher and flatter than the profile shown in the solid black curve, with no central depression. A noticeable trend is also seen with respect to stellar mass: lower-mass galaxies have cooler temperature profiles, with a difference of $\sim0.3$ dex between the two bins when the sample is split at the median stellar mass. This trend is expected, since both the galaxy stellar mass and halo virial temperature scale with halo mass.

For comparison, in the temperature-profiles panel, we include results from the Auriga simulations \citep[Au6 at $z=0$;][]{voort2021} -- these are shown in dashed dark gray curves. While Au6 reaches its peak temperature at roughly the same distance as the TNG50 median ($\sim 0.3 \RVIR$), its value is slightly higher ($\sim 0.3$ dex), although the virial radius of Au6 is close to the average virial radius of the TNG50 sample at $\sim 210$ kpc. We also show a profile from one of the FIRE-2 simulated galaxies \citep[m12b;][]{esmerian2021}.\footnote{The FIRE-2 profile is available at $z=0.25$, and is restricted to the hot halo only, by excluding sub-structure and low entropy gas ($K < 5$ keV cm$^{2}$).} Shown in dark gray dot-dashed curves, the gas temperature profile of this galaxy is higher in the inner halo ($\lesssim 0.45 \RVIR$) in comparison to the TNG50 median, possibly due to stronger heating from the supernovae feedback model of FIRE. 

We quantify trends of gas metallicity in the center-right panel of Figure~\ref{fig:radial_profiles}: gas close to the disk is more enriched than farther away into the halo, although the radial change is moderate, declining by $\sim 0.3$ dex from 0.15 to 1.0 $\RVIR$. On average, halos hosting galaxies with lower versus higher stellar mass have roughly the same metallicity out to a distance of $\sim0.4 \rm{R_{200c}}$, with more massive galaxies having greater gas-phase metallicity beyond this distance, up to a peak difference of $\sim0.2$ dex when splitting the sample into two bins at the median stellar mass. Although the comparison has caveats, the Au6 metallicity profile is qualitatively different than the median TNG50 behavior: it is steeper, with higher values within the inner halo ($\lesssim 0.5 \RVIR$), but lower values in the outer halo ($\gtrsim 0.5 \RVIR$). The Au6 profile is, nonetheless, bracketed within the population diversity of the TNG50 sample, as indicated by the scatter among the colored lines. The FIRE-2 galaxy (m12b) profile, on the other hand, exhibits a considerably metal-poorer CGM in comparison to the TNG50 median (a difference of $\sim 0.6 - 0.8$ dex, depending on distance): this is possibly because only hot gas is included, and/or is related to the differing details of metal mixing in the gaseous halos between the numerical techniques of FIRE-2 versus TNG.

Returning to TNG50, Figure~\ref{fig:radial_profiles} includes thermal pressure radial profiles (lower left panel). As expected given the rapid decline of density to large distance, pressure is also highest towards the halo center. Gas in the CGM of galaxies with lower stellar masses has lower thermal pressure, with an offset of $\sim0.1$ dex to $\sim0.2$ dex when splitting the sample into two bins at the median. The FIRE-2 galaxy thermal pressure profile is qualitatively different than the median TNG50 behavior: it is steeper, and so has higher pressures within the inner halo ($\lesssim 0.5 \RVIR$), but lower pressure in the outer halo ($\gtrsim 0.5 \RVIR$). Nonetheless, we can still identify a few TNG50 MW-like galaxies with similar pressure profiles.

Finally, the bottom-right panel of Figure~\ref{fig:radial_profiles} shows the behavior of gas entropy as a function of galactocentric distance. Moving away from the disk, entropy increases sharply out to a distance of around $0.3 \rm{R_{200c}}$, beyond which its value increases only gradually. As with temperature, the drop at small distance is due to the dominance of cold gas in the inner halo. In the dotted black curve, we show the median trend including only gas hotter than $10^5$\,K. In the inner halo, this curve is once again higher than the profile shown in the solid black curve, and asymptotically approaches the latter at large distances ($\gtrsim 0.5 \RVIR$). Lower-mass galaxies have relatively lower entropy profiles, with differences of $\sim0.3$ dex to $\sim0.4$ dex across the sample. Except for the inner halo ($\lesssim 0.3 \RVIR$), the FIRE-2 galaxy profile is lower by $\sim 0.4$ dex than the median TNG50 profile.

In each of the above panels, except for marginal cases, most halos behave similar to the median except with an overall normalization offset, and albeit with a large scatter.

\subsubsection{Radial profiles of the gas beyond azimuthal averages}

The previous analysis of radial profiles has one major limitation: spherically-averaged quantities (at a given distance) do not carry information about non-radial structure, i.e. they remove all detail about the diverse nature of gas at various angular positions at a given (galactocentric) distance. To explore this, Figure~\ref{fig:2d_radial_profiles} shows two-dimensional distributions of the same six physical quantities of Figure~\ref{fig:radial_profiles}, albeit only for one TNG50 MW-like galaxy (the same halo visualised in Figures~\ref{fig:theoryMwImages} and \ref{fig:obsMwImages}) -- averaging over the entire sample would hide any localized features. The different colors in each panel show the (relative) amount of gas mass, such that the full distribution function of each physical property at each radius is visible; black curves show median trends, namely the spherically-averaged quantity as in the thin curves of Figure~\ref{fig:radial_profiles}.

Some physical properties shown in Figure~\ref{fig:2d_radial_profiles}, notably total gas density and thermal pressure (top left and bottom left panels) are more or less continuous, meaning that the distribution of these quantities is unimodal and well characterized by the median radial value. However, this it not always the case.
For instance, gas at a distance of $\sim~0.25 \rm{R_{200c}}$ is separated into two different components: a cold phase making up the star-forming body of the galaxy, in the form of a thin, disk-like morphology, which extends in the plane of the disk to these distances, and a hot phase heated to the virial temperature outside of the galaxy itself, i.e. beyond the disk extent and/or just above or below its vertical extent. Similar features are visible in the HI density -- a significant amount of neutral hydrogen close to the disk, which is otherwise ionized in gas of higher temperatures, and entropy -- rotationally-supported cold gas has lower entropy compared to the hot gas in the halo. 

The distributions of gas temperature (center left) and entropy (lower right panel) show large scatter in values at fixed distance, namely large variations from one location within the CGM to another. For example, at $0.4 \RVIR$, the temperature of gas varies by over two orders of magnitude; on the other hand, the HI density can vary by $\sim$ 9 orders of magnitude, while the spread in metallicity ranges from super-solar to nearly pristine at fixed distance, indicative of extremely-incomplete metal mixing in the CGM as well as the accretion of hardly enriched gas. 

Figure~\ref{fig:2d_radial_profiles} clearly shows that, according to TNG50 and at fixed distance, different parts of the halo, i.e. at different angular positions, exist in very different physical states. This is the case not only for the specific galaxy depicted there, but applies to each MW-like galaxy in TNG50. For temperature and entropy, we again include dotted black curves showing the median trend when only gas hotter than $10^5$\,K is considered. These profiles diverge (i.e. solid versus dotted) in the inner halo as a result of the dominance of cold gas, but asymptotically converge in the outer halo.\footnote{Recall that satellites are excluded from our analysis, which would otherwise be visible as overdensity spikes at the radii where satellites are present.}

\begin{figure*}
\centering
\includegraphics[height=20.5cm]{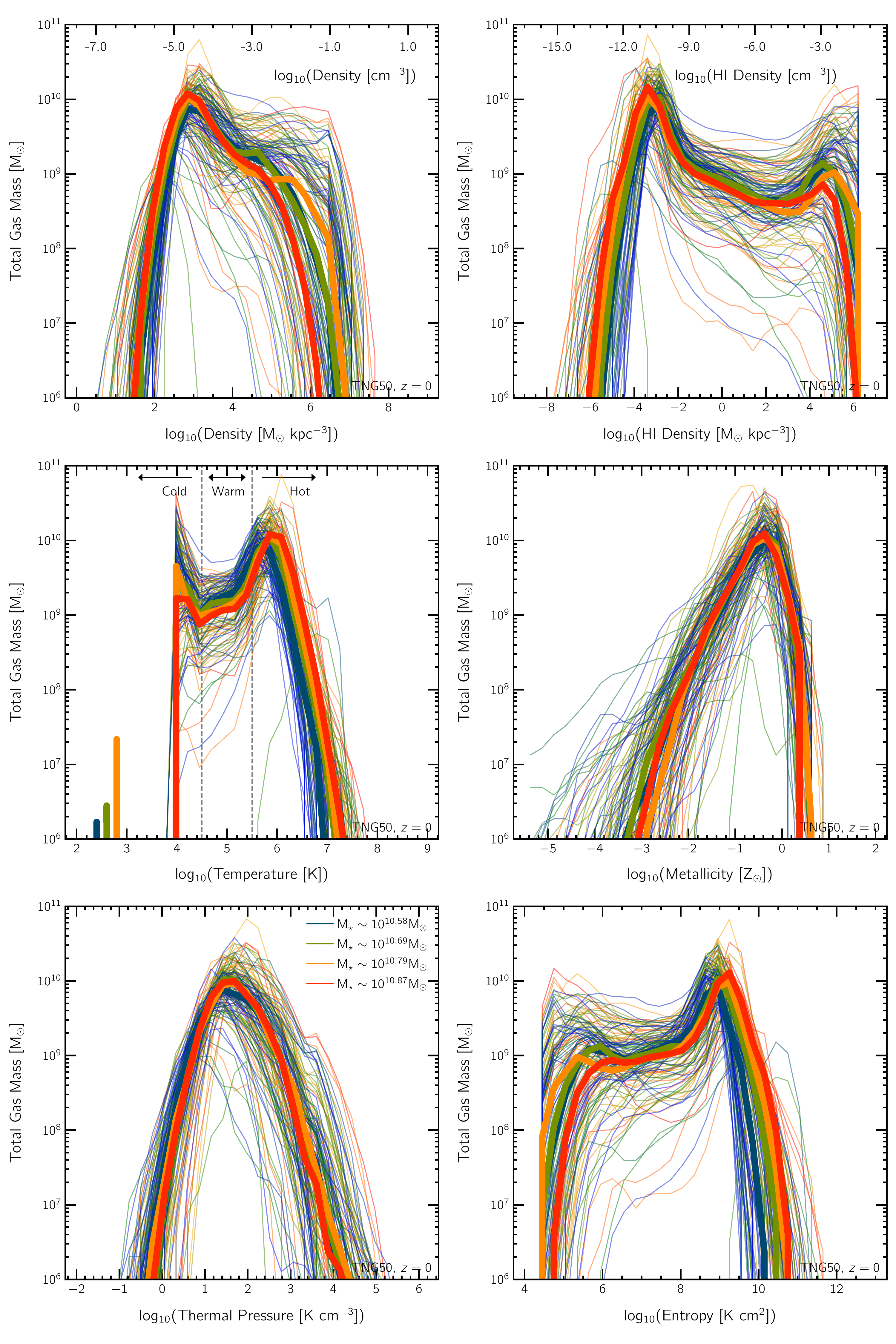}
\includegraphics[width=6cm]{figures/fig6Cbar.pdf}
\caption{Distributions of the physical properties of gas in the CGM of TNG50 MW-like galaxies at $z=0$ (in each case, histograms employ 30 bins, and extend between the visible tick marks. Each (thin) curve corresponds to a single galaxy/halo, color showing its stellar mass, and accounts, as per definition, only of the gas between $0.15$ and $1.00 \times \rm{R_{vir}}$. The thicker curves show medians across halos binned into four galaxy stellar mass quartiles. While the metallicity and thermal pressure distributions are unimodal, gas density, HI density, temperature, and entropy are multimodal, suggesting that the CGM hosts multi-phase gas that forms distinct populations in density, temperature, and entropy. Only temperature and entropy show peaks with strong monotonic trends as a function of stellar mass, with more more massive galaxies hosting gas at higher temperatures and higher entropy in their CGM.}
\label{fig:histPlots}
\end{figure*}

\subsection{Ranges of variation of TNG50 CGM properties}

To quantify the diversity of CGM properties around each galaxy of the TNG50 MW-like sample, Figure~\ref{fig:histPlots} shows histograms of the same six important properties of gas in the CGM: density, HI density, temperature, metallicity, thermal pressure, and entropy. In each panel, the thin curves correspond to individual galaxies, which are colored by the stellar mass of the galaxy. The four thicker curves show the binned behaviour of halos split into four quartiles of stellar mass, as indicated by their colors.

In the distribution of CGM gas density (Figure~\ref{fig:histPlots}, top left panel) a dominant peak occurs at $10^3~\rm{M_\odot}$ kpc$^{-3}$ for most galaxies. A few halos also show a prominent peak at $10^{6.5}~\rm{M_\odot}$ kpc$^{-3}$, which however is no longer visible in the binned curves. The first peak corresponds to a density of $\sim 10^{-4.5}~\rm{cm^{-3}}$ and is characteristic of densities in the outer halo ($\gtrsim \RVIR$), while the second peak ($10^{6.5} \rm{M}_\odot$ kpc$^{-3}$ $\gtrsim 0.1 \rm{cm^{-3}}$) traces star-forming gas, a negligible component of the CGM mass budget. No clear monotonic trend with stellar mass is visible in the binned curves. Similar features are apparent with HI density (top right panel), but in this case, two peaks are observed in the binned curves as well. While the peak at high densities presumably corresponds to largely neutral gas, the peak at lower values corresponds to largely ionized gas.

\begin{figure*}
\centering
\includegraphics[width=16cm]{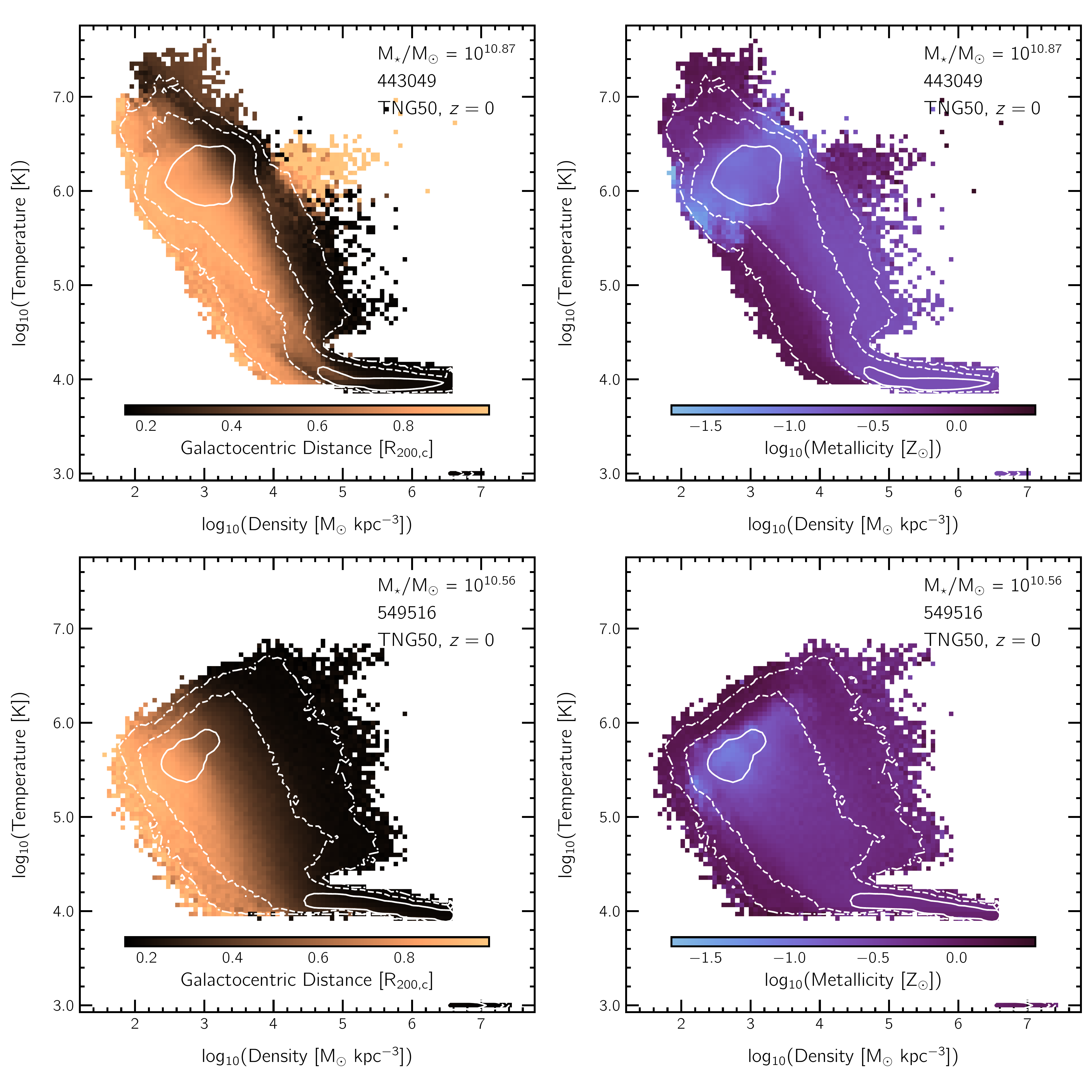}
\caption{Phase diagrams of temperature vs density for gas in the CGM of two particular TNG50 MW-like galaxies. We select these galaxies as characteristic of two rather different subsets of our sample: the supermassive black hole (SMBH) at $z=0$ is releasing energy in the low-accretion state kinetic mode (top row, subhalo ID 443049, which exhibits X-ray bubbles in its CGM: \citealt{pillepich2021}), or in the high-accretion state thermal mode (bottom row, subhalo ID 549516). Panels on the left show phase diagrams colored by galactocentric distance, while those on the right are colored by metallicity. In all cases, contours show where gas mass is actually located in this plane ([solid, dashed, dot-dashed] = [10\%, 1\%, 0.1\%]), outlining those pixels which contain the respective mass fractions, relative to the maximum. Cold, dense gas is preferentially located in the inner regions of the halo, as opposed to the diffuse, warm-hot phase that dominates at larger distances. The kinetic mode channel of SBMH feedback can heat gas to super-virial temperatures ($\gtrsim 10^7$ K), generating metal-rich components which can span the extent of the halo.}
\label{fig:phaseDiagrams}
\end{figure*}

The distributions of gas temperature (Figure~\ref{fig:histPlots}, center left panel) show a secondary peak at 10$^4$\,K for all halos, and a main peak near the virial temperature for halos of this mass ($\sim10^6$ K). Gas colder than 10$^4$\,K is not realized in the simulation (see Section~\ref{TNG}), hence the abrupt cutoff at low temperatures. On the other hand, we place star-forming gas at $\sim 10^3$\,K for visualization, separating the different mass bins horizontally. Notably, the median value for the highest stellar mass quartile is zero, and in general, star forming gas is negligible in comparison to the mass of the rest of the CGM. Integrating these distributions, we find that the mass of cold gas is typically about half an order of magnitude smaller than hot gas mass, consistent with Figures~\ref{fig:cgm_props_vs_mass} and \ref{fig:bary_comp}. The binned temperature distributions evolve monotonically with mass: halos with lower stellar mass have lower virial temperatures. The shape of these distributions is similar at cooler temperatures (irrespective of stellar mass), since this traces the physics of the radiative cooling function.

Observations of the Milky Way halo report a temperature component around $\sim 10^{6.3}$\,K \citep{bluem2022}, which is roughly consistent with the second peak observed in the TNG50 temperature distributions. In addition, observational studies infer gas phases as hot as $\sim 10^{6.5}$\,K \citep[southern galactic sky][]{kaaret2020} and $\sim 10^{6.9}$\,K \citep[northern galactic sky][]{bluem2022} in the Milky Way halo, albeit the abundance of these phases is uncertain. The temperature of gas within the eROSITA bubble(s) (see Introduction)  has  been inferred to be about $10^{6.5}$\,K with Suzaku \citep{kataoka2013} and $10^{6.6-6.7}$\,K from the modelling of OVII and OVIII emission lines \citep{miller2016}. Gas at such high temperatures or even higher exists in a large fraction of TNG50 MW-like galaxies, and is due, in TNG, to galactic-center bubble-like heating events from the SMBH \citep{pillepich2021}. 

Distributions of metallicity (Figure~\ref{fig:histPlots}, center right panel) are distinctly unimodal. They exhibit a peak slightly below solar metallicity ($\sim 10^{-0.4}\, \rm{Z_\odot}$), with the amount of gas mass dropping sharply on either side: on average, the fractional mass of super-solar gas is $\sim 6\%$, while low metallicity gas with $< 0.1\, \rm{Z_\odot}$ accounts for $\sim 11\%$ of the mass budget. No clear monotonic trends are predicted by TNG50 as a function of galaxy stellar mass. While metallicity estimates of gas in the Milky Way halo are largely uncertain, \cite{ponti2022} report a sub-solar value of $\sim 0.05 - 0.1$ Z$_\odot$ along a given line of sight. On the other end of the metallicity spectrum, observations of high velocity clouds reveal the possible existence of highly super-solar clouds with metallicities of $\sim 1.65~Z_\odot$ \citep{zech2008} and $\sim 2.08~Z_\odot$ \citep{yao2011}, hinting towards a wide range of metallicites in the Milky Way halo, similar to that of TNG50.

Thermal pressure distributions (Figure~\ref{fig:histPlots}, lower left panel) are also similarly unimodal, and peak at $\sim 10^2$ K cm$^{-3}$. Binned curves show no clear dependence across the four stellar mass quartile bins. We remind the reader here that the pressure for star forming gas is derived from the effective equation of state.

Given that entropy is a linear function of temperature, the corresponding distributions ((Figure~\ref{fig:histPlots}, lower left panel), lower right panel) behave similarly to those of temperature, as expected. There is a subtle peak at low entropy ($\sim~10^5$ K cm$^{2}$), followed by a dominant peak at $\sim~10^9$ K cm$^{2}$. On average, only $\sim 20\%$ of gas has an entropy lower than $\sim 10^7$ K cm$^{2}$. As with the temperature distributions, the binned curves of halos with larger stellar mass peak at larger values of entropy.

It is somewhat surprising that multimodal distributions are not seen in the thermal pressure distributions, given the presence of distinct density and temperature components in the CGM gas. This suggests that these different phases are in fact in pressure equilibrium. Similarly, the lack of multimodality in the metallicity distributions is a bit unexpected, since the center-right panel of Figure~\ref{fig:radial_profiles} suggests incomplete mixing of metals.

\subsubsection{Cross-correlations among CGM phases}

To explore the relationship between properties of CGM gas across different components, Figure~\ref{fig:phaseDiagrams} shows phase diagrams (i.e. two-dimensional distributions of temperature vs density) for two particular galaxies. We select these two galaxies to be in two different states with respect to their SMBH feedback modes. The first (top row) is in the low-accretion state, and therefore in the kinetic feedback mode of the TNG model: this galaxy exhibits multiple X-ray bubbles in its CGM \citep[as indicated by][top right panel of their Figs. 2 and 3]{pillepich2021}. The second galaxy (bottom row) is in the high-accretion state, and so in the thermal or `quasar' feedback mode \citep{weinberger2017}. Broadly speaking, the former case has had ongoing and recent effective SMBH feedback, while the second case is still dominated, in terms of feedback energy or feedback effects, by the supernovae explosions of stars. The former has a relatively high virial temperature, $\rm{T_{vir}} \sim 10^{6}$\,K, while the latter is slightly cooler at $\rm{T_{vir}} \sim 10^{5.6}$\,K, a difference which primarily arises due to the difference in halo virial masses of the two systems.

In the left panels of Figure~\ref{fig:phaseDiagrams}, we color the phase diagrams by gas galactocentric distance normalised by halo virial radius. We see that dense gas, near or past the star-formation threshold, is preferentially located close to the center, as expected. A diagonal gradient is evident, such that moving diagonally from the upper right to towards the lower left of the phase diagram, i.e. towards low density and low temperature, corresponds to moving outwards in radius. However, in the kinetic AGN feedback example (top), we also find dense (hot) gas at large galactocentric distances, driven away by the intense activity of the SMBH in the kinetic feedback mode. No such feature is present in the halo in the thermal mode AGN feedback.

In the right panels of Figure~\ref{fig:phaseDiagrams}, we instead color the gas phases by metallicity. Gas with average metallicities between $\sim 10^{-1} - 10^{-0.5}\, \rm{Z_\odot}$ is predominantly at the virial temperature, and at low densities of $\sim 10^3$ M$_\odot$ kpc$^{-3}$. At these densities, gas that is hotter than the virial temperature has higher metallicity, which is again presumably linked to feedback processes and outflows from the galaxy. We again note the presence of a gradient that extends diagonally down and to the right, from the dark cloud, towards the star-forming gas. This likely reflects a cooling channel, i.e. the path that gas takes through the phase diagram as it cools. At the high-density end, we note differences between the two galaxies: in the kinetic SMBH feedback example (top), we see that regions of high-density/high-temperature have relatively higher metallicities than the rest of the high-density regions, a signature of enriched outflows driven by the SMBH-driven outflows \citep[see also][]{pillepich2021}.

To visualize the relative abundance of gas mass across these phase diagrams, we also draw three white contours on each panel: the solid, dashed, and dot-dashed curves trace those pixels that have [10\%, 1\%, 0.1\%] mass fractions with respect to the maximum value in each diagram, respectively. For both galaxies, we note that the majority of the mass is in either (i) dense, cold gas with T$\sim 10^4$K, or (ii) in diffuse gas close to the virial temperature, which is consistent with the bimodal structure of the one-dimensional temperature distributions examined earlier.

\subsection{Kinematics and further connections to SMBH feedback}

\begin{figure*}
\centering
\hspace{1.1cm}
\includegraphics[width=6.4cm]{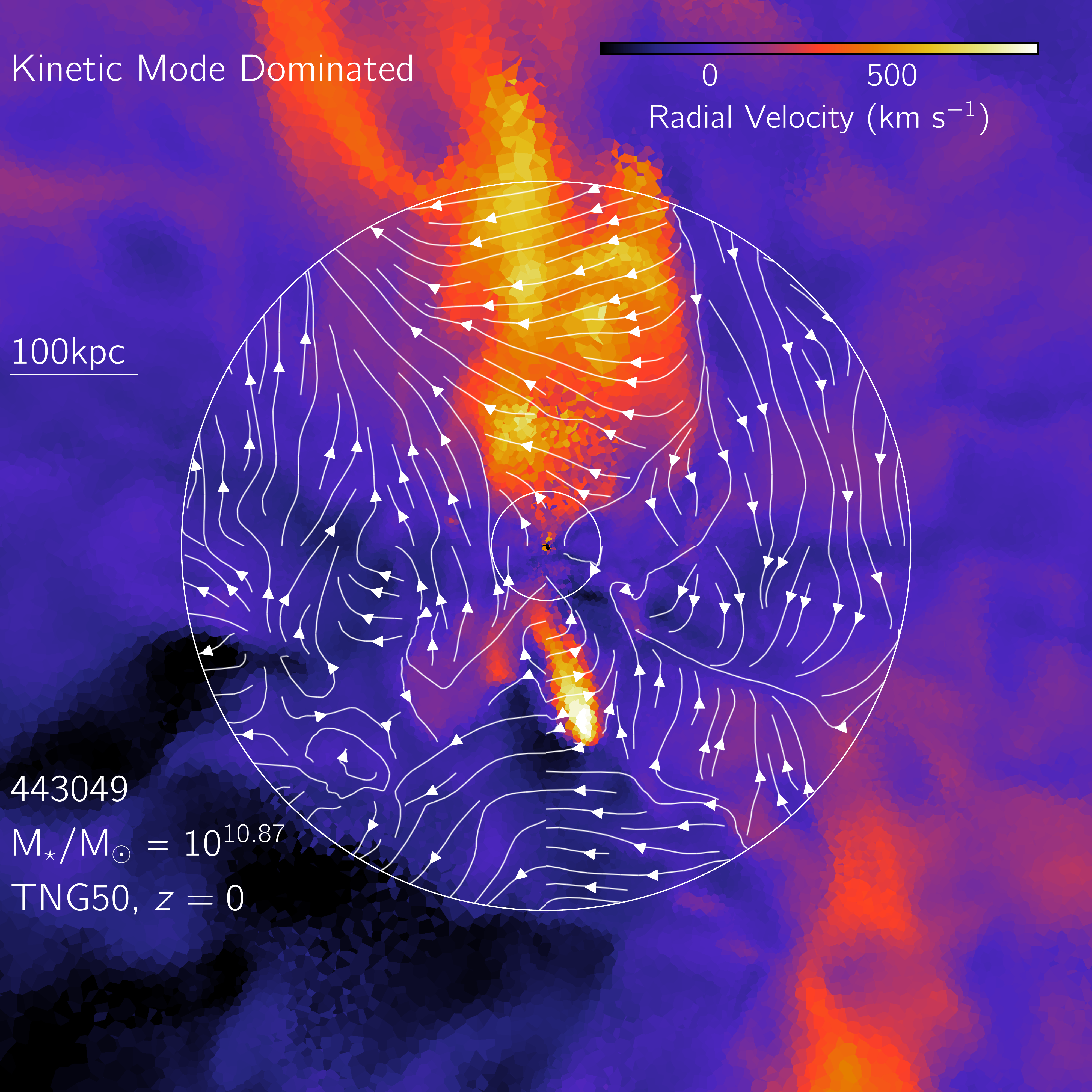}\hspace{1.6cm}
\includegraphics[width=6.4cm]{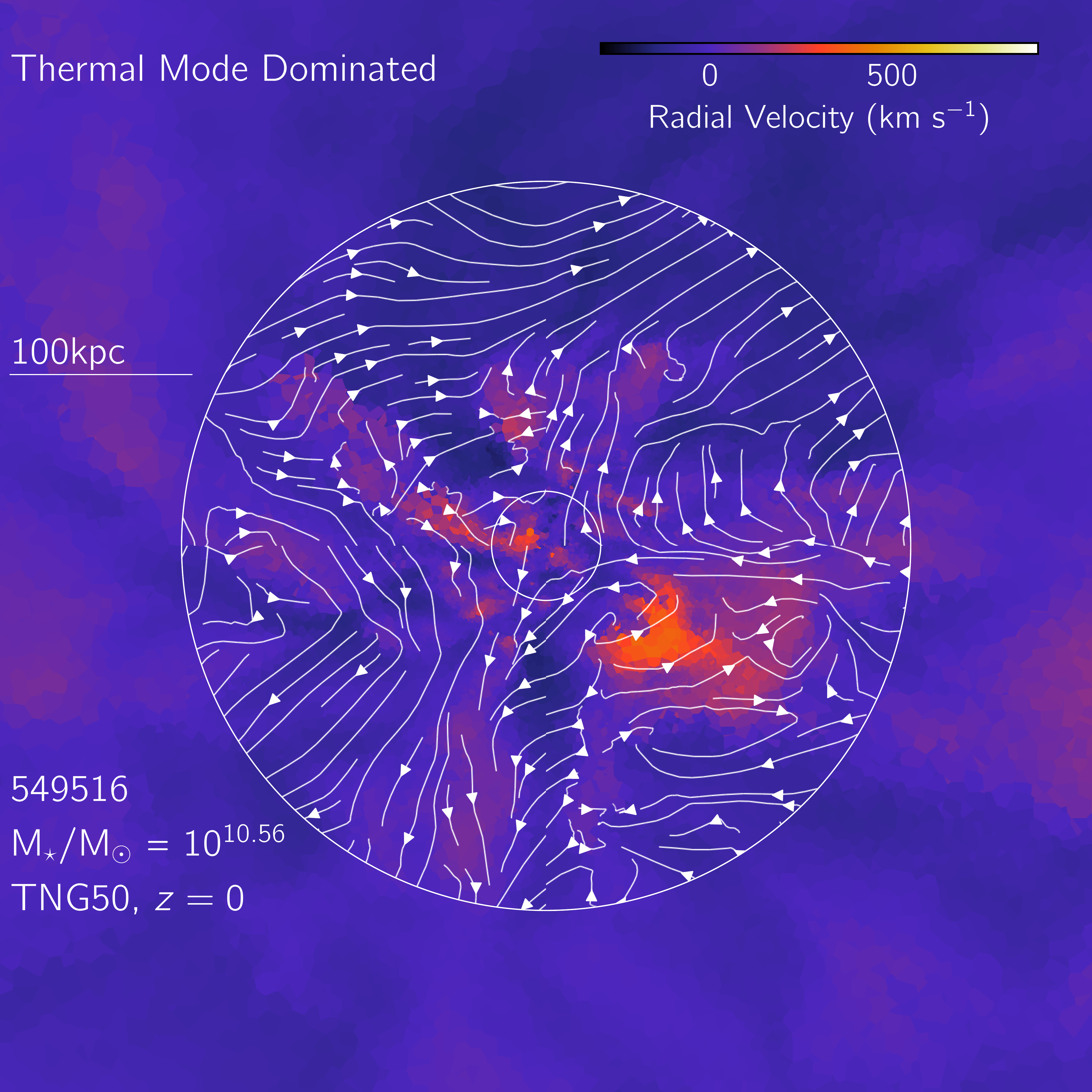}
\includegraphics[height=8cm]{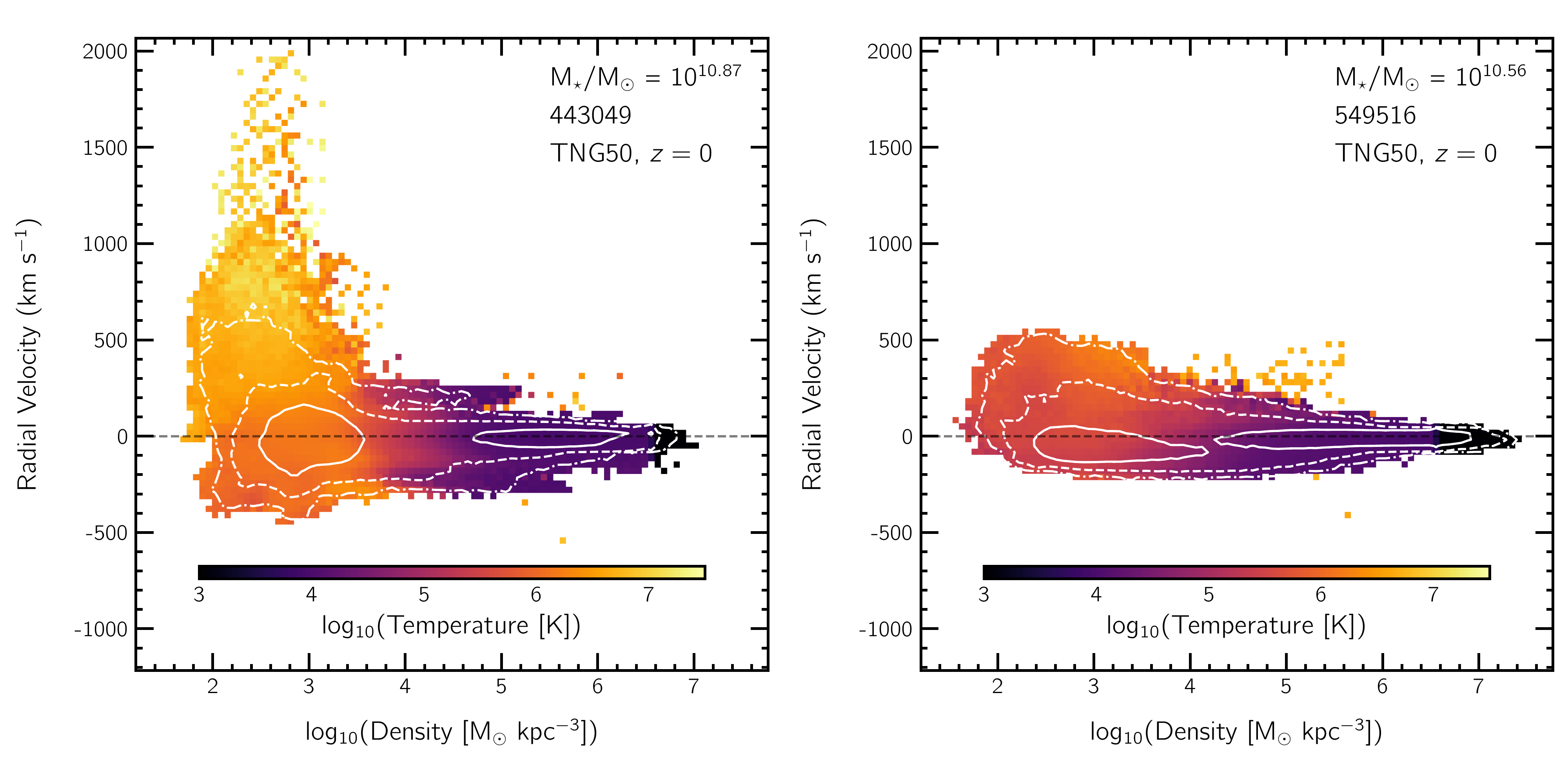}
\caption{Gas kinematics in the CGM and their relation to SMBH feedback from the galaxy's center according to TNG50 at $z=0$. The top panels visualize gas radial velocity (negative denoting inflow, and positive denoting outflow), in the frame of reference of the galaxy. We orient the galaxies edge-on, and take an infinitesimally thin slice through the midplane of the halo. The left and right columns show two different galaxies, the same as previously shown for the phase diagram analysis. Namely, the TNG50 MW-like galaxy on the left is an example dominated by SMBH feedback in the kinetic mode (low-accretion rate; subhalo ID 443049), whereas the TNG50 MW-like galaxy on the right is experiencing only SMBH feedback from the thermal mode (high-accretion rate; subhalo ID 549516). The bottom row shows phase diagrams of radial velocity versus density, colored by temperature. Contours show where gas mass is actually located in this plane ([solid, dashed, dot-dashed] = [10\%, 1\%, 0.1\%]), outlining those pixels which contain the respective mass fractions, relative to the maximum. The galaxy on the right has only modest outflows up to $\sim500$ km s$^{-1}$, with no strong directionality. In contrast, the galaxy on the left has outflows up to $\sim2000$ km s$^{-1}$, which extend to large distances and are rather asymmetric, and which are dominated by hot $\sim 10^7$ K gas.}
\label{fig:radVelSlice}
\end{figure*}

The preceding analyses reveal several examples of relationships predicted by the TNG model between the properties of CGM gas and properties of the central galaxy. Previous works have shown that the TNG model of kinetic SMBH feedback effectively drives high-velocity outflows in galaxies \citep{weinberger2017, nelson2019b, pillepich2021}. Here, we therefore further explore the connection of SMBH feedback with the kinematics of CGM gas in the case of TNG50 MW-like galaxies and quantify the complexity of the velocity fields in the CGM of MW-like galaxies. Specifically, we distinguish galaxies based on their modes of SMBH feedback in the TNG model: high-accretion thermal mode feedback vs low-accretion kinetic mode feedback. 

In Figure~\ref{fig:radVelSlice}, we first provide a visualization to motivate how the mode of SMBH feedback affects gas kinematics, at least in TNG50. The top panels show slices, with the galaxies having been rotated to an edge-on orientation, of radial velocities for two galaxies. On the left, we choose a prototypical TNG50 MW-like galaxy where the total (i.e. cumulative in time) energy injection by SMBH feedback has been dominated by the kinetic mode, and the SMBH is accreting gas at a sufficiently low Eddington ratio at the present day ($z=0$), that it is also in the kinetic mode. On the right, we choose a representative TNG50 MW-like galaxy in which the thermal mode has been the dominant channel for SMBH feedback energy injection. 

Two main differences are immediately apparent in Figure~\ref{fig:radVelSlice}: the (outflowing) radial velocities are much larger in the first case (left column), and this outflowing gas is preferentially located perpendicular to the disk. In the thermal mode dominated galaxy (right column), SMBH feedback does not produce, within the TNG implementation of thermal energy injection, such a high-velocity outflow, nor one with such a strong spatial dependence. We note, however, that other implementations of thermal energy injection from SMBHs, different from the TNG model, may be able to drive fast outflows: see the discussion in \cite{pillepich2021}.

As mentioned in Section \ref{TNG}, energy injection through both these modes is isotropic at the injection scale. The directionality of the outflows in the kinetic mode instead stems from the interaction of the outflows with gas in the disk and/or CGM \citep{nelson2019, peroux2020, pillepich2021, truong2021}. As discussed in previous works, the kinetic mode of SMBH feedback in the TNG model is an efficient quenching mechanism due to a combination of strong ejective feedback i.e. outflows, together with a preventative heating of the CGM \citep{nelson2018, zinger2020, terrazas2020}. The galaxy shown in the left panel is indeed in the process of quenching and departing from the main sequence, thereby moving into the green valley.

The bottom panels of Figure~\ref{fig:radVelSlice} show phase diagrams of radial velocity vs density, colored by temperature, for the same two galaxies. We see that radial velocities are larger in the galaxy undergoing kinetic mode SMBH feedback (left), where outflows up to $\sim2000$ km s$^{-1}$ are evident, although such high speeds are attained only by largely low-density gas ($\lesssim 10^4$ M$_\odot$ kpc$^{-3}$). This is in contrast to the second galaxy undergoing thermal mode SMBH feedback, where supernovae and stellar feedback in general are primarily responsible for driving galactic-scale outflows (right), where speeds reach only $\sim500$ km s$^{-1}$. In addition, gas temperatures are higher in the first case ($\gtrsim 10^7$ K), and this hot gas component outflows the fastest.

In both cases, we note that the majority of cold, dense gas, on average, is neither strongly outflowing nor inflowing, i.e. this is the material which makes up the rotationally-supported gaseous disk centered at zero radial velocity. However, it is believed, both in observations \citep{teodoro2020} and simulations \citep{schneider2020}, that non-negligible amounts of cold gas exist in outflows, although the mechanism through which cold gas is accelerated, or produced, in outflows remains an open question.

\begin{figure*}
\centering
\includegraphics[height=8cm]{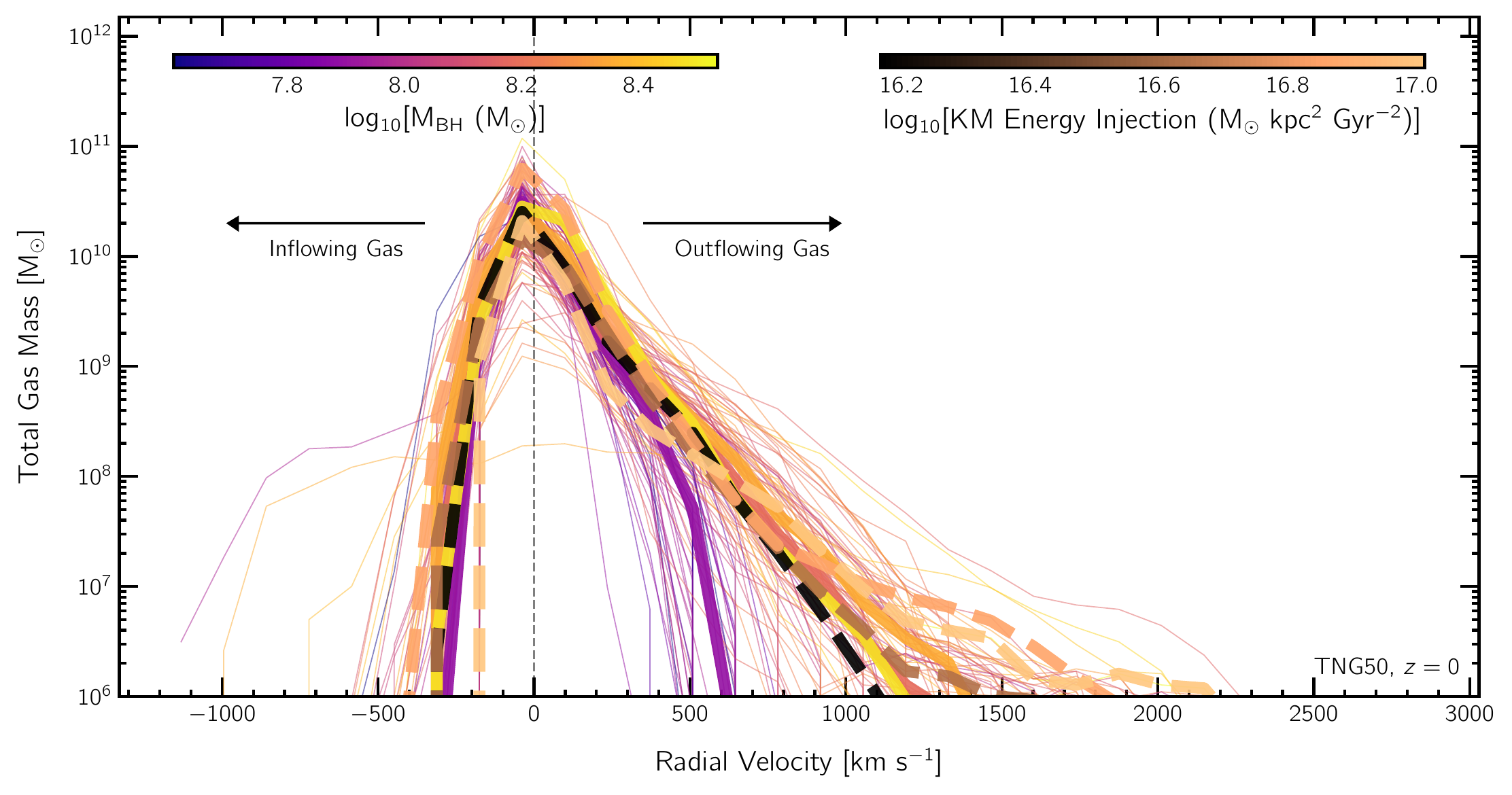}
\includegraphics[height=7.6cm]{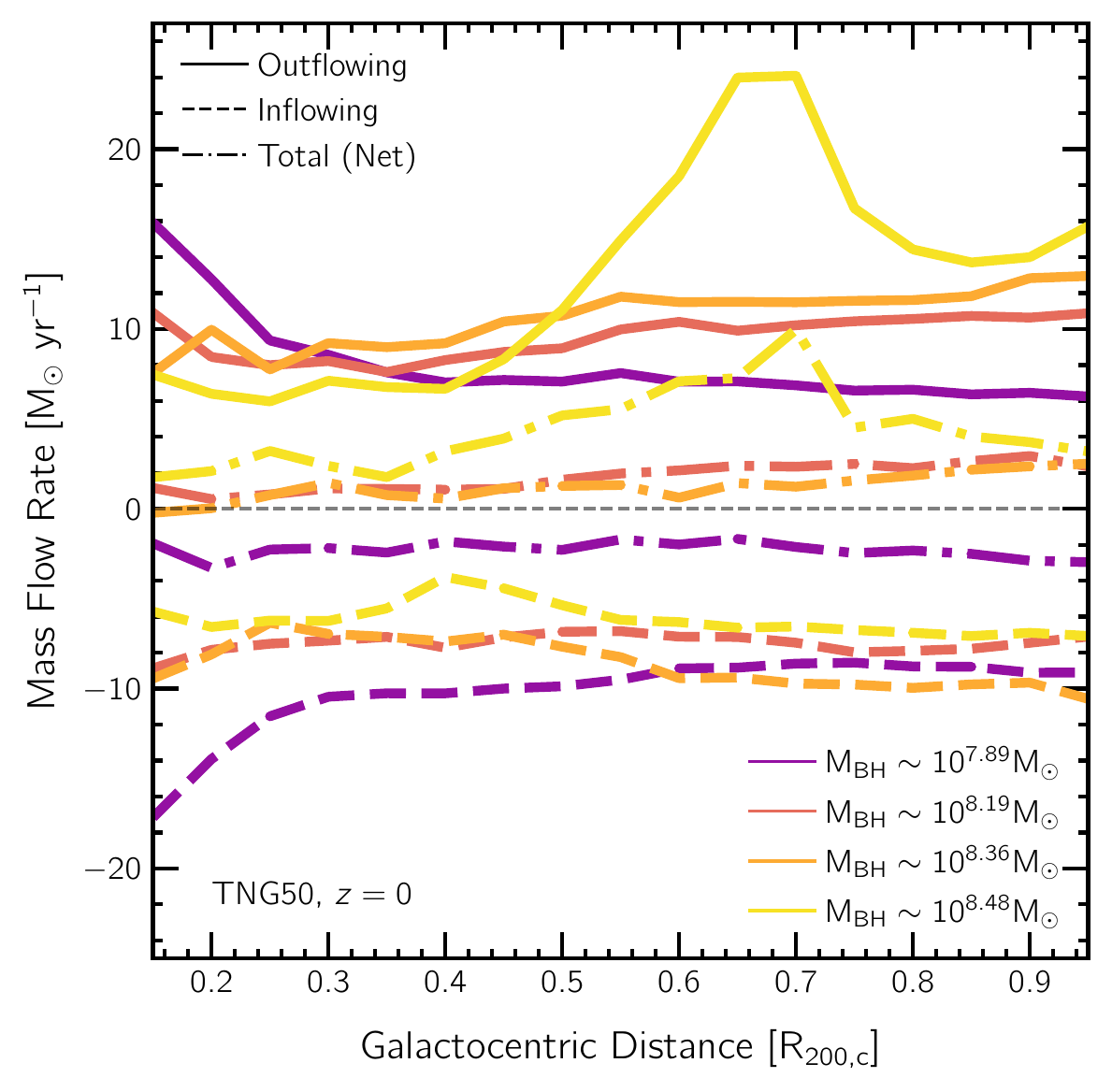}
\includegraphics[height=7.6cm]{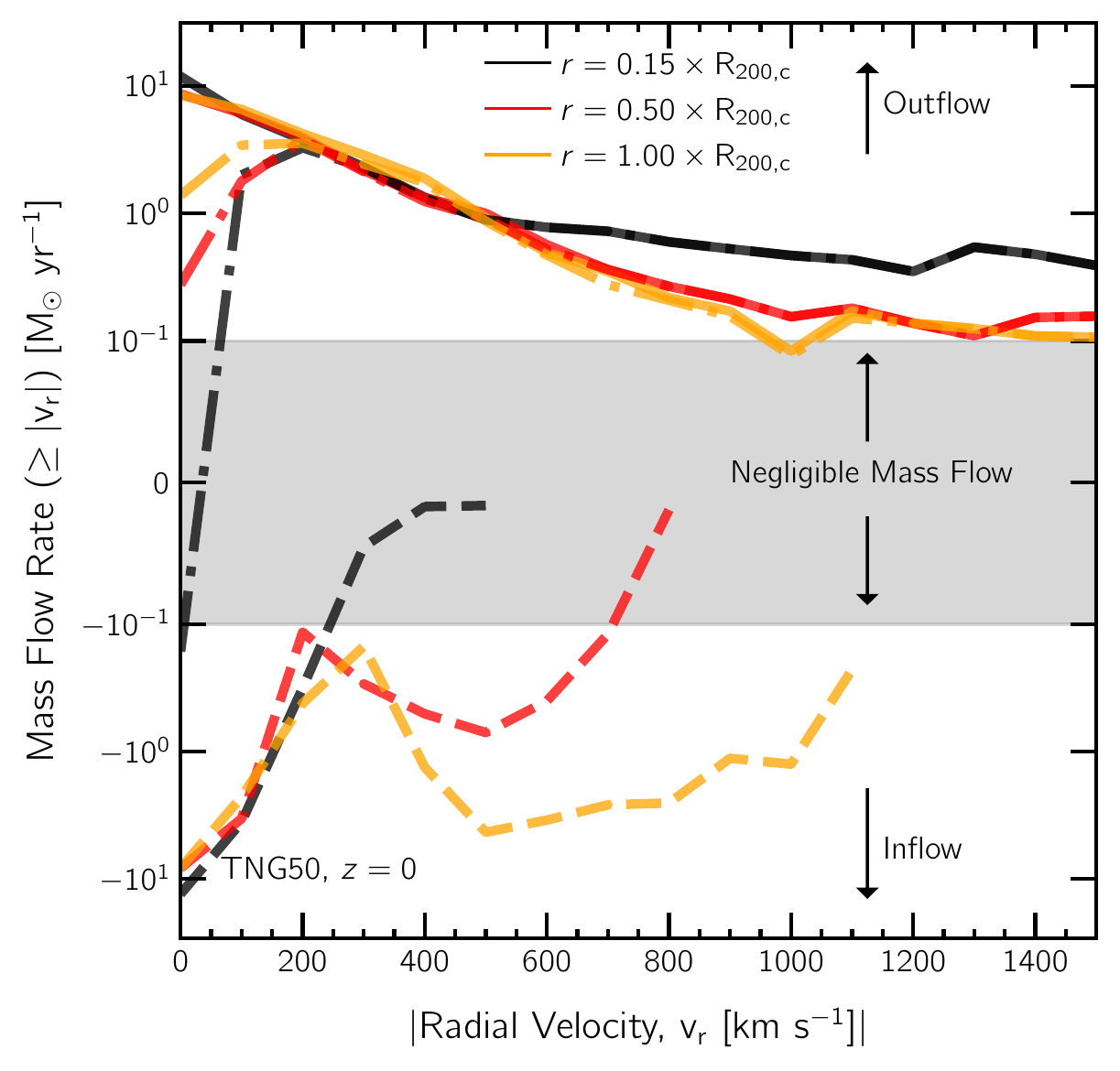}
\caption{The kinematics and velocity structure of gas flows, and the balance between inflows and outflows, in the CGM of $z=0$ MW-like galaxies in TNG50. In the top panel, we show distributions of gas radial velocity (histograms employ 30 bins, and extend (exactly) between -1100 and 3000 km s$^{-1}$). Thin curves correspond to individual galaxies, colored by SMBH mass (top left colorbar), while the two sets of thick curves are binned into four quartiles each: solid in bins of SMBH mass (top left colobar), and dashed in bins of total, time-integrated kinetic mode (KM) energy injection from the SMBH (top right colobar). We see that galaxies with larger total kinetic mode energy injection have tails of higher-velocity outflows, representing the integrated impact of multiple episodes of kinetic-mode energy injection. The bottom-left panel shows the (median) mass outflow rate as a function of galactocentric distance, with galaxies split into four quartiles based on their SMBH mass (four colors). We use solid, dashed and dot-dashed lines to represent outflowing, inflowing and all gas, respectively. We find that galaxies with the least massive  SMBHs are dominated by inflows at all distances, while the other three bins are dominated by outflows at almost all distances. The bottom-right panel shows mass flow rates of gas moving at a given absolute radial velocity, or faster (i.e. cumulative in velocity). Once again, we use solid, dashed and dot-dashed lines to represent outflowing, inflowing and all gas, respectively, averaged across all 132 MW-like galaxies. The black, red and orange lines refer to outflow rates across shells at radii 0.15, 0.5 and 1.0 R$_{\rm{200c}}$, respectively. We see that gas outflows at larger velocities at smaller galactocentric distances, while the opposite is true for inflows.}
\label{fig:outflowRates}
\end{figure*}

In Figure~\ref{fig:outflowRates}, we quantify halo gas kinematics and inflow/outflow behavior across our full sample of 132 TNG50 MW-like galaxies. The top panel shows distributions of radial velocity of CGM gas. Each thin curve corresponds to the CGM of an individual galaxy, colored by mass of the SMBH. The solid thick curves show median values for four quartiles, with the galaxy sample split based on SMBH mass. The dashed thick curves also show median lines for four quartiles, but with the sample split based on the total, time-integrated (i.e. cumulative in time) kinetic mode SMBH energy injection of the SMBH.

Most of the distributions are skewed towards positive radial velocities (outflowing gas), although the radial velocities of the bulk of all gas mass are close to zero, i.e. most gas is neither strongly inflowing nor outflowing. The solid median lines show that the galaxies with the least massive SMBHs have smaller high-velocity tails, in comparison to those with more massive SMBHs. However, this trend is not monotonic, i.e. the quartile corresponding to the most massive SMBHs does not have the highest velocity outflows. This arises because the connection between SMBH mass and energy injection in a given mode is not one-to-one. The thick dashed lines make this clear: they show a steady monotonic trend where galaxies whose SMBHs have injected progressively more total energy in the kinetic mode have the highest outflow velocity tails. Although not shown here, such a trend is not seen when galaxies are split into quartiles based on their total energy injection in the thermal quasar mode.

In order to also consider spatial information, the bottom left panel of Figure~\ref{fig:outflowRates} shows three distinct mass flow rates as a function of distance. In solid, dashed and dash-dot lines, we show the mass outflow-, mass inflow-, and net mass flow-rates, respectively, averaged across subsets of TNG50 MW-like galaxies. As above, we split the sample into quartiles based on SMBH mass (from least to most massive: purple, red, orange, yellow). We find that MW-like galaxies with less massive SMBHs have greater outflow rates at smaller distances (i.e. at the inner boundary of the CGM), with values decreasing towards larger distances. The trend reverses for galaxies with more massive SMBHs, where outflow rates are higher at larger distances. We interpret this as the effect of the transition from thermal to kinetic SMBH feedback mode: in TNG, less massive SMBHs are predominantly in the thermal mode, while more massive SMBHs are dominated by kinetic mode feedback.

Similarly, inflow rates (dashed lines) are larger for galaxies with less massive SMBHs at small galactocentric distances, with the inflow rate reducing towards the virial radius. The increased inflow rate close to the disk for galaxies in the lowest quartile indicates a cooling flow regime, wherein cooling turns more efficient at high densities. For galaxies with more massive SMBHs, however, no strong trends with respect to distance are observed, with the gas inflow rate remaining rather flat through most of the CGM. Finally, the net flow rates (dot-dashed lines) are low, of order a few solar masses per year, and have little dependence on galactocentric distance. Only galaxies with the least massive SMBHs have a net (median) inflow rate at all distances. The other three bins have positive net mass flow rates at almost all distances; galaxies with more massive SMBHs have slighly larger net outflow rates throughout the CGM. Overall, the majority of circumgalactic gas in Milky-Way like halos is neither strongly outflowing nor inflowing. 

The bottom right panel of Figure~\ref{fig:outflowRates} shows mass flow rates as a function of absolute radial velocity, or faster (i.e. cumulative in velocity). Once again, in solid, dashed and dash-dot lines, we show the mass outflow-, mass inflow-, and net mass flow-rates, respectively. However, in this case, we show median lines for the entire sample (i.e. without splitting into quartiles). Instead, we show the outflow rate at three radii: $\{0.15, 0.5, 1.0\} \times \rm{R}_{\rm{200c}}$ -- black, red, and orange, respectively. Similar to the top panel, we observe that most gas is moving slowly, with both inflow and outflow rates decaying quickly towards high velocities, at all distances. For instance, at a distance of $0.15 \RVIR$, the inflow rate is $\sim 10~\rm{M_\odot~yr^{-1}}$ when all inflowing gas is considered (i.e. all gas with $\rm{v_r} < 0$ km s$^{-1}$), dropping quickly to $\sim 1~\rm{M_\odot~yr^{-1}}$ when only gas inflowing faster than 200 km s$^{-1}$ is considered. Similar numbers hold for outflows as well.

For outflows, we see that the rate of slowly outflowing gas is roughly independent of distance. At high speeds ($\gtrsim 600$ km s$^{-1}$), outflow rates begin to depend on distance: such outflows are faster at smaller distances. For instance, the outflow rate for gas with $v_{\rm{rad}} > 600$ km s$^{-1}$ at a distance of $0.15 \RVIR$ is $\sim 1~\rm{M_\odot~yr^{-1}}$, but is only $\sim 0.3~\rm{M_\odot~yr^{-1}}$ at the virial radius. At even higher speeds ($\gtrsim 1200$ km s$^{-1}$), the outflow rate at large distances is negligible. There is little gas moving at such large velocities so far from the galaxy itself.

Similarly, the rate of gas inflow at small velocities is roughly independent of distance, possibly indicating the existence of significant sub-centrifugal rotation throughout the CGM, resulting in deviations from local hydrostatic equilibrium \citep{oppenheimer2018}. However, inflow rates at large velocities ($\gtrsim 250$ km s$^{-1}$) decrease towards smaller radii; only gas at large distances from the centre has large inflow speeds. We interpret this as the interaction of accreting gas with gas in the halo, which slows down due to drag and/or interactions with outflows. Finally, the median net flow rate is almost always positive (i.e. gas is outflowing), except for slowly moving material with $\lesssim 100$ km s$^{-1}$, where the net flow rate drops essentially to zero. In conjunction with the fact that slowly moving material dominates the mass budget of the CGM (top panel), we conclude that a majority of gas is in quasi-static equilibrium.

This quantification of the kinematics of CGM gas makes it clear that the gaseous halo of TNG50 MW-like galaxies hosts not only a complex multi-phase structure, but also a complex dynamical structure that is closely linked to energetic feedback events from the central galaxy and its SMBH.

\begin{figure*}
\centering
\hspace{0.7cm}
\includegraphics[width=6.5cm]{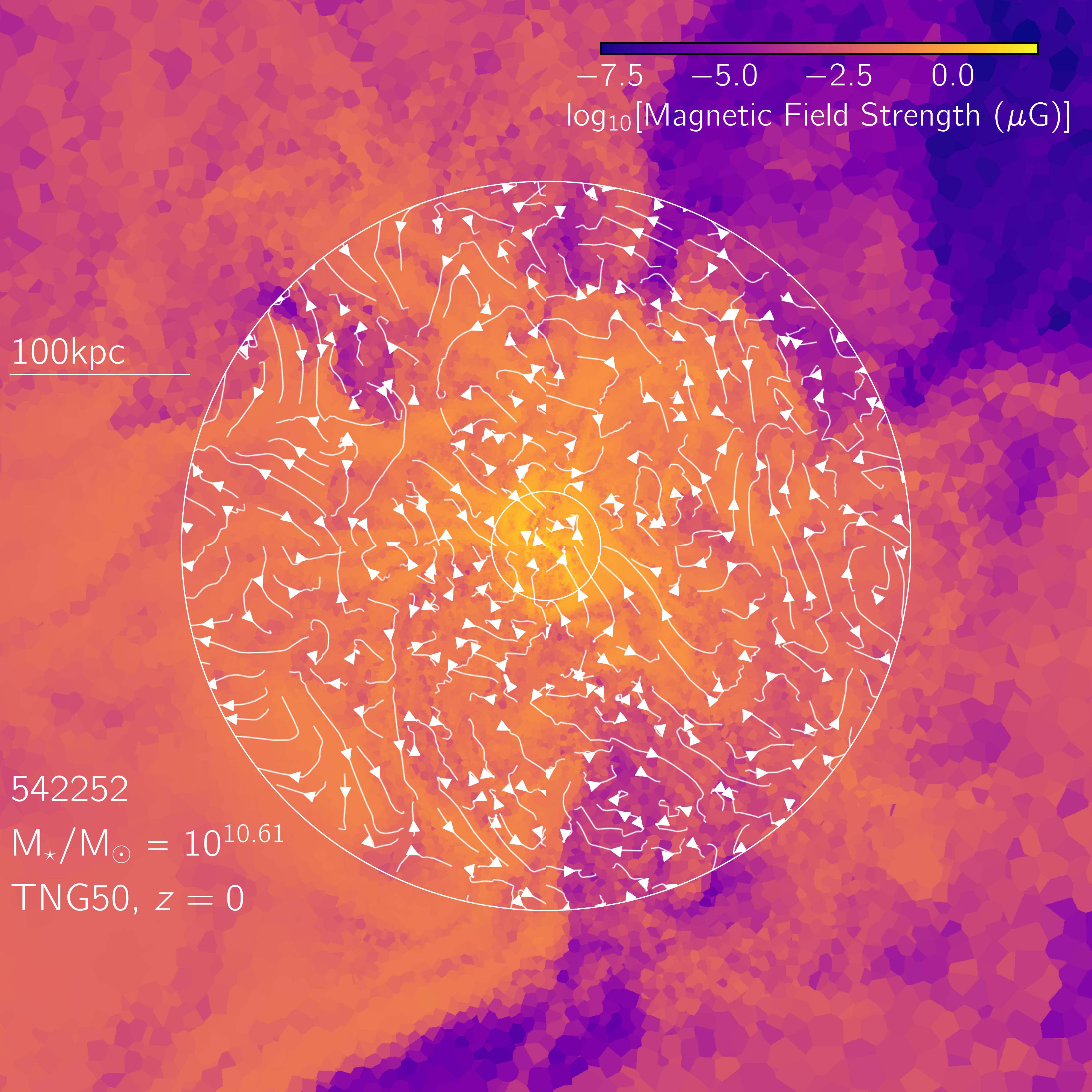}\hspace{1.2cm}
\includegraphics[width=6.5cm]{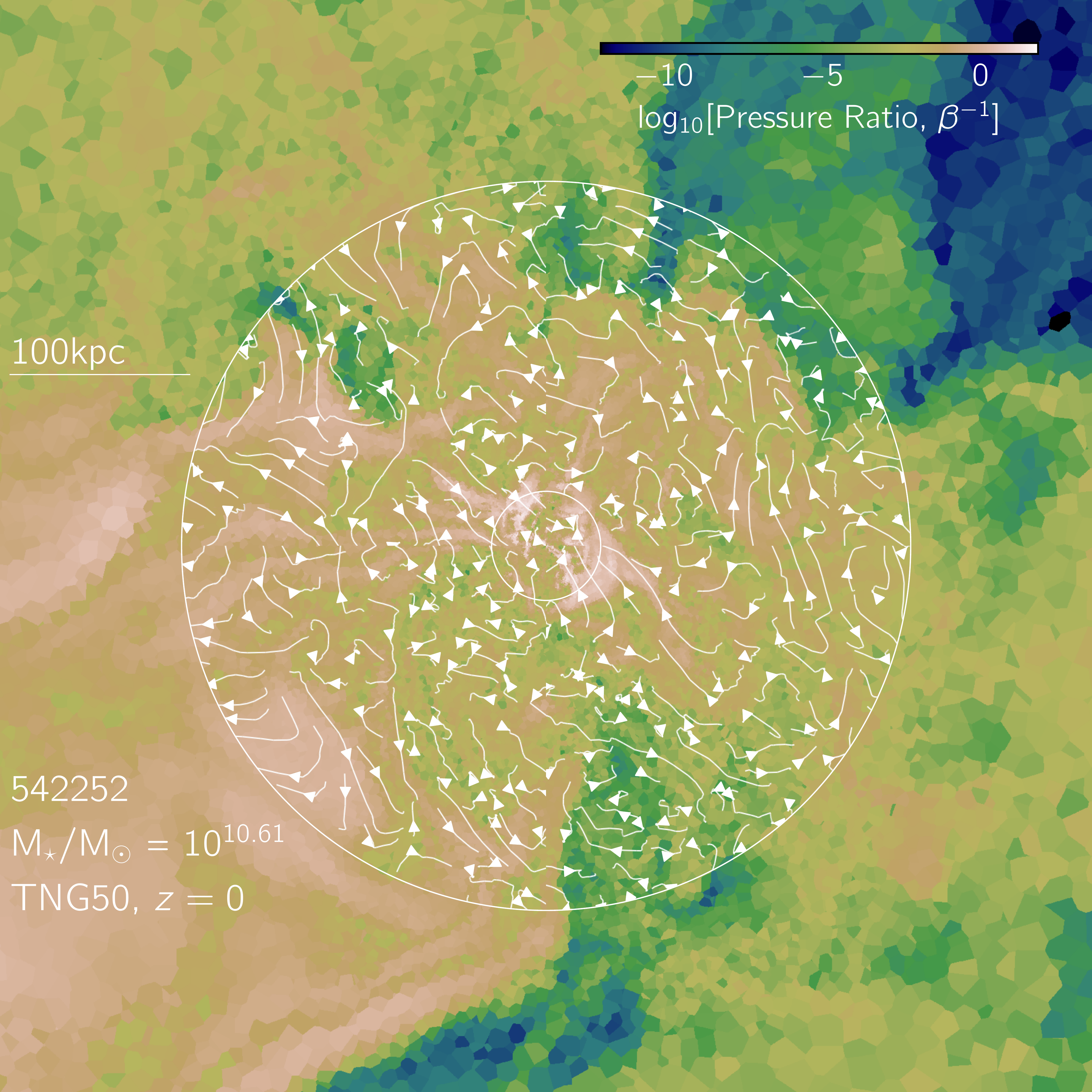}
\includegraphics[width=15.5cm]{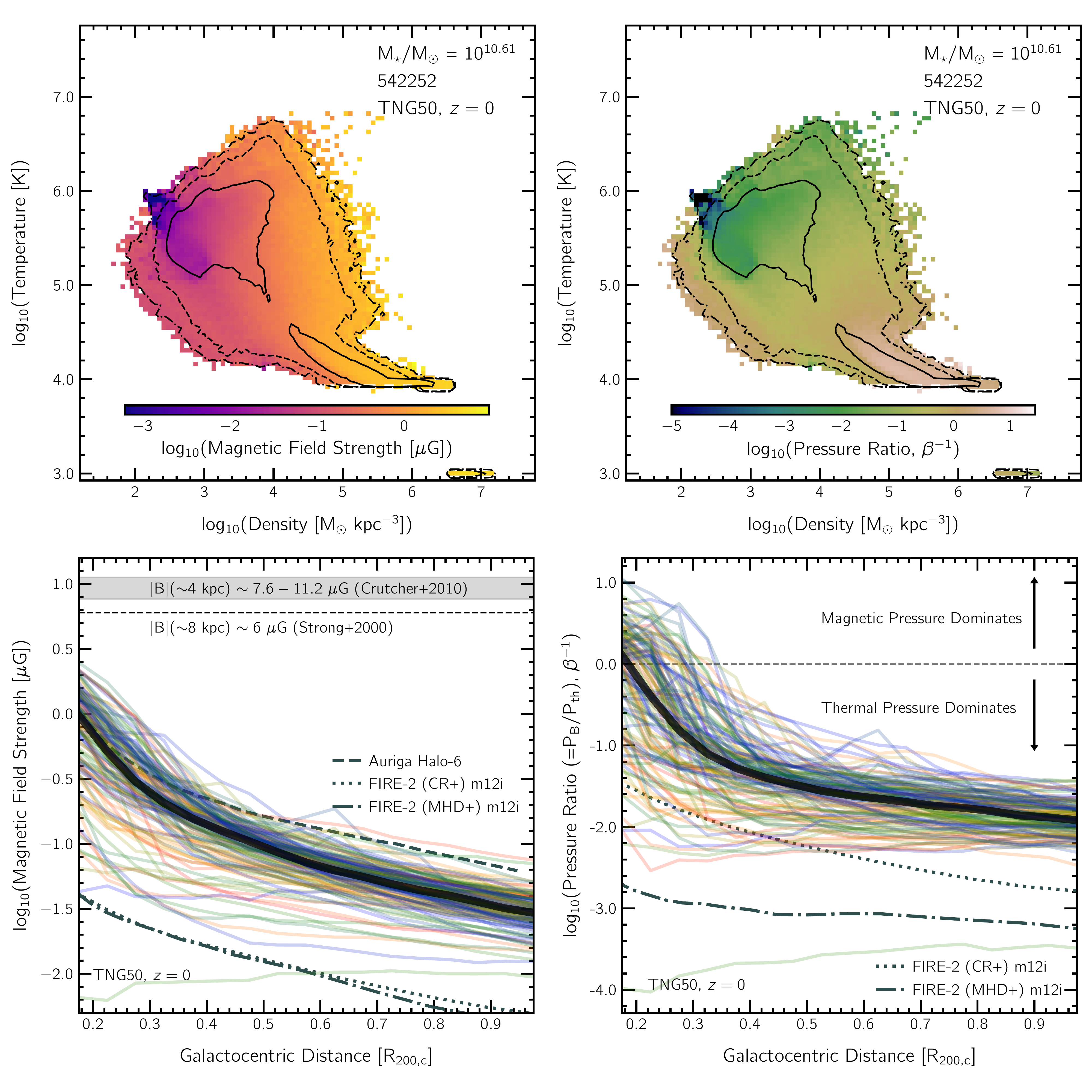}
\caption{The structure, strength, and importance of magnetic fields in the CGM of MW-like galaxies in TNG50. Top and middle panels: magnetic fields properties of the same MW-like galaxy previously visualized in detail (subhalo ID 542252). In the top, we show slices of magnetic field strength (left) and pressure ratio (right). For these projections, the galaxy is oriented edge-on and streamlines show the x-/z-components of the B-fields, i.e. the local direction of the magnetic fields in the plane of the image. The middle panels show phase diagrams in the temperature-density plane, colored by the same two B field quantities. Both the magnetic field strength $|B|$ and pressure ratio ($\beta^{-1}$) are highest in the inner CGM. Bottom panels: spherically-averaged radial profiles of magnetic field strength and magnetic-to-thermal ratio for all the 132 MW-like galaxies in TNG50. Color represents stellar mass, as previously.}
\label{fig:bField}
\end{figure*}

\subsection{Magnetic fields in the CGM of TNG50 MW-like galaxies}

A modeling element that sets apart the TNG simulations from many other similar projects is the inclusion of MHD. Therefore, we close this analysis by quantifying the existence and properties of magnetic fields in the CGM of MW-like galaxies. In the TNG simulations, a uniform (weak) primordial magnetic field ($10^{-14}$ comoving Gauss) is seeded at the start of the simulations ($z=127$), which is then amplified due to the process of structure formation and feedback \citep{marinacci2018}. By low redshift, magnetic fields in halos obtain significantly higher strengths, which could influence the evolution and dynamics of halo gas \citep{pakmor2020}, possibly by playing a role in suppressing fluid instabilities \citep[e.g.][]{berlok2019, sparre2020}. Here we focus on two related properties: magnetic field strength $|B|$ and the pressure ratio $\beta^{-1}$ of magnetic to thermal pressure components in the gas.

Figure~\ref{fig:bField} shows our key results for the magnetic fields in the halos of MW-like galaxies from TNG50 at $z=0$. In the top and middle panels, we visualize the magnetic field properties for a single galaxy, which is oriented edge-on. This is the same system shown in eight different visualized quadrants in Figures~\ref{fig:theoryMwImages} and \ref{fig:obsMwImages}. In the top, we show slices of magnetic field strength (left) and pressure ratio (right): the magnetic field strength is generally higher closer to the disk/inner region of the CGM, as a result of which the magnetic pressure is also the highest in the same regions. However, the structure at fixed distance is complex, in terms of variation in magnetic field strength.

The middle panels of Figure~\ref{fig:bField} show phase diagrams for the same galaxy as the top panels. On the left, we color by the magnetic field strength, while we color by the pressure ratio on the right.\footnote{As before, contours show where gas mass is actually located in this plane ([solid, dashed, dot-dashed] = [10\%, 1\%, 0.1\%]), outlining those pixels which contain the respective mass fractions, relative to the maximum.} We see that cold, dense gas has the strongest magnetic fields, while diffuse, hot gas has the weakest. A diagonal gradient is present in the bottom left panel: as one moves diagonally from high- temperature and high-density gas (top right) to lower-temperature and lower-density gas (lower left), the magnetic field strength decreases, tracing the increase in galactocentric distance. Although weak, the reverse trend is present in the bottom right panel: the pressure ratio increases gradually diagonally towards lower-temperature and lower-density gas (lower left), since this is the direction in which thermal pressure tends to decrease.

Finally, the bottom panels of Figure~\ref{fig:bField} give the spherically-averaged radial profiles of the same two magnetic-field quantities as a function of galactocentric distance, for all 132 TNG50 MW-like galaxies: thin curves, colored by galaxy stellar mass as in the previous Figures. These are mass-weighted average magnetic field properties at the given galactocentric distance. These plots quantify the visual impressions above: the median magnetic field strength is roughly one and a half orders of magnitude greater at the inner boundary of the CGM as compared to the outer boundary, dropping from $\sim 1 \mu$G at $0.2 \RVIR$ to $\sim 0.1 \mu$G at $0.5 \RVIR$, to $\sim 0.03 \mu$G at the virial radius (in the median). In addition, the magnetic field strength, on average, is larger in the CGM of galaxies with lower stellar masses (at all distances).

We compare to other simulation results for the magnetic field profile around a MW-like galaxy from the Auriga simulations \citep[Au6;][]{voort2021}. While the median value of TNG50 MW-like galaxies is similar to the Au6 profile (dark gray dashed line) close to the inner boundary of the CGM, differences begin to arise at $\sim 0.25 \rm{R_{200c}}$, beyond which the TNG50 median curve is moderately lower, by up to $\sim0.3$ dex. However, the Au6 profile is within the galaxy-to-galaxy variation predicted by TNG50 for MW-like galaxies. While Auriga is also run with \textsc{AREPO}, and utilises the same MHD solver as TNG, along with the same primordial seed strength, differences in the galaxy physics model, i.e. feedback physics, could contribute to the offset of the Au6 profile with respect to the TNG50 median. This is, further, a comparison of a single halo versus the TNG50 sample, and halo to halo diversity exists. In addition, Au6 is simulated at a slightly higher resolution ($m_{\rm{gas}} \sim 5.4 \times 10^4~\rm{M_\odot}$, plus added refinement to maintain a spatial resolution of $\sim1$ kpc in the CGM) with respect to TNG50 ($m_{\rm{gas}} \sim 8.4 \times 10^4~\rm{M_\odot}$). Slightly stronger magnetic field strengths are therefore expected, given that the late time strengths increase slightly with better numerical resolution \cite{marinacci2015}.

We also show results \citep{ponnada2022} from one of the simulations based on the FIRE-2 model, where magnetic fields were also included (MHD+). We note here that the MHD solver utilized in the FIRE-2 simulations is quite different from the one used in TNG, and so is the mass resolution of gas ($m_{\rm{gas}} \sim 7 \times 10^3~\rm{M_\odot}$). The magnetic field strength in the FIRE-2 MHD+ m12i run (dark gray dot-dashed line) is offset by $\sim -1.5$ dex with respect to the TNG50 median at all distances. We also include the curve corresponding to the FIRE-2 CR+ m12i run at $z=0$, in which cosmic rays were included in addition to magnetic fields (effective diffusion coefficient: $\kappa=3 \times 10^{29}$). This case (dark gray dotted line) is very similar to the MHD+ run, expect at large galactocentric distances ($\gtrsim 0.65 \times \RVIR$), where the CR+ curve is slightly higher with respect to MHD+.

Owing to their low strength, magnetic fields are difficult to measure observationally within the Milky Way halo. While \cite{strong2000} report a value of $\sim 6~\mu$G around the solar neighbourhood ($\sim 8$ kpc), and \cite{crutcher2010} estimate a slightly higher value of $\sim 7.6 - 11.2 ~\mu$G close to the galactic disk ($\sim 4$ kpc), magnetic field strengths remain poorly constrained in the Milky Way halo. The above two constraints are shown through the black dashed line and gray shaded region, respectively, in the centre left panel. While we do not show radial profiles of TNG50 at these galactocentric distances, it is visible through the slice visualisation (top left panel) that the typical magnetic field strength in the central regions of TNG50 MW-likes is of order $\sim 10~\mu$G, broadly consistent with available observational constraints.

As a result of lower magnetic field strengths in the CGM, magnetic pressure is generally subdominant in the CGM in comparison to thermal pressure \citep{nelson2020}, as can be seen in the radial profile of $P_{\rm{B}}/P_{\rm{th}}$ (Figure~\ref{fig:bField}, lower right panel), although there are regions where magnetic pressure dominates, as is visible in the top right panel. The median TNG50 ratio declines from $\sim 1$ at $0.2 \RVIR$ to $\sim 0.01$ at the virial radius. An analysis of this magnetic to thermal pressure ratio has also been made with the same set of FIRE-2 simulations discussed above \citep{hopkins2020}. The pressure ratio in the FIRE-2 MHD+ m12i run at $z=0$ run (dark gray dot-dashed line) is significantly lower than the median value from TNG50 throughout the CGM: by $2 - 3$ dex, depending on distance. In the case of the FIRE-2 CR+ m12i run (dark gray dotted line), the pressure ratio is closer to the TNG50 result, but still smaller at all distances: a difference of $\sim1.2$ dex in the inner regions of the halo, and $\sim 0.4$ dex at a distance of $\sim 0.7 \RVIR$. As in our comparison with Au6 above, the difference between TNG and FIRE-2 is likely due to a combination of differences in sample selection, ISM/feedback models, and numerical techniques.

While recent theoretical studies of the CGM have begun exploring the importance of magnetic fields in the evolution of MW-like galaxies, different simulations currently produce impressively different results. Comparisons with the real Milky Way halo will provide important constraints. The recent observation of magnetic fields in an external $M\rm{_\star} \sim 10^{10.69}~\rm{M_\odot}$ galaxy at $z \sim 0.36$ by \cite{prochaska2019} suggests that observational inference of halo magnetic field properties is a promising future direction \citep{heald2020}.


\section{Summary and Conclusions}\label{summary}

In this paper, we have quantified the circumgalactic medium (CGM) at $z=0$ of MW-like galaxies in TNG50, a cosmological magnetohydrodynamical simulation hosting 132 such galaxies. These have been selected based on their stellar mass ($10^{10.5-10.9}\,\MSUN$), stellar diskyness, and Mpc-scale environment. They are situated within halos whose $\rm{M_{200c}}$ is in the range $\sim 10^{11.7-12.5}~\MSUN$. Throughout, we define the CGM as the region between $0.15 \RVIR~\rm{and}~1.0 \RVIR$ of each halo, removing satellites.

The key result of our analysis is that, according to TNG50, the physical properties of gas in the CGM -- density, temperature, pressure, entropy, HI content, metallicity, and X-ray emission -- are diverse both across different MW-like galaxies and across the CGM of the same galaxy, as are the magnetic field properties, and the kinematics of inflows and outflows. The galaxy-to-galaxy diversity is connected, according to the TNG model, to the properties of the galaxy itself and to recent feedback activity, particularly from the central SMBH.

We summarise our specific findings as follows:

\begin{enumerate}
    \item The resolved two-dimensional structure of physical gas properties such as density, temperature, and metallicity reveal enormous diversity across the sample of MW-like halos, as well as within individual halos, as a function of distance, and due to inhomogeneities on small $\sim$kpc scales in different regions of the CGM (Figures~\ref{fig:collageMW}, \ref{fig:theoryMwImages} and \ref{fig:obsMwImages}).\\
    
    \item Integral properties of the CGM such as total gas mass, gas mass in the cold phase, and mass of metals in gas show a large scatter across the sample, but show no significant trend as a function of galaxy stellar mass. Total X-ray luminosity does increase rapidly with stellar mass, with more massive galaxies emitting more. Large galaxy-to-galaxy variation is present, and this scatter is explained by a strong correlation with specific star formation rate (sSFR) of the galaxy, which in turn depends on SMBH feedback, with higher sSFR galaxies hosting most massive and X-ray luminous CGM components (Figure~\ref{fig:cgm_props_vs_mass}).\\    
    
    \item The amount of gas in the CGM varies significantly across the TNG50 MW-like sample: while the median fractional mass of gas in the CGM with respect to $\rm{f_b M_{200c}}$ is $\sim 29\%$, the 16$^{\rm{th}}$ and 84$^{\rm{th}}$ percentiles are $\sim 15\%$ to $\sim 41\%$, respectively. The median fractions of mass for cold gas, warm gas, hot gas and metals (all in the CGM) are $\sim$ $5^{+7}_{-4}$, $6^{+5}_{-4}$, $14^{+8}_{-6}$ and $17^{+8}_{-10}$, respectively. Importantly, the hot gas component (T $>10^{5.5}$K) dominates the CGM mass budget of MW-like galaxies (Figure~\ref{fig:bary_comp}). \\
    
    \item Spherically-averaged radial profiles of temperature and entropy indicate that the (population-wide median) value of these two quantities is smaller in the inner regions of the CGM and increases with distance. However, the spherically-averaged gas density, HI density, metallicity and thermal pressure decline monotonically outwards (Figure~\ref{fig:radial_profiles}). Two-dimensional radial distributions reveal the existence of multimodal distributions at fixed distance, particularly for temperature and entropy, although the overall homogeneity of thermal pressure indicates that different components are roughly in pressure equilibrium (Figure~\ref{fig:2d_radial_profiles}). According to TNG50, within such a pressure-equilibrated CGM, different phases of gas co-exist, ranging across $>4$ orders of magnitude in density, temperature, metallicity, and entropy (Figure~\ref{fig:histPlots}).\\
    
    \item In TNG50, feedback from the central SMBH has an important effect not only on the properties (Figure~\ref{fig:cgm_props_vs_mass}) but also on the kinematics of gas in the CGM (Figure~\ref{fig:radVelSlice}). Kinetic mode SMBH feedback can produce gas outflows in MW-like galaxies with large velocities (\mbox{$\gtrsim 500-2000$ km s$^{-1}$}) and super-virial temperatures (\mbox{$> 10^{6.5-7}$ K}). As a result, galaxies with more massive SMBHs suppress inflow through the CGM and shift the net flow through the CGM from inflow to outflow. The majority of gas in the CGM is slowly moving ($\lesssim 100$ km s$^{-1}$), tracing a quasi-static halo (Figure~\ref{fig:outflowRates}).\\
    
    \item Finally, the CGM of TNG50 MW-like galaxies has a rich magnetic field structure: complex on small scales yet coherent across the halo. The magnetic field strength is as high as $\sim\,1\mu$G in the halo, but with a steep radial profile: largest near the center and declining with radius. As a result, magnetic pressure dominates over thermal pressure only in the inner regions of the CGM of MW-like galaxies, although there do exist local regions of gas throughout the halo in which magnetic pressure is the dominant component (Figure~\ref{fig:bField}).
\end{enumerate}

In this work, we have explored several important physical properties of the CGM of Milky Way-like galaxies as predicted by the TNG50 simulation. However, several linked processes and topics have been excluded: for instance, the impact of mergers and satellite galaxies and, in general, the time evolution of the CGM and its state at $z>0$ for the progenitors of the present-day Milky Way. The origin of cold-phase gas in the CGM, and small-scale cold clouds in particular, remains a topic for future work. While the TNG simulations include magnetic fields and complex turbulent motions, they neglect other non-thermal components, namely cosmic rays, which may play a role in the evolution of the CGM. Future simulations and analyses will explore these questions together with their observable signatures in the gaseous halo of the Milky Way.

\section*{Data Availability}

The IllustrisTNG simulations, including TNG50, are publicly available and accessible at \url{www.tng-project.org/data} \citep{nelson2019b}. Post-processing catalogs of star formation rates and molecular hydrogen fractions used in this paper are also available on the same website. Data directly related to this publication is available upon reasonable request.

\section*{Acknowledgements}

RR and DN acknowledge funding from the Deutsche Forschungsgemeinschaft (DFG) through an Emmy Noether Research Group (grant number NE 2441/1-1). AP acknowledges funding by the DFG -- Project-ID 138713538 -- SFB 881 (``The Milky Way System'', subprojects A01 and A06). RR is a Fellow of the International Max Planck Research School for Astronomy and Cosmic Physics at the University of Heidelberg (IMPRS-HD). The TNG50 simulation was run with compute time granted by the Gauss Centre for Supercomputing (GCS) under Large-Scale Projects GCS-DWAR on the GCS share of the supercomputer Hazel Hen at the High Performance Computing Center Stuttgart (HLRS). GCS is the alliance of the three national supercomputing centres HLRS (Universit{\"a}t Stuttgart), JSC (Forschungszentrum J{\"u}lich), and LRZ (Bayerische Akademie der Wissenschaften), funded by the German Federal Ministry of Education and Research (BMBF) and the German State Ministries for Research of Baden-W{\"u}rttemberg (MWK), Bayern (StMWFK) and Nordrhein-Westfalen (MIWF). This analysis has been carried out on the VERA supercomputer of the Max Planck Institute for Astronomy (MPIA), operated by the Max Planck Computational Data Facility (MPCDF). This research has made use of NASA's Astrophysics Data System Bibliographic Services. The authors thank the anonymous referee for constructive feedback, and Philip Hopkins and Sam Ponnada for sharing relevant data points.

\bibliographystyle{mnras}
\bibliography{references}

\begin{thebibliography}{}
\makeatletter
\relax
\def\mn@urlcharsother{\let\do\@makeother \do\$\do\&\do\#\do\^\do\_\do\%\do\~}
\def\mn@doi{\begingroup\mn@urlcharsother \@ifnextchar [ {\mn@doi@}
  {\mn@doi@[]}}
\def\mn@doi@[#1]#2{\def\@tempa{#1}\ifx\@tempa\@empty \href
  {http://dx.doi.org/#2} {doi:#2}\else \href {http://dx.doi.org/#2} {#1}\fi
  \endgroup}
\def\mn@eprint#1#2{\mn@eprint@#1:#2::\@nil}
\def\mn@eprint@arXiv#1{\href {http://arxiv.org/abs/#1} {{\tt arXiv:#1}}}
\def\mn@eprint@dblp#1{\href {http://dblp.uni-trier.de/rec/bibtex/#1.xml}
  {dblp:#1}}
\def\mn@eprint@#1:#2:#3:#4\@nil{\def\@tempa {#1}\def\@tempb {#2}\def\@tempc
  {#3}\ifx \@tempc \@empty \let \@tempc \@tempb \let \@tempb \@tempa \fi \ifx
  \@tempb \@empty \def\@tempb {arXiv}\fi \@ifundefined
  {mn@eprint@\@tempb}{\@tempb:\@tempc}{\expandafter \expandafter \csname
  mn@eprint@\@tempb\endcsname \expandafter{\@tempc}}}

\bibitem[\protect\citeauthoryear{{Anderson}, {Gaspari}, {White}, {Wang}  \&
  {Dai}}{{Anderson} et~al.}{2015}]{anderson2015}
{Anderson} M.~E.,  {Gaspari} M.,  {White} S. D.~M.,  {Wang} W.,   {Dai} X.,
  2015, \mn@doi [\mnras] {10.1093/mnras/stv437}, \href
  {https://ui.adsabs.harvard.edu/abs/2015MNRAS.449.3806A} {449, 3806}

\bibitem[\protect\citeauthoryear{Ashley, Fox, Jenkins, Wakker, Bordoloi,
  Lockman, Savage  \& Karim}{Ashley et~al.}{2020}]{ashley2020}
Ashley T.,  Fox A.~J.,  Jenkins E.~B.,  Wakker B.~P.,  Bordoloi R.,  Lockman
  F.~J.,  Savage B.~D.,   Karim T.,  2020, \mn@doi [ApJ]
  {10.3847/1538-4357/ab9ff8}, 898, 128

\bibitem[\protect\citeauthoryear{{Ashley}, {Fox}, {Cashman}, {Lockman},
  {Bordoloi}, {Jenkins}, {Wakker}  \& {Karim}}{{Ashley}
  et~al.}{2022}]{ashley2022}
{Ashley} T.,  {Fox} A.~J.,  {Cashman} F.~H.,  {Lockman} F.~J.,  {Bordoloi} R.,
  {Jenkins} E.~B.,  {Wakker} B.~P.,   {Karim} T.,  2022, arXiv e-prints, \href
  {https://ui.adsabs.harvard.edu/abs/2022arXiv220708838A} {p. arXiv:2207.08838}

\bibitem[\protect\citeauthoryear{{Ayromlou}, {Nelson}, {Yates}, {Kauffmann}  \&
  {White}}{{Ayromlou} et~al.}{2019}]{ayromlou2019}
{Ayromlou} M.,  {Nelson} D.,  {Yates} R.~M.,  {Kauffmann} G.,   {White} S.
  D.~M.,  2019, \mn@doi [\mnras] {10.1093/mnras/stz1549}, \href
  {https://ui.adsabs.harvard.edu/abs/2019MNRAS.487.4313A} {487, 4313}

\bibitem[\protect\citeauthoryear{{Berlok} \& {Pfrommer}}{{Berlok} \&
  {Pfrommer}}{2019}]{berlok2019}
{Berlok} T.,  {Pfrommer} C.,  2019, \mn@doi [\mnras] {10.1093/mnras/stz2347},
  \href {https://ui.adsabs.harvard.edu/abs/2019MNRAS.489.3368B} {489, 3368}

\bibitem[\protect\citeauthoryear{{Bluem} et~al.,}{{Bluem}
  et~al.}{2022}]{bluem2022}
{Bluem} J.,  et~al., 2022, arXiv e-prints, \href
  {https://ui.adsabs.harvard.edu/abs/2022arXiv220802477B} {p. arXiv:2208.02477}

\bibitem[\protect\citeauthoryear{{Bogd{\'a}n} et~al.,}{{Bogd{\'a}n}
  et~al.}{2013a}]{bogdan2013b}
{Bogd{\'a}n} {\'A}.,  et~al., 2013a, \mn@doi [\apj]
  {10.1088/0004-637X/772/2/97}, \href
  {https://ui.adsabs.harvard.edu/abs/2013ApJ...772...97B} {772, 97}

\bibitem[\protect\citeauthoryear{{Bogd{\'a}n}, {Forman}, {Kraft}  \&
  {Jones}}{{Bogd{\'a}n} et~al.}{2013b}]{bogdan.2013a}
{Bogd{\'a}n} {\'A}.,  {Forman} W.~R.,  {Kraft} R.~P.,   {Jones} C.,  2013b,
  \mn@doi [\apj] {10.1088/0004-637X/772/2/98}, \href
  {https://ui.adsabs.harvard.edu/abs/2013ApJ...772...98B} {772, 98}

\bibitem[\protect\citeauthoryear{Bordoloi et~al.,}{Bordoloi
  et~al.}{2017}]{bordoloi2017}
Bordoloi R.,  et~al., 2017, \mn@doi [ApJ] {10.3847/1538-4357/834/2/191}, 834,
  191

\bibitem[\protect\citeauthoryear{{Br{\"u}ns} et~al.,}{{Br{\"u}ns}
  et~al.}{2005}]{bruns2005}
{Br{\"u}ns} C.,  et~al., 2005, \mn@doi [\aap] {10.1051/0004-6361:20040321},
  \href {https://ui.adsabs.harvard.edu/abs/2005A&A...432...45B} {432, 45}

\bibitem[\protect\citeauthoryear{{Burchett} et~al.,}{{Burchett}
  et~al.}{2019}]{burchett2019}
{Burchett} J.~N.,  et~al., 2019, \mn@doi [\apjl] {10.3847/2041-8213/ab1f7f},
  \href {https://ui.adsabs.harvard.edu/abs/2019ApJ...877L..20B} {877, L20}

\bibitem[\protect\citeauthoryear{{Butsky} \& {Quinn}}{{Butsky} \&
  {Quinn}}{2018}]{butsky2018}
{Butsky} I.~S.,  {Quinn} T.~R.,  2018, \mn@doi [\apj]
  {10.3847/1538-4357/aaeac2}, \href
  {https://ui.adsabs.harvard.edu/abs/2018ApJ...868..108B} {868, 108}

\bibitem[\protect\citeauthoryear{{Butsky}, {Fielding}, {Hayward}, {Hummels},
  {Quinn}  \& {Werk}}{{Butsky} et~al.}{2020}]{butsky2020}
{Butsky} I.~S.,  {Fielding} D.~B.,  {Hayward} C.~C.,  {Hummels} C.~B.,  {Quinn}
  T.~R.,   {Werk} J.~K.,  2020, \mn@doi [\apj] {10.3847/1538-4357/abbad2},
  \href {https://ui.adsabs.harvard.edu/abs/2020ApJ...903...77B} {903, 77}

\bibitem[\protect\citeauthoryear{{Byrohl} et~al.,}{{Byrohl}
  et~al.}{2021}]{byrohl2021}
{Byrohl} C.,  et~al., 2021, \mn@doi [\mnras] {10.1093/mnras/stab1958}, \href
  {https://ui.adsabs.harvard.edu/abs/2021MNRAS.506.5129B} {506, 5129}

\bibitem[\protect\citeauthoryear{{Cai} et~al.,}{{Cai} et~al.}{2017}]{cai2017}
{Cai} Z.,  et~al., 2017, \mn@doi [\apj] {10.3847/1538-4357/aa6a1a}, \href
  {https://ui.adsabs.harvard.edu/abs/2017ApJ...839..131C} {839, 131}

\bibitem[\protect\citeauthoryear{{Chadayammuri}, {Bogdan}, {Oppenheimer},
  {Kraft}, {Forman}  \& {Jones}}{{Chadayammuri}
  et~al.}{2022}]{chadayammuri2022}
{Chadayammuri} U.,  {Bogdan} A.,  {Oppenheimer} B.,  {Kraft} R.,  {Forman} W.,
   {Jones} C.,  2022, arXiv e-prints, \href
  {https://ui.adsabs.harvard.edu/abs/2022arXiv220301356C} {p. arXiv:2203.01356}

\bibitem[\protect\citeauthoryear{{Chen}, {Zahedy}, {Johnson}, {Pierce},
  {Huang}, {Weiner}  \& {Gauthier}}{{Chen} et~al.}{2018}]{chen2018}
{Chen} H.-W.,  {Zahedy} F.~S.,  {Johnson} S.~D.,  {Pierce} R.~M.,  {Huang}
  Y.-H.,  {Weiner} B.~J.,   {Gauthier} J.-R.,  2018, \mn@doi [\mnras]
  {10.1093/mnras/sty1541}, \href
  {https://ui.adsabs.harvard.edu/abs/2018MNRAS.479.2547C} {479, 2547}

\bibitem[\protect\citeauthoryear{{Clark}, {Bordoloi}  \& {Fox}}{{Clark}
  et~al.}{2022}]{clark2022}
{Clark} S.,  {Bordoloi} R.,   {Fox} A.~J.,  2022, \mn@doi [\mnras]
  {10.1093/mnras/stac504}, \href
  {https://ui.adsabs.harvard.edu/abs/2022MNRAS.512..811C} {512, 811}

\bibitem[\protect\citeauthoryear{{Comparat} et~al.,}{{Comparat}
  et~al.}{2022}]{comparat2022}
{Comparat} J.,  et~al., 2022, arXiv e-prints, \href
  {https://ui.adsabs.harvard.edu/abs/2022arXiv220105169C} {p. arXiv:2201.05169}

\bibitem[\protect\citeauthoryear{{Corlies}, {Peeples}, {Tumlinson}, {O'Shea},
  {Lehner}, {Howk}, {O'Meara}  \& {Smith}}{{Corlies}
  et~al.}{2020}]{corlies2020}
{Corlies} L.,  {Peeples} M.~S.,  {Tumlinson} J.,  {O'Shea} B.~W.,  {Lehner} N.,
   {Howk} J.~C.,  {O'Meara} J.~M.,   {Smith} B.~D.,  2020, \mn@doi [\apj]
  {10.3847/1538-4357/ab9310}, \href
  {https://ui.adsabs.harvard.edu/abs/2020ApJ...896..125C} {896, 125}

\bibitem[\protect\citeauthoryear{{Crutcher}, {Wandelt}, {Heiles}, {Falgarone}
  \& {Troland}}{{Crutcher} et~al.}{2010}]{crutcher2010}
{Crutcher} R.~M.,  {Wandelt} B.,  {Heiles} C.,  {Falgarone} E.,   {Troland}
  T.~H.,  2010, \mn@doi [\apj] {10.1088/0004-637X/725/1/466}, \href
  {https://ui.adsabs.harvard.edu/abs/2010ApJ...725..466C} {725, 466}

\bibitem[\protect\citeauthoryear{{Damle} et~al.,}{{Damle}
  et~al.}{2022}]{damle2022}
{Damle} M.,  et~al., 2022, \mn@doi [\mnras] {10.1093/mnras/stac663}, \href
  {https://ui.adsabs.harvard.edu/abs/2022MNRAS.512.3717D} {512, 3717}

\bibitem[\protect\citeauthoryear{{Davies}, {Crain}, {Oppenheimer}  \&
  {Schaye}}{{Davies} et~al.}{2020}]{davies2020}
{Davies} J.~J.,  {Crain} R.~A.,  {Oppenheimer} B.~D.,   {Schaye} J.,  2020,
  \mn@doi [\mnras] {10.1093/mnras/stz3201}, \href
  {https://ui.adsabs.harvard.edu/abs/2020MNRAS.491.4462D} {491, 4462}

\bibitem[\protect\citeauthoryear{{Di Teodoro}, {McClure-Griffiths}, {Lockman},
  {Denbo}, {Endsley}, {Ford}  \& {Harrington}}{{Di Teodoro}
  et~al.}{2018}]{diteodoro2018}
{Di Teodoro} E.~M.,  {McClure-Griffiths} N.~M.,  {Lockman} F.~J.,  {Denbo}
  S.~R.,  {Endsley} R.,  {Ford} H.~A.,   {Harrington} K.,  2018, \mn@doi [\apj]
  {10.3847/1538-4357/aaad6a}, \href
  {https://ui.adsabs.harvard.edu/abs/2018ApJ...855...33D} {855, 33}

\bibitem[\protect\citeauthoryear{{Di Teodoro}, {McClure-Griffiths}, {Lockman}
  \& {Armillotta}}{{Di Teodoro} et~al.}{2020}]{teodoro2020}
{Di Teodoro} E.~M.,  {McClure-Griffiths} N.~M.,  {Lockman} F.~J.,
  {Armillotta} L.,  2020, \mn@doi [\nat] {10.1038/s41586-020-2595-z}, \href
  {https://ui.adsabs.harvard.edu/abs/2020Natur.584..364D} {584, 364}

\bibitem[\protect\citeauthoryear{{Donnari} et~al.,}{{Donnari}
  et~al.}{2019}]{donnari2019}
{Donnari} M.,  et~al., 2019, \mn@doi [\mnras] {10.1093/mnras/stz712}, \href
  {https://ui.adsabs.harvard.edu/abs/2019MNRAS.485.4817D} {485, 4817}

\bibitem[\protect\citeauthoryear{{Dutta}, {Sharma}  \& {Nelson}}{{Dutta}
  et~al.}{2022}]{dutta2022}
{Dutta} A.,  {Sharma} P.,   {Nelson} D.,  2022, \mn@doi [\mnras]
  {10.1093/mnras/stab3653}, \href
  {https://ui.adsabs.harvard.edu/abs/2022MNRAS.510.3561D} {510, 3561}

\bibitem[\protect\citeauthoryear{{Engler} et~al.,}{{Engler}
  et~al.}{2021}]{engler2021}
{Engler} C.,  et~al., 2021, \mn@doi [\mnras] {10.1093/mnras/stab2437}, \href
  {https://ui.adsabs.harvard.edu/abs/2021MNRAS.507.4211E} {507, 4211}

\bibitem[\protect\citeauthoryear{{Esmerian}, {Kravtsov}, {Hafen},
  {Faucher-Gigu{\`e}re}, {Quataert}, {Stern}, {Kere{\v{s}}}  \&
  {Wetzel}}{{Esmerian} et~al.}{2021}]{esmerian2021}
{Esmerian} C.~J.,  {Kravtsov} A.~V.,  {Hafen} Z.,  {Faucher-Gigu{\`e}re} C.-A.,
   {Quataert} E.,  {Stern} J.,  {Kere{\v{s}}} D.,   {Wetzel} A.,  2021, \mn@doi
  [\mnras] {10.1093/mnras/stab1281}, \href
  {https://ui.adsabs.harvard.edu/abs/2021MNRAS.505.1841E} {505, 1841}

\bibitem[\protect\citeauthoryear{{Faerman}, {Sternberg}  \& {McKee}}{{Faerman}
  et~al.}{2017}]{faerman2017}
{Faerman} Y.,  {Sternberg} A.,   {McKee} C.~F.,  2017, \mn@doi [\apj]
  {10.3847/1538-4357/835/1/52}, \href
  {https://ui.adsabs.harvard.edu/abs/2017ApJ...835...52F} {835, 52}

\bibitem[\protect\citeauthoryear{{Faerman}, {Sternberg}  \& {McKee}}{{Faerman}
  et~al.}{2020}]{faerman2020}
{Faerman} Y.,  {Sternberg} A.,   {McKee} C.~F.,  2020, \mn@doi [\apj]
  {10.3847/1538-4357/ab7ffc}, \href
  {https://ui.adsabs.harvard.edu/abs/2020ApJ...893...82F} {893, 82}

\bibitem[\protect\citeauthoryear{{Fang}, {Bullock}  \& {Boylan-Kolchin}}{{Fang}
  et~al.}{2013}]{fang2013}
{Fang} T.,  {Bullock} J.,   {Boylan-Kolchin} M.,  2013, \mn@doi [\apj]
  {10.1088/0004-637X/762/1/20}, \href
  {https://ui.adsabs.harvard.edu/abs/2013ApJ...762...20F} {762, 20}

\bibitem[\protect\citeauthoryear{{Faucher-Gigu{\`e}re}, {Lidz}, {Zaldarriaga}
  \& {Hernquist}}{{Faucher-Gigu{\`e}re} et~al.}{2009}]{fg2009}
{Faucher-Gigu{\`e}re} C.-A.,  {Lidz} A.,  {Zaldarriaga} M.,   {Hernquist} L.,
  2009, \mn@doi [\apj] {10.1088/0004-637X/703/2/1416}, \href
  {https://ui.adsabs.harvard.edu/abs/2009ApJ...703.1416F} {703, 1416}

\bibitem[\protect\citeauthoryear{{Ferland} et~al.,}{{Ferland}
  et~al.}{2013}]{ferland2013}
{Ferland} G.~J.,  et~al., 2013, \rmxaa, \href
  {https://ui.adsabs.harvard.edu/abs/2013RMxAA..49..137F} {49, 137}

\bibitem[\protect\citeauthoryear{{Fern{\'a}ndez}, {Joung}  \&
  {Putman}}{{Fern{\'a}ndez} et~al.}{2012}]{fernandez2012}
{Fern{\'a}ndez} X.,  {Joung} M.~R.,   {Putman} M.~E.,  2012, \mn@doi [\apj]
  {10.1088/0004-637X/749/2/181}, \href
  {https://ui.adsabs.harvard.edu/abs/2012ApJ...749..181F} {749, 181}

\bibitem[\protect\citeauthoryear{{Fielding}, {Quataert}, {McCourt}  \&
  {Thompson}}{{Fielding} et~al.}{2017}]{fielding2017}
{Fielding} D.,  {Quataert} E.,  {McCourt} M.,   {Thompson} T.~A.,  2017,
  \mn@doi [\mnras] {10.1093/mnras/stw3326}, \href
  {https://ui.adsabs.harvard.edu/abs/2017MNRAS.466.3810F} {466, 3810}

\bibitem[\protect\citeauthoryear{{Fielding} et~al.,}{{Fielding}
  et~al.}{2020}]{fielding2020}
{Fielding} D.~B.,  et~al., 2020, \mn@doi [\apj] {10.3847/1538-4357/abbc6d},
  \href {https://ui.adsabs.harvard.edu/abs/2020ApJ...903...32F} {903, 32}

\bibitem[\protect\citeauthoryear{Fox et~al.,}{Fox et~al.}{2015}]{fox2015}
Fox A.~J.,  et~al., 2015, \mn@doi [ApJ] {10.1088/2041-8205/799/1/l7}, 799, L7

\bibitem[\protect\citeauthoryear{Genel et~al.,}{Genel et~al.}{2014}]{genel2014}
Genel S.,  et~al., 2014, Monthly Notices of the Royal Astronomical Society,
  445, 175

\bibitem[\protect\citeauthoryear{{Gnedin} \& {Kravtsov}}{{Gnedin} \&
  {Kravtsov}}{2011}]{gnedin2011}
{Gnedin} N.~Y.,  {Kravtsov} A.~V.,  2011, \mn@doi [\apj]
  {10.1088/0004-637X/728/2/88}, \href
  {https://ui.adsabs.harvard.edu/abs/2011ApJ...728...88G} {728, 88}

\bibitem[\protect\citeauthoryear{{Goulding} et~al.,}{{Goulding}
  et~al.}{2016}]{goulding2016}
{Goulding} A.~D.,  et~al., 2016, \mn@doi [\apj] {10.3847/0004-637X/826/2/167},
  \href {https://ui.adsabs.harvard.edu/abs/2016ApJ...826..167G} {826, 167}

\bibitem[\protect\citeauthoryear{{Gronke} \& {Oh}}{{Gronke} \&
  {Oh}}{2020}]{gronke2020}
{Gronke} M.,  {Oh} S.~P.,  2020, \mn@doi [\mnras] {10.1093/mnras/stz3332},
  \href {https://ui.adsabs.harvard.edu/abs/2020MNRAS.492.1970G} {492, 1970}

\bibitem[\protect\citeauthoryear{{Gupta}, {Mathur}, {Krongold}, {Nicastro}  \&
  {Galeazzi}}{{Gupta} et~al.}{2012}]{gupta2012}
{Gupta} A.,  {Mathur} S.,  {Krongold} Y.,  {Nicastro} F.,   {Galeazzi} M.,
  2012, \mn@doi [\apjl] {10.1088/2041-8205/756/1/L8}, \href
  {https://ui.adsabs.harvard.edu/abs/2012ApJ...756L...8G} {756, L8}

\bibitem[\protect\citeauthoryear{{Hafen} et~al.,}{{Hafen}
  et~al.}{2019}]{hafen2019}
{Hafen} Z.,  et~al., 2019, \mn@doi [\mnras] {10.1093/mnras/stz1773}, \href
  {https://ui.adsabs.harvard.edu/abs/2019MNRAS.488.1248H} {488, 1248}

\bibitem[\protect\citeauthoryear{{Hani}, {Ellison}, {Sparre}, {Grand},
  {Pakmor}, {Gomez}  \& {Springel}}{{Hani} et~al.}{2019}]{hani2019}
{Hani} M.~H.,  {Ellison} S.~L.,  {Sparre} M.,  {Grand} R. J.~J.,  {Pakmor} R.,
  {Gomez} F.~A.,   {Springel} V.,  2019, \mn@doi [\mnras]
  {10.1093/mnras/stz1708}, \href
  {https://ui.adsabs.harvard.edu/abs/2019MNRAS.488..135H} {488, 135}

\bibitem[\protect\citeauthoryear{{Heald} et~al.,}{{Heald}
  et~al.}{2020}]{heald2020}
{Heald} G.,  et~al., 2020, \mn@doi [Galaxies] {10.3390/galaxies8030053}, \href
  {https://ui.adsabs.harvard.edu/abs/2020Galax...8...53H} {8, 53}

\bibitem[\protect\citeauthoryear{{Henley}, {Shelton}, {Kwak}, {Joung}  \& {Mac
  Low}}{{Henley} et~al.}{2010}]{henley2010}
{Henley} D.~B.,  {Shelton} R.~L.,  {Kwak} K.,  {Joung} M.~R.,   {Mac Low}
  M.-M.,  2010, \mn@doi [\apj] {10.1088/0004-637X/723/1/935}, \href
  {https://ui.adsabs.harvard.edu/abs/2010ApJ...723..935H} {723, 935}

\bibitem[\protect\citeauthoryear{{Hennawi}, {Prochaska}, {Cantalupo}  \&
  {Arrigoni-Battaia}}{{Hennawi} et~al.}{2015}]{hennawi2015}
{Hennawi} J.~F.,  {Prochaska} J.~X.,  {Cantalupo} S.,   {Arrigoni-Battaia} F.,
  2015, \mn@doi [Science] {10.1126/science.aaa5397}, \href
  {https://ui.adsabs.harvard.edu/abs/2015Sci...348..779H} {348, 779}

\bibitem[\protect\citeauthoryear{{Hodges-Kluck}, {Miller}  \&
  {Bregman}}{{Hodges-Kluck} et~al.}{2016}]{kluck2016}
{Hodges-Kluck} E.~J.,  {Miller} M.~J.,   {Bregman} J.~N.,  2016, \mn@doi [\apj]
  {10.3847/0004-637X/822/1/21}, \href
  {https://ui.adsabs.harvard.edu/abs/2016ApJ...822...21H} {822, 21}

\bibitem[\protect\citeauthoryear{{Hopkins}}{{Hopkins}}{2015}]{hopkins2015}
{Hopkins} P.~F.,  2015, \mn@doi [\mnras] {10.1093/mnras/stv195}, \href
  {https://ui.adsabs.harvard.edu/abs/2015MNRAS.450...53H} {450, 53}

\bibitem[\protect\citeauthoryear{{Hopkins} et~al.,}{{Hopkins}
  et~al.}{2020}]{hopkins2020}
{Hopkins} P.~F.,  et~al., 2020, \mn@doi [\mnras] {10.1093/mnras/stz3321}, \href
  {https://ui.adsabs.harvard.edu/abs/2020MNRAS.492.3465H} {492, 3465}

\bibitem[\protect\citeauthoryear{{Huang}, {Jiang}  \& {Davis}}{{Huang}
  et~al.}{2022}]{huang2022}
{Huang} X.,  {Jiang} Y.-f.,   {Davis} S.~W.,  2022, \mn@doi [\apj]
  {10.3847/1538-4357/ac69dc}, \href
  {https://ui.adsabs.harvard.edu/abs/2022ApJ...931..140H} {931, 140}

\bibitem[\protect\citeauthoryear{{Ji} et~al.,}{{Ji} et~al.}{2020}]{ji2020}
{Ji} S.,  et~al., 2020, \mn@doi [\mnras] {10.1093/mnras/staa1849}, \href
  {https://ui.adsabs.harvard.edu/abs/2020MNRAS.496.4221J} {496, 4221}

\bibitem[\protect\citeauthoryear{{Joung}, {Bryan}  \& {Putman}}{{Joung}
  et~al.}{2012}]{juong2012}
{Joung} M.~R.,  {Bryan} G.~L.,   {Putman} M.~E.,  2012, \mn@doi [\apj]
  {10.1088/0004-637X/745/2/148}, \href
  {https://ui.adsabs.harvard.edu/abs/2012ApJ...745..148J} {745, 148}

\bibitem[\protect\citeauthoryear{{Kaaret} et~al.,}{{Kaaret}
  et~al.}{2020}]{kaaret2020}
{Kaaret} P.,  et~al., 2020, \mn@doi [Nature Astronomy]
  {10.1038/s41550-020-01215-w}, \href
  {https://ui.adsabs.harvard.edu/abs/2020NatAs...4.1072K} {4, 1072}

\bibitem[\protect\citeauthoryear{{Kataoka} et~al.,}{{Kataoka}
  et~al.}{2013}]{kataoka2013}
{Kataoka} J.,  et~al., 2013, \mn@doi [\apj] {10.1088/0004-637X/779/1/57}, \href
  {https://ui.adsabs.harvard.edu/abs/2013ApJ...779...57K} {779, 57}

\bibitem[\protect\citeauthoryear{{Kauffmann}, {Nelson}, {Borthakur}, {Heckman},
  {Hernquist}, {Marinacci}, {Pakmor}  \& {Pillepich}}{{Kauffmann}
  et~al.}{2019}]{kauffmann2019}
{Kauffmann} G.,  {Nelson} D.,  {Borthakur} S.,  {Heckman} T.,  {Hernquist} L.,
  {Marinacci} F.,  {Pakmor} R.,   {Pillepich} A.,  2019, \mn@doi [\mnras]
  {10.1093/mnras/stz1029}, \href
  {https://ui.adsabs.harvard.edu/abs/2019MNRAS.486.4686K} {486, 4686}

\bibitem[\protect\citeauthoryear{{Kere{\v{s}}} \& {Hernquist}}{{Kere{\v{s}}} \&
  {Hernquist}}{2009}]{keres2009}
{Kere{\v{s}}} D.,  {Hernquist} L.,  2009, \mn@doi [\apjl]
  {10.1088/0004-637X/700/1/L1}, \href
  {https://ui.adsabs.harvard.edu/abs/2009ApJ...700L...1K} {700, L1}

\bibitem[\protect\citeauthoryear{{Leclercq} et~al.,}{{Leclercq}
  et~al.}{2017}]{leclercq2017}
{Leclercq} F.,  et~al., 2017, \mn@doi [\aap] {10.1051/0004-6361/201731480},
  \href {https://ui.adsabs.harvard.edu/abs/2017A&A...608A...8L} {608, A8}

\bibitem[\protect\citeauthoryear{{Leclercq} et~al.,}{{Leclercq}
  et~al.}{2020}]{leclercq2020}
{Leclercq} F.,  et~al., 2020, \mn@doi [\aap] {10.1051/0004-6361/201937339},
  \href {https://ui.adsabs.harvard.edu/abs/2020A&A...635A..82L} {635, A82}

\bibitem[\protect\citeauthoryear{{Lehner} et~al.,}{{Lehner}
  et~al.}{2020}]{lehner2020}
{Lehner} N.,  et~al., 2020, \mn@doi [\apj] {10.3847/1538-4357/aba49c}, \href
  {https://ui.adsabs.harvard.edu/abs/2020ApJ...900....9L} {900, 9}

\bibitem[\protect\citeauthoryear{{Li} \& {Tonnesen}}{{Li} \&
  {Tonnesen}}{2020}]{li2020}
{Li} M.,  {Tonnesen} S.,  2020, \mn@doi [\apj] {10.3847/1538-4357/ab9f9f},
  \href {https://ui.adsabs.harvard.edu/abs/2020ApJ...898..148L} {898, 148}

\bibitem[\protect\citeauthoryear{{Li}, {Decourchelle}, {Miceli}, {Vink}  \&
  {Bocchino}}{{Li} et~al.}{2016}]{Li.2016}
{Li} J.-T.,  {Decourchelle} A.,  {Miceli} M.,  {Vink} J.,   {Bocchino} F.,
  2016, \mn@doi [\mnras] {10.1093/mnras/stw1640}, \href
  {https://ui.adsabs.harvard.edu/abs/2016MNRAS.462..158L} {462, 158}

\bibitem[\protect\citeauthoryear{{Li}, {Hopkins}, {Squire}  \& {Hummels}}{{Li}
  et~al.}{2020}]{liZ2020}
{Li} Z.,  {Hopkins} P.~F.,  {Squire} J.,   {Hummels} C.,  2020, \mn@doi
  [\mnras] {10.1093/mnras/stz3567}, \href
  {https://ui.adsabs.harvard.edu/abs/2020MNRAS.492.1841L} {492, 1841}

\bibitem[\protect\citeauthoryear{{Lim}, {Barnes}, {Vogelsberger}, {Mo},
  {Nelson}, {Pillepich}, {Dolag}  \& {Marinacci}}{{Lim} et~al.}{2021}]{lim2021}
{Lim} S.~H.,  {Barnes} D.,  {Vogelsberger} M.,  {Mo} H.~J.,  {Nelson} D.,
  {Pillepich} A.,  {Dolag} K.,   {Marinacci} F.,  2021, \mn@doi [\mnras]
  {10.1093/mnras/stab1172}, \href
  {https://ui.adsabs.harvard.edu/abs/2021MNRAS.504.5131L} {504, 5131}

\bibitem[\protect\citeauthoryear{{Marinacci}, {Vogelsberger}, {Mocz}  \&
  {Pakmor}}{{Marinacci} et~al.}{2015}]{marinacci2015}
{Marinacci} F.,  {Vogelsberger} M.,  {Mocz} P.,   {Pakmor} R.,  2015, \mn@doi
  [\mnras] {10.1093/mnras/stv1692}, \href
  {https://ui.adsabs.harvard.edu/abs/2015MNRAS.453.3999M} {453, 3999}

\bibitem[\protect\citeauthoryear{Marinacci et~al.,}{Marinacci
  et~al.}{2018}]{marinacci2018}
Marinacci F.,  et~al., 2018, Monthly Notices of the Royal Astronomical Society,
  480, 5113

\bibitem[\protect\citeauthoryear{{McCourt}, {O'Leary}, {Madigan}  \&
  {Quataert}}{{McCourt} et~al.}{2015}]{mccourt2015}
{McCourt} M.,  {O'Leary} R.~M.,  {Madigan} A.-M.,   {Quataert} E.,  2015,
  \mn@doi [\mnras] {10.1093/mnras/stv355}, \href
  {https://ui.adsabs.harvard.edu/abs/2015MNRAS.449....2M} {449, 2}

\bibitem[\protect\citeauthoryear{{McMillan}}{{McMillan}}{2011}]{mcmillan2011}
{McMillan} P.~J.,  2011, \mn@doi [\mnras] {10.1111/j.1365-2966.2011.18564.x},
  \href {https://ui.adsabs.harvard.edu/abs/2011MNRAS.414.2446M} {414, 2446}

\bibitem[\protect\citeauthoryear{{Miller} \& {Bregman}}{{Miller} \&
  {Bregman}}{2015}]{miller2015}
{Miller} M.~J.,  {Bregman} J.~N.,  2015, \mn@doi [\apj]
  {10.1088/0004-637X/800/1/14}, \href
  {https://ui.adsabs.harvard.edu/abs/2015ApJ...800...14M} {800, 14}

\bibitem[\protect\citeauthoryear{{Miller}, {Hodges-Kluck}  \&
  {Bregman}}{{Miller} et~al.}{2016}]{miller2016}
{Miller} M.~J.,  {Hodges-Kluck} E.~J.,   {Bregman} J.~N.,  2016, \mn@doi [\apj]
  {10.3847/0004-637X/818/2/112}, \href
  {https://ui.adsabs.harvard.edu/abs/2016ApJ...818..112M} {818, 112}

\bibitem[\protect\citeauthoryear{Naiman et~al.,}{Naiman
  et~al.}{2018}]{naiman2018}
Naiman J.~P.,  et~al., 2018, Monthly Notices of the Royal Astronomical Society,
  477, 1206

\bibitem[\protect\citeauthoryear{{Nelson}, {Genel}, {Vogelsberger}, {Springel},
  {Sijacki}, {Torrey}  \& {Hernquist}}{{Nelson} et~al.}{2015}]{nelson2015}
{Nelson} D.,  {Genel} S.,  {Vogelsberger} M.,  {Springel} V.,  {Sijacki} D.,
  {Torrey} P.,   {Hernquist} L.,  2015, \mn@doi [\mnras]
  {10.1093/mnras/stv017}, \href
  {https://ui.adsabs.harvard.edu/abs/2015MNRAS.448...59N} {448, 59}

\bibitem[\protect\citeauthoryear{Nelson et~al.,}{Nelson
  et~al.}{2018a}]{nelson2018}
Nelson D.,  et~al., 2018a, Monthly Notices of the Royal Astronomical Society,
  475, 624

\bibitem[\protect\citeauthoryear{{Nelson} et~al.,}{{Nelson}
  et~al.}{2018b}]{nelson2018b}
{Nelson} D.,  et~al., 2018b, \mn@doi [\mnras] {10.1093/mnras/sty656}, \href
  {https://ui.adsabs.harvard.edu/abs/2018MNRAS.477..450N} {477, 450}

\bibitem[\protect\citeauthoryear{{Nelson} et~al.,}{{Nelson}
  et~al.}{2019a}]{nelson2019b}
{Nelson} D.,  et~al., 2019a, \mn@doi [Computational Astrophysics and Cosmology]
  {10.1186/s40668-019-0028-x}, \href
  {https://ui.adsabs.harvard.edu/abs/2019ComAC...6....2N} {6, 2}

\bibitem[\protect\citeauthoryear{{Nelson} et~al.,}{{Nelson}
  et~al.}{2019b}]{nelson2019}
{Nelson} D.,  et~al., 2019b, \mn@doi [\mnras] {10.1093/mnras/stz2306}, \href
  {https://ui.adsabs.harvard.edu/abs/2019MNRAS.490.3234N} {490, 3234}

\bibitem[\protect\citeauthoryear{{Nelson} et~al.,}{{Nelson}
  et~al.}{2020}]{nelson2020}
{Nelson} D.,  et~al., 2020, \mn@doi [\mnras] {10.1093/mnras/staa2419}, \href
  {https://ui.adsabs.harvard.edu/abs/2020MNRAS.498.2391N} {498, 2391}

\bibitem[\protect\citeauthoryear{{Nicastro}, {Senatore}, {Krongold}, {Mathur}
  \& {Elvis}}{{Nicastro} et~al.}{2016}]{nicastro2016}
{Nicastro} F.,  {Senatore} F.,  {Krongold} Y.,  {Mathur} S.,   {Elvis} M.,
  2016, \mn@doi [\apjl] {10.3847/2041-8205/828/1/L12}, \href
  {https://ui.adsabs.harvard.edu/abs/2016ApJ...828L..12N} {828, L12}

\bibitem[\protect\citeauthoryear{{Oppenheimer}}{{Oppenheimer}}{2018}]{oppenheimer2018}
{Oppenheimer} B.~D.,  2018, \mn@doi [\mnras] {10.1093/mnras/sty1918}, \href
  {https://ui.adsabs.harvard.edu/abs/2018MNRAS.480.2963O} {480, 2963}

\bibitem[\protect\citeauthoryear{{Oppenheimer} et~al.,}{{Oppenheimer}
  et~al.}{2020}]{oppenheimer2020}
{Oppenheimer} B.~D.,  et~al., 2020, \mn@doi [\mnras] {10.1093/mnras/stz3124},
  \href {https://ui.adsabs.harvard.edu/abs/2020MNRAS.491.2939O} {491, 2939}

\bibitem[\protect\citeauthoryear{{Pakmor}, {Marinacci}  \& {Springel}}{{Pakmor}
  et~al.}{2014}]{pakmor2014}
{Pakmor} R.,  {Marinacci} F.,   {Springel} V.,  2014, \mn@doi [\apjl]
  {10.1088/2041-8205/783/1/L20}, \href
  {https://ui.adsabs.harvard.edu/abs/2014ApJ...783L..20P} {783, L20}

\bibitem[\protect\citeauthoryear{{Pakmor} et~al.,}{{Pakmor}
  et~al.}{2020}]{pakmor2020}
{Pakmor} R.,  et~al., 2020, \mn@doi [\mnras] {10.1093/mnras/staa2530}, \href
  {https://ui.adsabs.harvard.edu/abs/2020MNRAS.498.3125P} {498, 3125}

\bibitem[\protect\citeauthoryear{{Peek} et~al.,}{{Peek}
  et~al.}{2011}]{peek2011}
{Peek} J.~E.~G.,  et~al., 2011, \mn@doi [\apjs] {10.1088/0067-0049/194/2/20},
  \href {https://ui.adsabs.harvard.edu/abs/2011ApJS..194...20P} {194, 20}

\bibitem[\protect\citeauthoryear{{Peeples}, {Werk}, {Tumlinson}, {Oppenheimer},
  {Prochaska}, {Katz}  \& {Weinberg}}{{Peeples} et~al.}{2014}]{peeples2014}
{Peeples} M.~S.,  {Werk} J.~K.,  {Tumlinson} J.,  {Oppenheimer} B.~D.,
  {Prochaska} J.~X.,  {Katz} N.,   {Weinberg} D.~H.,  2014, \mn@doi [\apj]
  {10.1088/0004-637X/786/1/54}, \href
  {https://ui.adsabs.harvard.edu/abs/2014ApJ...786...54P} {786, 54}

\bibitem[\protect\citeauthoryear{{Peeples} et~al.,}{{Peeples}
  et~al.}{2019}]{peeples2019}
{Peeples} M.~S.,  et~al., 2019, \mn@doi [\apj] {10.3847/1538-4357/ab0654},
  \href {https://ui.adsabs.harvard.edu/abs/2019ApJ...873..129P} {873, 129}

\bibitem[\protect\citeauthoryear{{P{\'e}roux}, {Nelson}, {van de Voort},
  {Pillepich}, {Marinacci}, {Vogelsberger}  \& {Hernquist}}{{P{\'e}roux}
  et~al.}{2020}]{peroux2020}
{P{\'e}roux} C.,  {Nelson} D.,  {van de Voort} F.,  {Pillepich} A.,
  {Marinacci} F.,  {Vogelsberger} M.,   {Hernquist} L.,  2020, \mn@doi [\mnras]
  {10.1093/mnras/staa2888}, \href
  {https://ui.adsabs.harvard.edu/abs/2020MNRAS.499.2462P} {499, 2462}

\bibitem[\protect\citeauthoryear{Pillepich et~al.,}{Pillepich
  et~al.}{2018a}]{pillepich2018a}
Pillepich A.,  et~al., 2018a, Monthly Notices of the Royal Astronomical
  Society, 473, 4077

\bibitem[\protect\citeauthoryear{Pillepich et~al.,}{Pillepich
  et~al.}{2018b}]{pillepich2018b}
Pillepich A.,  et~al., 2018b, Monthly Notices of the Royal Astronomical
  Society, 475, 648

\bibitem[\protect\citeauthoryear{{Pillepich} et~al.,}{{Pillepich}
  et~al.}{2019}]{pillepich2019}
{Pillepich} A.,  et~al., 2019, \mn@doi [\mnras] {10.1093/mnras/stz2338}, \href
  {https://ui.adsabs.harvard.edu/abs/2019MNRAS.490.3196P} {490, 3196}

\bibitem[\protect\citeauthoryear{{Pillepich}, {Nelson}, {Truong}, {Weinberger},
  {Martin-Navarro}, {Springel}, {Faber}  \& {Hernquist}}{{Pillepich}
  et~al.}{2021}]{pillepich2021}
{Pillepich} A.,  {Nelson} D.,  {Truong} N.,  {Weinberger} R.,  {Martin-Navarro}
  I.,  {Springel} V.,  {Faber} S.~M.,   {Hernquist} L.,  2021, \mn@doi [\mnras]
  {10.1093/mnras/stab2779}, \href
  {https://ui.adsabs.harvard.edu/abs/2021MNRAS.508.4667P} {508, 4667}

\bibitem[\protect\citeauthoryear{{Planck Collaboration} et~al.,}{{Planck
  Collaboration} et~al.}{2016}]{planck2016}
{Planck Collaboration} et~al., 2016, Astronomy \& Astrophysics, 594, A13

\bibitem[\protect\citeauthoryear{{Ponnada} et~al.,}{{Ponnada}
  et~al.}{2022}]{ponnada2022}
{Ponnada} S.~B.,  et~al., 2022, \mn@doi [\mnras] {10.1093/mnras/stac2448},
  \href {https://ui.adsabs.harvard.edu/abs/2022MNRAS.516.4417P} {516, 4417}

\bibitem[\protect\citeauthoryear{{Ponti} et~al.,}{{Ponti}
  et~al.}{2022}]{ponti2022}
{Ponti} G.,  et~al., 2022, arXiv e-prints, \href
  {https://ui.adsabs.harvard.edu/abs/2022arXiv221003133P} {p. arXiv:2210.03133}

\bibitem[\protect\citeauthoryear{{Popping} et~al.,}{{Popping}
  et~al.}{2019}]{popping2019}
{Popping} G.,  et~al., 2019, \mn@doi [\apj] {10.3847/1538-4357/ab30f2}, \href
  {https://ui.adsabs.harvard.edu/abs/2019ApJ...882..137P} {882, 137}

\bibitem[\protect\citeauthoryear{Predehl et~al.,}{Predehl
  et~al.}{2020}]{predehl2020}
Predehl P.,  et~al., 2020, \mn@doi [Nature] {10.1038/s41586-020-2979-0}, 588,
  227

\bibitem[\protect\citeauthoryear{{Prochaska} et~al.,}{{Prochaska}
  et~al.}{2017}]{prochaska2017}
{Prochaska} J.~X.,  et~al., 2017, \mn@doi [\apj] {10.3847/1538-4357/aa6007},
  \href {https://ui.adsabs.harvard.edu/abs/2017ApJ...837..169P} {837, 169}

\bibitem[\protect\citeauthoryear{{Prochaska} et~al.,}{{Prochaska}
  et~al.}{2019}]{prochaska2019}
{Prochaska} J.~X.,  et~al., 2019, \mn@doi [Science] {10.1126/science.aay0073},
  \href {https://ui.adsabs.harvard.edu/abs/2019Sci...366..231P} {366, 231}

\bibitem[\protect\citeauthoryear{{Putman}, {Bland-Hawthorn}, {Veilleux},
  {Gibson}, {Freeman}  \& {Maloney}}{{Putman} et~al.}{2003}]{putman2003}
{Putman} M.~E.,  {Bland-Hawthorn} J.,  {Veilleux} S.,  {Gibson} B.~K.,
  {Freeman} K.~C.,   {Maloney} P.~R.,  2003, \mn@doi [\apj] {10.1086/378555},
  \href {https://ui.adsabs.harvard.edu/abs/2003ApJ...597..948P} {597, 948}

\bibitem[\protect\citeauthoryear{{Putman}, {Peek}  \& {Joung}}{{Putman}
  et~al.}{2012}]{putman2012}
{Putman} M.~E.,  {Peek} J.~E.~G.,   {Joung} M.~R.,  2012, \mn@doi [\araa]
  {10.1146/annurev-astro-081811-125612}, \href
  {https://ui.adsabs.harvard.edu/abs/2012ARA&A..50..491P} {50, 491}

\bibitem[\protect\citeauthoryear{{Qu} \& {Bregman}}{{Qu} \&
  {Bregman}}{2019}]{qu2019}
{Qu} Z.,  {Bregman} J.~N.,  2019, \mn@doi [\apj] {10.3847/1538-4357/ab2a0b},
  \href {https://ui.adsabs.harvard.edu/abs/2019ApJ...880...89Q} {880, 89}

\bibitem[\protect\citeauthoryear{{Qu}, {Bregman}, {Hodges-Kluck}, {Li}  \&
  {Lindley}}{{Qu} et~al.}{2020}]{qu2020}
{Qu} Z.,  {Bregman} J.~N.,  {Hodges-Kluck} E.,  {Li} J.-T.,   {Lindley} R.,
  2020, \mn@doi [\apj] {10.3847/1538-4357/ab774e}, \href
  {https://ui.adsabs.harvard.edu/abs/2020ApJ...894..142Q} {894, 142}

\bibitem[\protect\citeauthoryear{{Qu}, {Lindley}  \& {Bregman}}{{Qu}
  et~al.}{2022}]{qu2022}
{Qu} Z.,  {Lindley} R.,   {Bregman} J.~N.,  2022, \mn@doi [\apj]
  {10.3847/1538-4357/ac35cd}, \href
  {https://ui.adsabs.harvard.edu/abs/2022ApJ...924...86Q} {924, 86}

\bibitem[\protect\citeauthoryear{{Rahmati}, {Pawlik}, {Rai{\v{c}}evi{\'c}}  \&
  {Schaye}}{{Rahmati} et~al.}{2013}]{rahmati2013}
{Rahmati} A.,  {Pawlik} A.~H.,  {Rai{\v{c}}evi{\'c}} M.,   {Schaye} J.,  2013,
  \mn@doi [\mnras] {10.1093/mnras/stt066}, \href
  {https://ui.adsabs.harvard.edu/abs/2013MNRAS.430.2427R} {430, 2427}

\bibitem[\protect\citeauthoryear{{Salem}, {Besla}, {Bryan}, {Putman}, {van der
  Marel}  \& {Tonnesen}}{{Salem} et~al.}{2015}]{salem2015}
{Salem} M.,  {Besla} G.,  {Bryan} G.,  {Putman} M.,  {van der Marel} R.~P.,
  {Tonnesen} S.,  2015, \mn@doi [\apj] {10.1088/0004-637X/815/1/77}, \href
  {https://ui.adsabs.harvard.edu/abs/2015ApJ...815...77S} {815, 77}

\bibitem[\protect\citeauthoryear{{Sanchez}, {Werk}, {Tremmel}, {Pontzen},
  {Christensen}, {Quinn}  \& {Cruz}}{{Sanchez} et~al.}{2019}]{sanchez2019}
{Sanchez} N.~N.,  {Werk} J.~K.,  {Tremmel} M.,  {Pontzen} A.,  {Christensen}
  C.,  {Quinn} T.,   {Cruz} A.,  2019, \mn@doi [\apj]
  {10.3847/1538-4357/ab3045}, \href
  {https://ui.adsabs.harvard.edu/abs/2019ApJ...882....8S} {882, 8}

\bibitem[\protect\citeauthoryear{{Schneider}, {Ostriker}, {Robertson}  \&
  {Thompson}}{{Schneider} et~al.}{2020}]{schneider2020}
{Schneider} E.~E.,  {Ostriker} E.~C.,  {Robertson} B.~E.,   {Thompson} T.~A.,
  2020, \mn@doi [\apj] {10.3847/1538-4357/ab8ae8}, \href
  {https://ui.adsabs.harvard.edu/abs/2020ApJ...895...43S} {895, 43}

\bibitem[\protect\citeauthoryear{{Segers}, {Oppenheimer}, {Schaye}  \&
  {Richings}}{{Segers} et~al.}{2017}]{segers2017}
{Segers} M.~C.,  {Oppenheimer} B.~D.,  {Schaye} J.,   {Richings} A.~J.,  2017,
  \mn@doi [\mnras] {10.1093/mnras/stx1633}, \href
  {https://ui.adsabs.harvard.edu/abs/2017MNRAS.471.1026S} {471, 1026}

\bibitem[\protect\citeauthoryear{{Sharma}, {McCourt}, {Quataert}  \&
  {Parrish}}{{Sharma} et~al.}{2012}]{sharma2012}
{Sharma} P.,  {McCourt} M.,  {Quataert} E.,   {Parrish} I.~J.,  2012, \mn@doi
  [\mnras] {10.1111/j.1365-2966.2011.20246.x}, \href
  {https://ui.adsabs.harvard.edu/abs/2012MNRAS.420.3174S} {420, 3174}

\bibitem[\protect\citeauthoryear{Sijacki, Vogelsberger, Genel, Springel,
  Torrey, Snyder, Nelson  \& Hernquist}{Sijacki et~al.}{2015}]{sijacki2015}
Sijacki D.,  Vogelsberger M.,  Genel S.,  Springel V.,  Torrey P.,  Snyder
  G.~F.,  Nelson D.,   Hernquist L.,  2015, Monthly Notices of the Royal
  Astronomical Society, 452, 575

\bibitem[\protect\citeauthoryear{{Smith}, {Brickhouse}, {Liedahl}  \&
  {Raymond}}{{Smith} et~al.}{2001}]{smith2001}
{Smith} R.~K.,  {Brickhouse} N.~S.,  {Liedahl} D.~A.,   {Raymond} J.~C.,  2001,
  \mn@doi [\apjl] {10.1086/322992}, \href
  {https://ui.adsabs.harvard.edu/abs/2001ApJ...556L..91S} {556, L91}

\bibitem[\protect\citeauthoryear{{Soko{\l}owska}, {Mayer}, {Babul}, {Madau}  \&
  {Shen}}{{Soko{\l}owska} et~al.}{2016}]{sokolowska2016}
{Soko{\l}owska} A.,  {Mayer} L.,  {Babul} A.,  {Madau} P.,   {Shen} S.,  2016,
  \mn@doi [\apj] {10.3847/0004-637X/819/1/21}, \href
  {https://ui.adsabs.harvard.edu/abs/2016ApJ...819...21S} {819, 21}

\bibitem[\protect\citeauthoryear{{Sotillo-Ramos} et~al.,}{{Sotillo-Ramos}
  et~al.}{2022}]{sotillo2022}
{Sotillo-Ramos} D.,  et~al., 2022, \mn@doi [\mnras] {10.1093/mnras/stac2586},
  \href {https://ui.adsabs.harvard.edu/abs/2022MNRAS.516.5404S} {516, 5404}

\bibitem[\protect\citeauthoryear{{Sparre}, {Pfrommer}  \& {Ehlert}}{{Sparre}
  et~al.}{2020}]{sparre2020}
{Sparre} M.,  {Pfrommer} C.,   {Ehlert} K.,  2020, \mn@doi [\mnras]
  {10.1093/mnras/staa3177}, \href
  {https://ui.adsabs.harvard.edu/abs/2020MNRAS.499.4261S} {499, 4261}

\bibitem[\protect\citeauthoryear{Springel}{Springel}{2010}]{springel2010}
Springel V.,  2010, Monthly Notices of the Royal Astronomical Society, 401, 791

\bibitem[\protect\citeauthoryear{{Springel} \& {Hernquist}}{{Springel} \&
  {Hernquist}}{2003}]{springel2003}
{Springel} V.,  {Hernquist} L.,  2003, \mn@doi [\mnras]
  {10.1046/j.1365-8711.2003.06206.x}, \href
  {https://ui.adsabs.harvard.edu/abs/2003MNRAS.339..289S} {339, 289}

\bibitem[\protect\citeauthoryear{{Springel}, {White}, {Tormen}  \&
  {Kauffmann}}{{Springel} et~al.}{2001}]{springel2001}
{Springel} V.,  {White} S. D.~M.,  {Tormen} G.,   {Kauffmann} G.,  2001,
  \mn@doi [\mnras] {10.1046/j.1365-8711.2001.04912.x}, \href
  {https://ui.adsabs.harvard.edu/abs/2001MNRAS.328..726S} {328, 726}

\bibitem[\protect\citeauthoryear{Springel et~al.,}{Springel
  et~al.}{2018}]{springel2018}
Springel V.,  et~al., 2018, Monthly Notices of the Royal Astronomical Society,
  475, 676

\bibitem[\protect\citeauthoryear{{Stern}, {Hennawi}, {Prochaska}  \&
  {Werk}}{{Stern} et~al.}{2016}]{stern2016}
{Stern} J.,  {Hennawi} J.~F.,  {Prochaska} J.~X.,   {Werk} J.~K.,  2016,
  \mn@doi [\apj] {10.3847/0004-637X/830/2/87}, \href
  {https://ui.adsabs.harvard.edu/abs/2016ApJ...830...87S} {830, 87}

\bibitem[\protect\citeauthoryear{{Stern}, {Fielding}, {Faucher-Gigu{\`e}re}  \&
  {Quataert}}{{Stern} et~al.}{2019}]{stern2019}
{Stern} J.,  {Fielding} D.,  {Faucher-Gigu{\`e}re} C.-A.,   {Quataert} E.,
  2019, \mn@doi [\mnras] {10.1093/mnras/stz1859}, \href
  {https://ui.adsabs.harvard.edu/abs/2019MNRAS.488.2549S} {488, 2549}

\bibitem[\protect\citeauthoryear{{Stocke}, {Keeney}, {Danforth}, {Shull},
  {Froning}, {Green}, {Penton}  \& {Savage}}{{Stocke}
  et~al.}{2013}]{stocke2013}
{Stocke} J.~T.,  {Keeney} B.~A.,  {Danforth} C.~W.,  {Shull} J.~M.,  {Froning}
  C.~S.,  {Green} J.~C.,  {Penton} S.~V.,   {Savage} B.~D.,  2013, \mn@doi
  [\apj] {10.1088/0004-637X/763/2/148}, \href
  {https://ui.adsabs.harvard.edu/abs/2013ApJ...763..148S} {763, 148}

\bibitem[\protect\citeauthoryear{{Strong}, {Moskalenko}  \& {Reimer}}{{Strong}
  et~al.}{2000}]{strong2000}
{Strong} A.~W.,  {Moskalenko} I.~V.,   {Reimer} O.,  2000, \mn@doi [\apj]
  {10.1086/309038}, \href
  {https://ui.adsabs.harvard.edu/abs/2000ApJ...537..763S} {537, 763}

\bibitem[\protect\citeauthoryear{Su, Slatyer  \& Finkbeiner}{Su
  et~al.}{2010}]{su2010}
Su M.,  Slatyer T.~R.,   Finkbeiner D.~P.,  2010, \mn@doi [ApJ]
  {10.1088/0004-637x/724/2/1044}, 724, 1044

\bibitem[\protect\citeauthoryear{{Terrazas} et~al.,}{{Terrazas}
  et~al.}{2020}]{terrazas2020}
{Terrazas} B.~A.,  et~al., 2020, \mn@doi [\mnras] {10.1093/mnras/staa374},
  \href {https://ui.adsabs.harvard.edu/abs/2020MNRAS.493.1888T} {493, 1888}

\bibitem[\protect\citeauthoryear{{Truong} et~al.,}{{Truong}
  et~al.}{2020}]{truong2020}
{Truong} N.,  et~al., 2020, \mn@doi [\mnras] {10.1093/mnras/staa685}, \href
  {https://ui.adsabs.harvard.edu/abs/2020MNRAS.494..549T} {494, 549}

\bibitem[\protect\citeauthoryear{{Truong}, {Pillepich}  \& {Werner}}{{Truong}
  et~al.}{2021a}]{truong2021a}
{Truong} N.,  {Pillepich} A.,   {Werner} N.,  2021a, \mn@doi [\mnras]
  {10.1093/mnras/staa3880}, \href
  {https://ui.adsabs.harvard.edu/abs/2021MNRAS.501.2210T} {501, 2210}

\bibitem[\protect\citeauthoryear{{Truong}, {Pillepich}, {Nelson}, {Werner}  \&
  {Hernquist}}{{Truong} et~al.}{2021b}]{truong2021}
{Truong} N.,  {Pillepich} A.,  {Nelson} D.,  {Werner} N.,   {Hernquist} L.,
  2021b, \mn@doi [\mnras] {10.1093/mnras/stab2638}, \href
  {https://ui.adsabs.harvard.edu/abs/2021MNRAS.508.1563T} {508, 1563}

\bibitem[\protect\citeauthoryear{{Tumlinson}, {Peeples}  \& {Werk}}{{Tumlinson}
  et~al.}{2017}]{tumlinson2017}
{Tumlinson} J.,  {Peeples} M.~S.,   {Werk} J.~K.,  2017, \mn@doi [\araa]
  {10.1146/annurev-astro-091916-055240}, \href
  {https://ui.adsabs.harvard.edu/abs/2017ARA&A..55..389T} {55, 389}

\bibitem[\protect\citeauthoryear{Vogelsberger et~al.,}{Vogelsberger
  et~al.}{2014a}]{vogelsberger2014b}
Vogelsberger M.,  et~al., 2014a, Monthly Notices of the Royal Astronomical
  Society, 444, 1518

\bibitem[\protect\citeauthoryear{Vogelsberger et~al.,}{Vogelsberger
  et~al.}{2014b}]{vogelsberger2014a}
Vogelsberger M.,  et~al., 2014b, Nature, 509, 177

\bibitem[\protect\citeauthoryear{{Voit}, {Meece}, {Li}, {O'Shea}, {Bryan}  \&
  {Donahue}}{{Voit} et~al.}{2017}]{voit2017}
{Voit} G.~M.,  {Meece} G.,  {Li} Y.,  {O'Shea} B.~W.,  {Bryan} G.~L.,
  {Donahue} M.,  2017, \mn@doi [\apj] {10.3847/1538-4357/aa7d04}, \href
  {https://ui.adsabs.harvard.edu/abs/2017ApJ...845...80V} {845, 80}

\bibitem[\protect\citeauthoryear{{Wakker} \& {van Woerden}}{{Wakker} \& {van
  Woerden}}{1997}]{wakker1997}
{Wakker} B.~P.,  {van Woerden} H.,  1997, \mn@doi [\araa]
  {10.1146/annurev.astro.35.1.217}, \href
  {https://ui.adsabs.harvard.edu/abs/1997ARA&A..35..217W} {35, 217}

\bibitem[\protect\citeauthoryear{{Weinberger} et~al.,}{{Weinberger}
  et~al.}{2017}]{weinberger2017}
{Weinberger} R.,  et~al., 2017, \mn@doi [\mnras] {10.1093/mnras/stw2944}, \href
  {https://ui.adsabs.harvard.edu/abs/2017MNRAS.465.3291W} {465, 3291}

\bibitem[\protect\citeauthoryear{{Werk} et~al.,}{{Werk}
  et~al.}{2016}]{werk2016}
{Werk} J.~K.,  et~al., 2016, \mn@doi [\apj] {10.3847/1538-4357/833/1/54}, \href
  {https://ui.adsabs.harvard.edu/abs/2016ApJ...833...54W} {833, 54}

\bibitem[\protect\citeauthoryear{{Werk} et~al.,}{{Werk}
  et~al.}{2019}]{werk2019}
{Werk} J.~K.,  et~al., 2019, \mn@doi [\apj] {10.3847/1538-4357/ab54cf}, \href
  {https://ui.adsabs.harvard.edu/abs/2019ApJ...887...89W} {887, 89}

\bibitem[\protect\citeauthoryear{{Yao}, {Shull}  \& {Danforth}}{{Yao}
  et~al.}{2011}]{yao2011}
{Yao} Y.,  {Shull} J.~M.,   {Danforth} C.~W.,  2011, \mn@doi [\apjl]
  {10.1088/2041-8205/728/1/L16}, \href
  {https://ui.adsabs.harvard.edu/abs/2011ApJ...728L..16Y} {728, L16}

\bibitem[\protect\citeauthoryear{{Yun} et~al.,}{{Yun} et~al.}{2019}]{yun2019}
{Yun} K.,  et~al., 2019, \mn@doi [\mnras] {10.1093/mnras/sty3156}, \href
  {https://ui.adsabs.harvard.edu/abs/2019MNRAS.483.1042Y} {483, 1042}

\bibitem[\protect\citeauthoryear{{Zech}, {Lehner}, {Howk}, {Dixon}  \&
  {Brown}}{{Zech} et~al.}{2008}]{zech2008}
{Zech} W.~F.,  {Lehner} N.,  {Howk} J.~C.,  {Dixon} W. V.~D.,   {Brown} T.~M.,
  2008, \mn@doi [\apj] {10.1086/587135}, \href
  {https://ui.adsabs.harvard.edu/abs/2008ApJ...679..460Z} {679, 460}

\bibitem[\protect\citeauthoryear{{Zhu} et~al.,}{{Zhu} et~al.}{2014}]{zhu2014}
{Zhu} G.,  et~al., 2014, \mn@doi [\mnras] {10.1093/mnras/stu186}, \href
  {https://ui.adsabs.harvard.edu/abs/2014MNRAS.439.3139Z} {439, 3139}

\bibitem[\protect\citeauthoryear{{Zinger} et~al.,}{{Zinger}
  et~al.}{2020}]{zinger2020}
{Zinger} E.,  et~al., 2020, \mn@doi [\mnras] {10.1093/mnras/staa2607}, \href
  {https://ui.adsabs.harvard.edu/abs/2020MNRAS.499..768Z} {499, 768}

\bibitem[\protect\citeauthoryear{{de Graaff}, {Cai}, {Heymans}  \&
  {Peacock}}{{de Graaff} et~al.}{2019}]{degraff2019}
{de Graaff} A.,  {Cai} Y.-C.,  {Heymans} C.,   {Peacock} J.~A.,  2019, \mn@doi
  [\aap] {10.1051/0004-6361/201935159}, \href
  {https://ui.adsabs.harvard.edu/abs/2019A&A...624A..48D} {624, A48}

\bibitem[\protect\citeauthoryear{{van de Voort}, {Bieri}, {Pakmor},
  {G{\'o}mez}, {Grand}  \& {Marinacci}}{{van de Voort}
  et~al.}{2021}]{voort2021}
{van de Voort} F.,  {Bieri} R.,  {Pakmor} R.,  {G{\'o}mez} F.~A.,  {Grand} R.
  J.~J.,   {Marinacci} F.,  2021, \mn@doi [\mnras] {10.1093/mnras/staa3938},
  \href {https://ui.adsabs.harvard.edu/abs/2021MNRAS.501.4888V} {501, 4888}

\makeatother
\end{thebibliography}

\end{document}